\documentclass[11pt]{article}

\usepackage{amssymb,amsmath,amsfonts,amssymb,numbysec}
\usepackage{graphics,graphicx,color}
\hsize \textwidth
\advance \hsize by -\marginparwidth
\evensidemargin \oddsidemargin
\usepackage{amssymb}
\advance\hoffset by 5mm

\makeatletter \@addtoreset{equation}{section}

\def\@abssec#1{\vspace{.05in}\footnotesize \parindent .2in
{\bf #1. }\ignorespaces}
\def\proof{\par{\it Proof}. \ignorespaces}
\def\endproof{{\ \vbox{\hrule\hbox{%
   \vrule height1.3ex\hskip0.8ex\vrule}\hrule
  }}\par}

\newtheorem{theorem}{Theorem}[section]
\newtheorem{lemma}[theorem]{Lemma}
\newtheorem{proposition}[theorem]{Proposition}
\newtheorem{corollary}[theorem]{Corollary}

\newtheorem{remark}[theorem]{Remark}

\def \Rm {\mathbb R}

\newcommand{\eps}{\varepsilon}

\newcommand{\dsum}{\displaystyle\sum}
\newcommand{\dint}{\displaystyle\int}
\newcommand{\disp}{\displaystyle}
\newcommand{\pdr}[2]{\dfrac{\partial{#1}}{\partial{#2}}}

\newcommand{\dr}[2]{\dfrac{d{#1}}{d{#2}}}

\newcommand{\bk}{\mathbf k} 
\newcommand{\bx}{\mathbf x} 
\newcommand{\vu}{\mathbf u}\newcommand{\vy}{\mathbf y}
\newcommand{\vz}{\mathbf z}
\newcommand{\bz}{\mathbf z} \newcommand{\vx}{\mathbf x}
\newcommand{\be}{\mathbf e} 
\newcommand{\ve}{{\bf e}}
\newcommand{\bp}{\mathbf p} \newcommand{\bq}{\mathbf q}
\newcommand{\bu}{\mathbf u} \newcommand{\bv}{\mathbf v}
\newcommand{\bD}{\mathbf D} 
 
 \newcommand{\bQ}{\mathbf Q}
\newcommand{\bX}{\mathbf X} \newcommand{\bK}{\mathbf K}
\newcommand{\vX}{{\bf X}}\newcommand{\vb}{{\bf b}}
\newcommand{\bZ}{\mathbf Z}

\newcommand{\bxi}{\boldsymbol \xi}\newcommand{\vxi}{\boldsymbol \xi}
\newcommand{\vhatk}{\hat{\bk}}
\newcommand{\tr}[1]{\mbox{Tr}({#1})}

\newcommand{\commentout}[1]{}
\newcommand{\bes}{\begin{displaymath}}
\newcommand{\ees}{\end{displaymath}}
\newcommand{\ba}{\begin{eqnarray}}
\newcommand{\ea}{\end{eqnarray}}
\newcommand{\bas}{\begin{eqnarray*}}
\newcommand{\eas}{\end{eqnarray*}}

\newcommand{\bcE}{{\bf E}}

\newcommand{\bB}{{\bf B}}
\newcommand{\bl}{{\bml}}

\newcommand{\bze}{{\bf 0}}

\newcommand{\bt}{\beta}

\newcommand{\bby}{{\bf y}}
\newcommand{\by}{{\bmy}}

\newcommand{\bbi}{{\bf i}}
\newcommand{\bbj}{{\bf j}}
\newcommand{\bbl}{{\bf l}}

\newcommand{\R}{{\mathbb R}}
\newcommand{\bbP}{{\mathbb P}}
\newcommand{\E}{{\mathbb E}}

\newcommand{\bbz}{{\bf z}}

\newcommand{\bbk}{{\bf k}}

\newcommand{\vphi}{\varphi}

\newcommand{\si}{\sigma}
\newcommand{\ga}{\gamma}
\newcommand{\al}{\alpha}

\newcommand{\la}{\lambda}

\newcommand{\bbx}{{\bf x}}
\newcommand{\ep}{\varepsilon}

\newcommand{\da}{\downarrow0}

\newcommand{\bE}{\mathbb{E}}

\newcommand{\vl}{{\bf l}}

\newcommand{\Om}{\Omega}
\newcommand{\om}{\omega}

\newcommand{\nwc}{\newcommand}

\nwc{\beq}{\begin{eqnarray}} \nwc{\beqn}{\begin{eqnarray*}}
\nwc{\beqast}{\begin{eqnarray*}} \nwc{\bm}{\boldmath}
\nwc{\eeq}{\end{eqnarray}} \nwc{\eeqn}{\end{eqnarray*}}
\nwc{\eeqast}{\end{eqnarray*}}

\nwc{\veps}{\varepsilon} \nwc{\ie}{\mbox{e}} \nwc{\ibi}{\mbox{i}}

\nwc{\m}{\mbox} \nwc{\re}{\hbox{Re}}
\nwc{\lamb}{\lambda_\varepsilon} \nwc{\ls}{\stackrel{<}{\sim}}
\nwc{\gs}{\stackrel{>}{\sim}} \nwc{\ubm}{\unboldmath}
\nwc{\cls}{{\cal L}^s} \nwc{\mt}{\bar{t}} \nwc{\bla}{\m{\bm
$\lambda$\ubm}} \nwc{\bxsi}{\m{\bm $\xi$\ubm}} \nwc{\bpsi}{\m{\bm
$\psi$\ubm}} \nwc{\bmeta}{\m{\bm $\eta$\ubm}} \nwc{\bma}{\m{\bm
$a$\ubm}} \nwc{\bmb}{\m{\bm $b$\ubm}} \nwc{\bmc}{\m{\bm $c$\ubm}}
\nwc{\bmd}{\m{\bm $d$\ubm}} \nwc{\bme}{\m{\bm $e$\ubm}}\nwc{\bmL}{\m{\bm $L$\ubm}}
\nwc{\bmf}{\m{\bm $f$\ubm}} \nwc{\bmg}{\m{\bm $g$\ubm}}
\nwc{\bmh}{\m{\bm $h$\ubm}} \nwc{\bmi}{\m{\bm $i$\ubm}}
\nwc{\bmj}{\m{\bm $j$\ubm}} \nwc{\bmk}{\m{\bm $k$\ubm}}
\nwc{\bml}{\m{\bm $l$\ubm}} \nwc{\bmn}{\m{\bm $n$\ubm}}
\nwc{\bmo}{\m{\bm $o$\ubm}} \nwc{\bmp}{\m{\bm $p$\ubm}}
\nwc{\bmq}{\m{\bm $q$\ubm}} \nwc{\bmr}{\m{\bm $r$\ubm}}
\nwc{\bms}{\m{\bm $s$\ubm}} \nwc{\bmt}{\m{\bm $t$\ubm}}
\nwc{\bmu}{\m{\bm $u$\ubm}} \nwc{\bmv}{\m{\bm $v$\ubm}}
\nwc{\bmw}{\m{\bm $w$\ubm}} \nwc{\bmx}{\m{\bm $x$\ubm}}
\nwc{\bmxt}{\m{\bm $x$\ubm}^\varepsilon (t)} \nwc{\bmy}{\m{\bm
$y$\ubm}} \nwc{\bmz}{\m{\bm $z$\ubm}}

\nwc{\bmX}{\m{\bm $X$\ubm}} \nwc{\bmR}{\m{\bm $R$\ubm}}
\nwc{\bmU}{\m{\bm $U$\ubm}} \nwc{\bmE}{\m{\bm $E$\ubm}}
\nwc{\bmF}{\m{\bm $F$\ubm}} \nwc{\bmH}{\m{\bm $H$\ubm}}
\nwc{\bmI}{\m{\bm $I$\ubm}} \nwc{\bmP}{\m{\bm $P$\ubm}}
\nwc{\bmM}{\m{\bm $M$\ubm}} \nwc{\bmJ}{\m{\bm $J$\ubm}}
\nwc{\bmA}{\m{\bm $A$\ubm}} \nwc{\bmD}{\m{\bm $D$\ubm}}
\nwc{\bS}{{\bf S}}
\nwc{\bmtheta}{\m{\bm $\theta$\ubm}} \nwc{\bmnu}{\m{\bm
$\nu$\ubm}} \nwc{\bmomega}{\m{\bm $\omega$\ubm}}
\nwc{\bmsigma}{\m{\bm $\sigma$\ubm}} \nwc{\bmnabla}{\m{\bm
$\nabla$\ubm}} \nwc{\bmLambda}{\m{\bm $\Lambda$\ubm}}
\nwc{\bmlambda}{\m{\bm $\lambda$\ubm}} \nwc{\bmtau}{\m{\bm
$\tau$\ubm}} \nwc{\bmPhi}{\m{\bm $\Phi$\ubm}} \nwc{\bmphi}{\m{\bm
$\phi$\ubm}} \nwc{\bmGamma}{\m{\bm $\Gamma$\ubm}}
\nwc{\bmgamma}{\m{\bm $\gamma$\ubm}}

\nwc{\ca}{{\cal A}}
\nwc{\ii}{{\mbox{i}}} \nwc{\cao}{{\cal A}^{-1}} \nwc{\cb}{{\cal
B}} \nwc{\cc}{{\cal C}} \nwc{\cd}{{\cal D}} \nwc{\bone}{{\bf 1}}
\nwc{\ce}{{\cal E}} \nwc{\cf}{{\cal F}} \nwc{\cg}{{\cal G}}
\nwc{\vV}{{\bf V}} \nwc{\ch}{{\cal H}} \nwc{\ci}{{\cal I}}
\nwc{\cj}{{\cal J}} \nwc{\ck}{{\cal K}} \nwc{\cl}{{\cal L}}
\nwc{\cle}{{\cal L}^\varepsilon} \nwc{\clu}{{\cal L}{\cal U}}
\nwc{\cm}{{\cal M}} \nwc{\cn}{{\cal N}} \nwc{\co}{{\cal O}}
\nwc{\cp}{{\cal P}} \nwc{\cpt}{{\cal P}^\varepsilon_t}
\nwc{\cq}{{\cal Q}} \nwc{\calr}{{\cal R}} \nwc{\cs}{{\cal S}}
\nwc{\ct}{{\cal T}} \nwc{\cu}{{\cal U}} \nwc{\cv}{{\cal V}}
\nwc{\cw}{{\cal W}} \nwc{\cy}{{\cal Y}} \nwc{\cz}{{\cal Z}}
\nwc{\ob}{{\cal B(\Om)}}
\nwc{\bbE}{\mathbb{E}}
\nwc{\bbZ}{\mathbb{Z}^d}

\nwc{\uP}{{\em \bf Proof: }} \nwc{\uT}{\underline{Theorem:}}

\renewcommand{\arraystretch}{1.5}

\title{Self-averaging of Wigner transforms in random media}

\author{Guillaume Bal\thanks{Department of Applied Physics \& Applied
    Mathematics, Columbia University, New York, NY 10027, USA;
    {e-mail: gb2030@columbia.edu}} \and Tomasz Komorowski
  \thanks{Institute of Mathematics, UMCS, pl. Marii Curie Sk\l
    odowskiej 1, 20-031 Lublin, Poland; {e-mail:
      komorow@golem.umcs.lublin.pl}} \and Lenya Ryzhik
  \thanks{Department of Mathematics, University of Chicago, Chicago,
    IL 60637, USA; e-mail: ryzhik@math.uchicago.edu}}

\begin{document}

\maketitle
\numberbysection

\begin{abstract}
We establish the self-averaging properties of the Wigner transform
of a mixture of states in the regime when the correlation length
of the random medium is much longer than the wave length but much
shorter than the propagation distance. The main ingredients in the
proof are the error estimates for the semiclassical approximation
of the Wigner transform by the solution of the Liouville
equations, and the limit theorem for two-particle motion along the
characteristics of the Liouville equations. The results are
applied to a mathematical model of the time-reversal experiments
for the acoustic waves, and self-averaging properties of the
re-transmitted wave are proved.
\end{abstract}


\renewcommand{\thefootnote}{\fnsymbol{footnote}}
\renewcommand{\thefootnote}{\arabic{footnote}}

\renewcommand{\arraystretch}{1.1}





\section{Introduction}
\label{sec:intro}

\subsection{The Wigner transform of mixtures of states}

The Wigner transform is a useful tool in the analysis of the
semi-classical limits of non-dissipative evolution equations as
well as in the high frequency wave propagation
\cite{GMMP,LP,RPK-WM}. It is defined as follows: given a family of
functions $f_\eps(t,\vx)$ uniformly bounded in
$L^\infty([0,T];L^2({\mathbb R}^d))$ its Wigner transform is
\begin{equation}
  \label{eq:defwigner}
\tilde W_\eps(t,\vx,\bk)=\int
e^{i\bk\cdot\vy}f_\eps(t,\vx-\frac{\eps\vy}{2})
f_\eps^*(t,\vx+\frac{\eps\vy}{2})\frac{d\vy}{(2\pi)^d}.
\end{equation}
The family $\tilde W_\eps$ is uniformly bounded in the space of
Schwartz distributions ${\cal S}'({\mathbb R}^d\times {\mathbb
R}^d)$, and all its limit points are non-negative measures of
bounded total mass \cite{GMMP,LP}. It is customary to interpret a
limit Wigner measure $W$ as the energy density in the phase space,
since the limit points of $n_\eps=|f_\eps|^2$ are of the form
$n(t,\vx)=\int W(t,\vx,\bk)d\bk$ provided that the family $f_\eps$
is $\eps$-oscillatory and compact at infinity \cite{GMMP}.
However, while neither $n_\eps$ nor its limit $n(t,\vx)$ satisfy a
closed equation, both $\tilde W_\eps$ and $W$ usually obey an
evolution equation when the family $f_\eps(t,\vx)$ arises from a
time-dependent PDE. This makes the Wigner transform a useful tool
in the study of semiclassical and high frequency limits,
especially in random media
\cite{BPR-Nonlin,BPR-SD,Erdos-Yau,Poupaud,RPK-WM}. However, a
priori bounds on the Wigner transform $\tilde W_\eps$ other than
those mentioned above are usually difficult to obtain. It has been
observed in \cite{LP} that the Wigner transform of a mixture of
states
\[
W_\eps(\vx,\bk)=\int e^{i\bk\cdot\vy}f_\eps(\vx-\frac{\eps\vy}{2};\zeta)
f_\eps^*(\vx+\frac{\eps\vy}{2};\zeta)\frac{d\vy d\mu(\zeta)}{(2\pi)^d},
\]
enjoys better regularity properties. The family $f_\eps$ above depends
on an additional ``state'' parameter $\zeta\in S$, where $S$ is a
state space equipped with a non-negative bounded measure
$d\mu(\zeta)$.  Typically this corresponds to introducing random
initial data for $f_\eps$ at $t=0$ and estimating the expectation of
$\tilde W_\eps$ with respect to this randomness. This improved
regularity has been used, for instance, in \cite{Poupaud,Spohn} in the
analysis of the average of the Wigner transform of mixtures of states
in random media and in \cite{LP} in order to obtain an asymptotic
expansion for the Wigner transform of a mixture of states.

The purpose of this paper is to analyze the self-averaging properties
of moments of the mixed Wigner transform of the form $\int
W_\eps(t,\vx,\bk)S(\bk)d\bk$, where $S(\bk)$ is a test function, and
the family $f_\eps(t,\vx;\zeta)$ satisfies the acoustic equations.
This problem arises naturally in the mathematical study of the
experiments in time-reversal of acoustic waves that we will describe
in detail below. However, apart from the time-reversal application,
the statistical stability of such moments provides an important key to
understanding the physical applicability of the limit equations for
the Wigner transform in random media in the situations when results
for each realization are more relevant than the statistically averaged
quantities.

We start with the wave equation in dimension $d\ge 3$
\begin{equation}\label{eq-wave}
\frac{1}{c^2(\vx)}\frac{\partial^2\phi}{\partial t^2}-\Delta\phi=0
\end{equation}
and assume that the wave speed has the form
$c(\vx)=c_0+\sqrt{\delta}c_1(\vx)$. Here $c_0>0$ is the constant sound
speed of the uniform background medium, while the small parameter
$\delta\ll 1$ measures the strength of the mean zero random
perturbation $c_1$. Rescaling the spatial and temporal variables
$\vx=\vx'/\delta$ and $t=t'/\delta$ we obtain (after dropping the
primes) equation (\ref{eq-wave}) with rapidly fluctuating wave speed
\begin{equation}\label{c-fluct-delta}
c_\delta(\vx)=c_0+\sqrt{\delta}c_1\left(\frac{\vx}{\delta}\right).
\end{equation}
It is convenient to re-write (\ref{eq-wave}) as the system of acoustic
equations for the ``pressure'' $p=\dfrac{1}{c}\phi_t$ and ``acoustic velocity''
$\vu=-\nabla \phi$:
\begin{eqnarray}\label{acoust-eq-symm}
&&\pdr{\vu}{t}+\nabla\left(c_\delta(\vx)p\right)=0\\
&&\pdr{p}{t}+
c_\delta(\vx)\nabla\cdot\vu=0.\nonumber
\end{eqnarray}
The energy density for (\ref{acoust-eq-symm}) is
$E(t,\vx)=|\vu|^2+p^2$: $\int E(t,\vx)d\vx=\hbox{const}$ is
independent of time. We will denote for brevity
$\bv=(\bu,p)\in{\mathbb C}^{d+1}$ and write (\ref{acoust-eq-symm})
in the more general form of a first order linear symmetric hyperbolic
system
\begin{equation}\label{symm-version-intro}
\pdr{\bv_{\eps}^{\delta}}{t}+A_\delta(\vx)D^j\pdr{}{x^j}
\left(A_\delta(\vx)\bv_{\eps}^{\delta}(\vx)\right)=0.
\end{equation}
In the present case, the symmetric matrices $A_\delta$ and $D^j$ are
defined by
\begin{equation}
  \label{eq:symmatADj}
  A_\delta(\bx)=\hbox{diag}(1,1,1,c_\delta(\vx)), \qquad \mbox{ and } \qquad
  D^j={\bf e}_j\otimes{\bf e}_{d+1}+{\bf e}_{d+1}\otimes{\bf e}_j,\,\,
  j=1,\dots,d.
\end{equation}
Notice that the matrices $D^j$ are independent of $\vx$.
Here ${\bf e}_m\in{\mathbb R}^{d+1}$ is the standard orthonormal
basis: $({\bf e}_m)_k=\delta_{mk}$. The dispersion matrix for
(\ref{symm-version-intro}) is
\begin{equation}\label{P0-intro}
P_0^\delta(\vx,\bk)=iA_\delta(\vx)k_jD^jA_\delta(\vx)=ic_\delta(\vx)k_jD^j=
ic_\delta(\vx)\left[\tilde\bk\otimes{\bf e}_{d+1}+{\bf
e}_{d+1}\otimes\tilde\bk\right],~~\tilde\bk=\sum_{j=1}^dk_j{\bf
e}_j.
\end{equation}
The self-adjoint matrix $(-iP_0^\delta)$  has an eigenvalue $\lambda_0=0$
of multiplicity $d-1$, and two simple eigenvalues
$\lambda_{\pm}^\delta(\vx,\bk)=\pm c_\delta(\vx)|\bk|$. The
corresponding eigenvectors are
\begin{equation}
\label{eq:eigbs}
\vb_{m}^0=\left(\bk_m^\perp,0\right),~m=1,\dots,d-1;~~
\vb_\pm=\frac{1}{\sqrt{2}}\left(\frac{\tilde\bk}{|\bk|}\pm{\bf
e}_{d+1}\right),
\end{equation}
where $\bk_m^\perp\in{\mathbb R}^d$ is the orthonormal basis of
vectors orthogonal to $\bk$.

We assume that the initial data
$\bv_0(\vx;\zeta)=\bv_\eps^\delta(0,\vx;\zeta)=
(-\eps\nabla\phi_0^\eps,1/c_\delta\dot\phi_0^\eps)$
for (\ref{symm-version-intro}) is an $\eps$-oscillatory and compact at
infinity family of functions uniformly bounded in $L^2({\mathbb R}^d)$
\cite{GMMP} for each ``realization'' $\zeta$ of the initial data. The
scale $\eps$ of oscillations is much smaller than the correlation
length $\delta$ of the medium: $\eps\ll \delta\ll 1$. The
$(d+1)\times(d+1)$ Wigner matrix of a mixture of solutions of
(\ref{symm-version-intro}) is defined by
\[
W_\eps^\delta(t,\vx,\bk)=\int\limits_{{\mathbb R}^d\times S}
e^{i\bk\cdot\vy}
\bv_\eps^\delta(t,\vx-\frac{\eps\vy}{2};\zeta)
\bv_\eps^{\delta*}(t,\vx+\frac{\eps\vy}{2};\zeta)
\frac{d\vy d\mu(\zeta)}{(2\pi)^d}.
\]
The non-negative measure $d\mu$ has bounded total mass:
$\int_Sd\mu(\zeta)<\infty$. It is well-known \cite{GMMP,LP} that for
each fixed $\delta>0$ (and even without introduction of a mixture of
states) one may pass to the limit $\eps\to 0$ and show that
$W_\eps^\delta$ converges weakly in ${\cal S}'({\mathbb
  R}^d\times{\mathbb R}^d)$ to
\[
\bar
W^\delta(t,\vx,\bk)=u_+^\delta(t,\vx,\bk)\vb_+(\bk)\otimes\vb_+(\bk)
+u_-^\delta(t,\vx,\bk)\vb_-(\bk)\otimes\vb_-(\bk).
\]
The scalar amplitudes $u_\pm^\delta$ satisfy the Liouville
equations:
\begin{eqnarray}\label{liouv-intro}
\pdr{u_\pm^\delta}{t}+\nabla_\bk\lambda_\pm^\delta\cdot\nabla_\vx
u_\pm^\delta-\nabla_\vx\lambda_\pm^\delta\cdot\nabla_\bk
u_\pm^\delta=0.
\end{eqnarray}
Furthermore, one may formally pass to the limit $\delta\to 0$ in
(\ref{liouv-intro}) and show that (see \cite{BR-CRAS})
$\E\left\{u_\pm^\delta\right\}$ converge to the solution of
\begin{equation}\label{diff-intro}
\pdr{\bar u_\pm}{t}\pm c_0\vhatk\cdot\nabla_\vx\bar
u_\pm=\pdr{}{k_m}\left[|\bk|^2D_{mn}(\vhatk)\pdr{\bar
u_\pm}{k_n}\right].
\end{equation}
Here $\vhatk =\bk/|\bk|$, and the diffusion matrix $D$ is given by
\begin{equation}\label{diff-matrix}
D_{mn}=\frac{1}{2}\int_{-\infty}^\infty\frac{\partial^2
R(c_0s\vhatk)}{\partial x_n\partial x_m}ds,
\end{equation}
where $R(\vx)$ is the correlation function of $c_1$:
$\E\left\{c_1(\vy)c_1(\vx+\vy)\right\}=R(\vx)$.

The purpose of this paper is to make the passage to the limit
$\eps,\delta\to 0$ rigorous for a mixture of states (and eliminate the
consecutive limits $\eps\to0$ then $\delta\to0$) and establish the
self-averaging properties of moments of the form
$s_\eps^\delta(t,\vx)=\int W_\eps^\delta(t,\vx,\bk)S(\bk)d\bk$, where
$S\in L^2({\mathbb R}^d)$ is a given test function.

The assumption that $\eps\ll\delta$ is formalized as follows. We let
${\cal K}_\mu=\left\{(\eps,\delta):~~\delta\ge
  |\ln\eps|^{-2/3+\mu}\right\}$, with $0<\mu<2/3$ and assume that
$(\eps,\delta)\in{\cal K}_\mu$ for some $\mu\in(0,2/3)$. From now on,
$\mu$ is a given fixed number in $(0,2/3)$.

\subsection{The random medium}\label{sec-rand-assump}

We make the following assumptions on the random field $c_1(\vx)$.
Let  $(\Om,{\cal C},
\mathbb P)$ be a certain probability space, and let $\bbE$ denote the
expectation with respect to $\bbP$ and $\|\cdot\|_p$ denote the
respective $L^p$ norm for any $p\in[1,+\infty]$. We suppose further
that $ c_1:\R^d\times\Om\rightarrow\R$ is a measurable, strictly
stationary, mean-zero random field, that is pathwise $C^4$-smooth and
satisfies
\begin{equation}\label{d-i}
D_i:=\mathop{\mbox{ess-sup}}\limits_{\om\in \Om}|\nabla_\bbx^i
c_1(\bbx;\om)|<+\infty,\quad i=0,1,\dots,4.
\end{equation}
We assume in addition that $c_1$ is exponentially $\phi$-mixing.
More precisely,  for any $R>0$ we let ${\cal
C}_{R}^i:=\si\{c_1(\bbx):\,|\bbx|\leq R\}$ and ${\cal
C}_{R}^e:=\si\{c_1(\bbx):\,|\bbx|\geq R\}$. We also define
\[
\phi(\rho):=\sup[\,|\mathbb P(B)-\mathbb P(B|A)|:\,R>0,\,A\in
{\cal C}_{R}^i,\,B\in {\cal C}_{R+\rho}^e\,],
\]
for all $\rho>0$. We suppose that there exists a constant $C_1>0$ such
that
\begin{equation}\label{DR}
\phi(\rho)\leq 2\ie^{-C_1\rho},\quad\forall\,\rho>0.
\end{equation}
We let also
\[
R(\bby)=\bbE[c_1(\bby)c_1(\bze)],\quad\bby\in\R^d
\]
be the covariance function of the field $c_1(\cdot)$ and
note that (\ref{DR}) implies that there exists a constant $C_2>0$
such
that
\begin{equation}\label{53102-intro}
|\nabla^m_\bby R(\bby)|\leq C_2\ie^{-|\bby|/C_2},
\quad\forall\,\bby\in\R^d,\,m=0,\cdots,4.
\end{equation}
Finally we assume that $R\in C^\infty({\mathbb R}^d)$, this condition
will be used only to establish the hypoellipticity of
(\ref{diff-intro}). Notice that sufficiently regular random fields
with finite correlation length satisfy the hypotheses of this section.
The exponential $\phi$-mixing assumption was used in \cite{kp} to
analyze the solutions of Liouville equations with random coefficients.
Their techniques lay at the core of our proof of the mixing properties
presented in our main result, Theorem \ref{theorem-main}, below.

\subsection{The main result}

We assume that the initial Wigner transform
$W_\eps^\delta(0,\vx,\bk)$ is uniformly bounded in $L^2({\mathbb
R}^d\times{\mathbb R}^d)$ and
\begin{equation}\label{assump-conv-intro}
\hbox{$W_\eps^\delta(0,\vx,\bk)\to W_0(\vx,\bk)$ strongly in
$L^2({\mathbb R}^d\times{\mathbb R}^d)$ as 
${\cal K}_\mu\ni(\eps,\delta)\to0$.}
\end{equation}
We also assume that the limit $W_0\in C_c({\mathbb R}^d\times{\mathbb
  R}^d)$ with a support that satisfies
\begin{equation}\label{assump-conv-2-intro}
\hbox{supp}~ W_0(\vx,\bk)\subseteq X=\left\{(\vx,\bk):~|\vx|\le
C,~~ C^{-1}\le |\bk|\le C\right\}.
\end{equation}
Note that (\ref{assump-conv-intro}) may not hold for a pure state
since $\|\tilde W_\eps\|_2 = (2\pi\eps)^{-d/2}\|f_\eps\|^2_2$
\cite{LP}. We will later present examples where it does hold for a
mixture of states. Furthermore, we assume that $W_0$ has the form
\begin{equation}
\label{assump-conv-3-intro}
W_0(\vx,\bk)=
u_+^0\vb_+\otimes\vb_+ +u_-^0\vb_-\otimes\vb_-
\end{equation}
and let
\begin{equation} \label{barW-intro}
\bar W(t,\vx,\bk)=\bar u_+(t,\vx,\bk)\vb_+(\bk)\otimes\vb_+(\bk)
+\bar u_-(t,\vx,\bk)\vb_-(\bk)\otimes\vb_-(\bk).
\end{equation}
The functions $\bar u_{\pm}$ satisfy the Fokker-Planck equation
(\ref{diff-intro}) with initial data $u_\pm^0$ as in
(\ref{assump-conv-3-intro}).  The main result of this paper is the
following theorem.
\begin{theorem}\label{theorem-main} Let us assume that the random
  field $c_1(\vx)$ satisfies the assumptions given in Section
  \ref{sec-rand-assump} and that the initial data
  $W_\eps^\delta(0,\vx,\bk)$ satisfies (\ref{assump-conv-intro}) and
  (\ref{assump-conv-2-intro}). Let $S(\bk)\in L^2({\mathbb R}^d)$ be a
  test function, and define the moments 
  \begin{displaymath}
    s_\eps^\delta(t,\vx)=\int
  W_\eps^\delta(t,\vx,\bk)S(\bk)d\bk \qquad \mbox{ and } \qquad
  \bar s(t,\vx)=\int \bar
  W(t,\vx,\bk)S(\bk)d\bk,
  \end{displaymath}
  where $\bar W$ is given by (\ref{barW-intro}). Then for each $t>0$
  we have
\begin{equation}
\E\left\{\int|s_\eps^\delta(t,\vx)-\bar{s}(t,\vx)|^2d\vx\right\}\to
0
\end{equation}
as ${\cal K}_\mu\ni(\eps,\delta)\to 0$.
\end{theorem}
Theorem \ref{theorem-main} means that the moments $s_\eps^\delta$
converges to a deterministic limit. The main application of Theorem
\ref{theorem-main} we have in mind is the mathematical modeling of
refocusing in the time-reversal experiments we present in Section
\ref{sec:TR}.

Our results may be generalized in a fairly straightforward manner
to other wave equations that may be written in the form
(\ref{symm-version-intro}), which include acoustic equations with
variable density and compressibility, electromagnetic and elastic
equations \cite{RPK-WM}.

The papers is organized as follows. The mathematical framework of
the time-reversal experiment as well as the main result concerning
the self-averaging properties of the time reversed signal, Theorem
\ref{thm-main-TR}, are presented in Section \ref{sec:TR}. Section
\ref{sec:ac-eq} contains the derivation of the Liouville equations
in the $L^2$-framework. Some straightforward but tedious calculations
from this Section are presented in Appendices \ref{app:lem-drop-eps},
\ref{app:lem-evol-weps} and \ref{sec:reg-liouv}. The limit theorem
for the two-point motion along the characteristics of the Liouville
equations, Theorem \ref{thm2-main}, is presented in Section
\ref{sec:random}. Theorem \ref{theorem-main} follows from this
result. The proof of Theorem \ref{thm2-main} is contained in Section
\ref{asec1}.

{\bf Acknowledgment.} The authors thank the organizers of the
Mathematical Geophysics Summer School at Stanford, where part of this
work was completed, for their hospitality. This work was supported in
part by ONR grant N00014-02-1-0089. GB was supported in part by NSF
Grants DMS-0072008 and DMS-0233549, TK was partially supported by
grant Nr. 2 PO3A 031 23 from the State Committee for Scientific
Research of Poland, and LR in part by NSF Grants DMS-9971742 and
DMS-0203537, and by an Alfred P. Sloan Fellowship.

\section{Refocusing in the time-reversal experiments}\label{sec:TR}

\subsection{Mathematical formulation of the time-reversal
experiments}

Refocusing of time-reversed acoustic waves is a remarkable
mathematical property of wave propagation in complex media that has
been discovered and intensively studied by experimentalists in the
last decade (see \cite{Fink-Prada-IP,KHS97} and also \cite{BPZ} for
further references to the physical literature). A typical experiment
may be described schematically as follows. A point source emits a
localized signal.  The signal is recorded in time by an array of
receivers. It is then reemitted into the medium {\it reversed in time}
so that the part of the signal recorded last is reemitted first and
vice versa. There are two striking experimental observations. First
the repropagated signal tightly refocuses at the location of the
original source when the medium is sufficiently heterogeneous even
with a recording array of small size. This is to be compared with the
extremely poor refocusing that would occur if the heterogeneous medium
were replaced by a homogeneous medium.  Second, the repropagated
signal is self-averaging. This means that the refocused signal is
essentially independent of the realization of a random medium with
given statistics, assuming that we model the heterogeneous medium as a
random medium.

The first mathematical study of a time-reversal experiment has been
performed in \cite{Fouque-Clouet} in the framework of one-dimensional
layered random media. The one-dimensional case has been further
studied in \cite{Solna}, and a three-dimensional layered medium was
considered in \cite{Fouque-Solna}. The time-reversal experiments in an
ergodic domain have been analyzed mathematically in
\cite{Bardos-Fink}.  The basic ideas that explain the role of
randomness in the refocusing beyond the one-dimensional case were
first outlined in \cite{BPZ} in the parabolic approximation of the
wave equation, that was further analyzed in \cite{BPR-SD,PRS}.
Time-reversal in the general framework of multidimensional wave
equations in random media has been studied formally in
\cite{BR-CRAS,BR-SIAM}.  One of the purposes of this paper is to
present the rigorous proof of some of the results announced in
\cite{BR-CRAS}.

The re-transmission scheme introduced in \cite{BR-CRAS,BR-SIAM} is as
follows. Consider the system of acoustic equations
(\ref{acoust-eq-symm}) (or, equivalently, (\ref{symm-version-intro})
for the pressure $p$ and the acoustic velocity $\bu(t,\vx)$).  The
initial data for (\ref{symm-version-intro}) is assumed to be localized
in space:
\begin{equation}\label{n.1.3}
\bv_\eps(0,\vx)=\bS_0\left(\frac{\vx-\vx_0}{\eps}\right)=
\left(-\nabla\phi_0\left(\frac{\vx-\vx_0}{\eps}\right),
\frac{1}{c_\delta}\dot\phi_0\left(\frac{\vx-\vx_0}{\eps}\right)\right).
\end{equation}
Here $\vx_0\in{\mathbb R}^d$ is the location of the original
source, and $\bS_0\in{\cal S}({\mathbb R}^d)$ is the source shape
function. The small parameter $\eps\ll 1$ measures the spatial
localization of the source.  The signal $\bv_\eps^\delta(t,T)$ is
recorded at some time $t=T$, processed at the recording array and
re-emitted into the medium. The new signal $\tilde\bv_\eps^\delta$
is the solution of (\ref{symm-version-intro}) on the time interval
$T\le t\le 2T$ with the Cauchy data
\begin{equation}\label{n.1.5}
\tilde\bv_\eps^\delta(T,\vx)=
\Gamma[f_\eps\star[\chi\bv_\eps^\delta(T)](\vx)\chi(\vx).
\end{equation}
The initial data (\ref{n.1.5}) reflects the process of recording
of the signal at the array and its smoothing by the recording
process. The kernel $f_\eps(\vx)={\eps^{-d}} f(\vx/\eps)$
represents the smoothing. The array function $\chi(\vx)$ is either
the characteristic function of the set of the receivers, or some
non-uniform function supported on this set. We will assume for
simplicity that $f(|\vy|)$ is radially symmetric, and, moreover,
\begin{equation}\label{n.1.6}
\chi\in C_c({\mathbb R}^d),~~ \hat f\in C_c({\mathbb R}^d),~~
\hbox{supp}\hat f(\bk)\subseteq\left\{0<C^{-1}\le|\bk|\le
C<\infty\right\}
\end{equation}
where
\[
\hat f(\bk)=\int e^{-i\bk\cdot\vy}f(\vy)d\vy
\]
is the Fourier transform of $f$.  The matrix $\Gamma$ corresponds to
the linear transformation of the signal. The pure time-reversal
corresponds to keeping pressure unchanged but reversing the acoustic
velocity so that $\Gamma=\Gamma_0:=\hbox{diag} (-1,-1,-1,1)$. However,
this is only one possible transformation, and while we restrict
$\Gamma$ to the above choice our results may be extended to more
general matrices $\Gamma$, or even allow $\Gamma$ be a
pseudo-differential operator of the form $\Gamma(\vx,\eps D)$.

The re-propagated field near the
source at time $t=2T$ is defined as a function of the local coordinate
$\vxi$ and of the source location $\vx_0$:
\[
\bv_\eps^{\delta,B}(\vxi;\vx_0)=\tilde\bv_\eps^\delta(2T,\vx_0+\eps\vxi).
\]

\subsection{The re-propagated signal and the Wigner transform}

Let us assume that the random field $c_1$ satisfies the
assumptions of Theorem \ref{theorem-main} outlined in Section
\ref{sec-rand-assump}. Then Theorem \ref{theorem-main} implies the
following result.
\begin{theorem}\label{thm-main-TR}
  Under the assumptions made above the re-propagated field
  $\bv_\eps^{\delta,B}(\vxi,\vx_0)$ converges as ${\cal
    K}_\mu\ni(\eps,\delta)\to 0$
\[
\bv^B(\vxi,\vx_0)=\int e^{i\bk\cdot\vxi}[ u_+(T,\vx_0,\bk)\langle\hat
S_0(\bk),\vb_-(\bk)\rangle\vb_+(\bk) +u_-(T,\vx_0,\bk)\langle\hat
S_0(\bk),\vb_+(\bk)\rangle\vb_-(\bk)] \frac{d\bk}{(2\pi)^d}
\]
in the sense that
\begin{equation}\label{eq-thm-main}
\sup_{\vxi\in{\mathbb R}^d} {\mathbb E}\left\{
\int|\bv_\eps^{\delta,B}(\vxi,\vx_0)-\bv^B(\vxi,\vx_0)|^2d\vx_0\right\}\to
0.
\end{equation}
The functions $u_\pm(t,\vx,\bk)$ are the solutions of the
Fokker-Planck equation (\ref{diff-intro}) with initial data
$a_\pm(0,\vx,\bk)=|\chi(\vx)|^2\hat f(\bk)$.
\end{theorem}
The proof of Theorem \ref{thm-main-TR} is based on Theorem
\ref{theorem-main} and a representation of the re-propagated
signal in terms of the Wigner transform of a mixture of solutions
of the acoustic wave equations. The latter arises as follows. Let
$\bQ_\eps^\delta(t,\vx;\bq)$ be the matrix-valued solution of
(\ref{symm-version-intro}) with initial data
\begin{equation}
  \label{eq:Q-init}
  \bQ_\eps^\delta(0,\vx;\bq)=\chi(\vx)e^{i\bq\cdot\vx/\eps}I,
\end{equation}
where $I$ is the $(d+1)\times(d+1)$ identity matrix, $\chi(\vx)$
is the array function, and $\bq\in{\mathbb R}^d$ is a fixed
vector. It plays the role of the "state" of the initial data.
Physically $\bQ_\eps^\delta$ describes evolution of a wave that is
emitted by the recorders-transducers with a wave vector $\bq$. The
Wigner transform of the family $\bQ_\eps^\delta(t,\vx;\bq)$ is
\begin{equation}\label{wig-single}
\tilde W_\eps^\delta(t,\vx,\bk;\bq)= \int
e^{i\bk\cdot\vy}\bQ_\eps^\delta(t,\vx-\frac{\eps\vy}{2};\bq)
\bQ_\eps^{\delta
*}(t,\vx+\frac{\eps\vy}{2};\bq)\frac{d\vy}{(2\pi)^d}.
\end{equation}
The corresponding ``mixed'' Wigner transform is
\begin{equation}\label{wig-mix}
W_\eps^\delta(t,\vx,\bk)=\int \tilde
W_\eps^\delta(t,\vx,\bk;\bq)\hat f(\bq)d\bq.
\end{equation}
Then the re-propagated signal is described as follows in terms of
$W_\eps^\delta$.
\begin{lemma}\label{lem:vB-Q}
The re-propagated signal may be expressed as
\begin{equation}\label{bv-wigner}
\bv_\eps^{\delta,B}(\vxi,\vx_0)=\int e^{i\bk\cdot(\vxi-\vy)}
W_\eps^\delta(T,\vx_0+\frac{\eps(\vxi+\vy)}{2},\bk)\Gamma_0\bS_0(\vy)
\frac{d\bk d\vy}{(2\pi)^d}.
\end{equation}
\end{lemma}
\begin{proof}
Let $G(t,\vx;\vy)$ be the Green's matrix of
(\ref{symm-version-intro}), that is, solution of (\ref{symm-version-intro})
with the initial data $G(0,\vx;\vy)=I\delta(\vx-\vy)$. Then the
signal arriving to the recorders-transducers array is
\[
\bv_\eps^\delta(T,\vx)=\int  G(T,\vx;\vy)\bv_\eps^\delta(0,\vy)d\vy= \int
G(T,\vx;\vy)\bS_0\left(\frac{\vy-\vx_0}{\eps}\right)d\vy
\]
and the re-emitted signal is
\[
\tilde\bv_\eps^\delta(T,\vz)=\int \Gamma_0f_\eps(\vz-\vz')\chi(\vz)\chi(\vz')
\bv_\eps^\delta(T,\vz')d\vz'.
\]
Therefore we obtain
\begin{eqnarray}\label{3.1}
&&\bv_\eps^{\delta,B}(\vxi,\vx_0)=
\int G(T,\vx_0+\eps\vxi;\vz)\tilde\bv_\eps(T,\vz)d\vz\\
&&= \int
G(T,\vx_0+\eps\vxi;\vz)\Gamma_0f\left(\frac{\vz-\vz'}{\eps}\right)
\chi(\vz)\chi(\vz')
G(T,\vz';\vy)\bS_0\left(\frac{\vy-\vx_0}{\eps}\right) \frac{d\vz
d\vz' d\vy}{\eps^d}.\nonumber
\end{eqnarray}
However, we also have
\begin{equation}\label{G-G*}
\Gamma_0G(t,\vx;\vy)\Gamma_0=G^*(t,\vy;\vx).
\end{equation}
This is seen as follows: a solution of (\ref{symm-version-intro})
satisfies
\[
\bv(t,\vx)=\int  G(t-s,\vx;\vy)\bv(s,\vy)d\vy
\]
for all $0\le s\le t$. Differentiating the above equation with
respect to $s$, using (\ref{symm-version-intro}) for $\bv(s,\vy)$
and integrating by parts we obtain
\[
0=\int\left(-\pdr{G(t-s,\vx;\vy)}{t}
+\pdr{}{y_j}\left(G(t-s,\vx;\vy)A_\delta(\vy)
\right)D^jA_\delta(\vy)\right)\bv(s,\vy).
\]
Passing to the limit $s\to 0$ and using the fact that the initial
data $\bv_0(\vy)$ is arbitrary  we obtain
\begin{equation}\label{G-G*1}
\pdr{G(t,\vy;\vx)}{t}-\pdr{}{x_j}\left(G(t,\vy;\vx)A_\delta(\vx)
\right)D^jA_\delta(\vx)=0.
\end{equation}
Furthermore, the matrix $G^*(t,\vx;\vy)$ satisfies
\[
\pdr{G^*(t,\vx;\vy)}{t}+\pdr{}{x_j}\left(G^*(t,\vx;\vy)A_\delta(\vx)
\right)D^jA_\delta(\vx)=0.
\]
Multiplying (\ref{G-G*1}) by $\Gamma_0$ on the left and on the
right, and using the commutation relations
\begin{equation}
\label{Gamma-commut}
  \Gamma_0 A_\delta=A_\delta\Gamma_0,~~\Gamma_0 D^j=-D^j\Gamma_0.
\end{equation}
we deduce (\ref{G-G*}). Then, since $\Gamma_0^2=I$, (\ref{3.1}) may be
re-written as
\begin{eqnarray}
&&\!\!\bv_\eps^{\delta,B}(\vxi,\vx_0)=\!\int\!
G(T,\vx_0+\eps\vxi;\vz)\chi(\vz)e^{i\bq\cdot\vz/\eps}
\chi(\vz')e^{-i\bq\cdot\vz'/\eps}G^*(T,\vx_0+\eps\vy;\vz')\Gamma_0\hat
f(\bq)
\bS_0(\vy)\frac{d\vz d\vz' d\vy d\bq}{(2\pi)^d}\nonumber\\
&&\label{3.2}=\int \bQ_\eps^\delta(T,\vx_0+\eps\vxi;\bq)
\bQ_\eps^{\delta*}(T,\vx_0+\eps\vy;\bq) \hat
f(\bq)\Gamma_0\bS_0(\vy)\frac{d\vy d\bq}{(2\pi)^d}.
\end{eqnarray}
and (\ref{bv-wigner}) follows.
\end{proof}

The following lemma allows us to drop the term of order $\eps$ in the
argument of $W_\eps$ in (\ref{bv-wigner}).
\begin{lemma}\label{lem-drop-eps}
Let us define
$\disp
\tilde \bv_\eps^{\delta,B}(\vxi;\vx_0)=\int e^{i\bk\cdot(\vxi-\vy)}
W_\eps^\delta(T,\vx_0,\bk)\Gamma_0\bS_0(\vy)
\frac{d\bk d\vy}{(2\pi)^d}.
$
There exists a deterministic function $C(\eps,\delta)$ so that
\begin{equation}\label{bvtilde}
\sup_{\vxi}\|\bv_\eps^{\delta,B}-\tilde\bv_\eps^{\delta,B}\|_{L_{\vx_0}^2}
\le C(\eps,\delta)
\end{equation}
and $C(\eps,\delta)\to 0$ as ${\cal K}_\mu\ni(\eps,\delta)\to 0$.
\end{lemma}
The proof of Lemma \ref{lem-drop-eps} is presented in Appendix
\ref{app:lem-drop-eps}. Note that Theorem \ref{theorem-main} may be
applied directly to the moment
\[
\tilde \bv_\eps^\delta(\vxi;\vx_0)=\int e^{i\bk\cdot\vxi}
W_\eps^\delta(T,\vx_0,\bk)\Gamma_0\hat\bS_0(\bk)
\frac{d\bk}{(2\pi)^d}
\]
and the conclusion of Theorem \ref{thm-main-TR} follows.

\section{The high frequency analysis}
\label{sec:ac-eq}

In this section we study the deterministic high-frequency behavior of
the Wigner transform and estimate the error between the Wigner
transform and the solution of the Liouville equations.  It is well
known \cite{GMMP,LP,RPK-WM} that in the high frequency regime the weak
limit of the Wigner transform as $\eps\to 0$ and $\delta>0$ is fixed,
is described by the classical Liouville equations in the phase space.
Here, we do not pass to the limit $\eps\to 0$ at $\delta$ fixed but
rather control the error introduced by the semi-classical
approximation.  As explained in the introduction, this is possible
because we are dealing with the Wigner transform of a mixture of
states that may have strong limits \cite{LP} rather than the Wigner
transform of pure states, which converges only weakly.

\subsection{Convergence on the initial data}
\label{sec:vb-wig}

We first show that the assumptions on the convergence of the initial
data in Theorem \ref{theorem-main} are purely academic, and in
particular are satisfied in the time-reversal application. We note
that the $L^2$-norm of a pure Wigner transform $\tilde
W_\eps(t,\vx,\bk;\bq)$ of a single wave function, such as
(\ref{wig-single}), blows up as $\eps\to 0$ in $L^2({\mathbb R}^d)$,
because
\begin{eqnarray}
&&\|\tilde W_\eps(t;\zeta)\|_{L^2({\mathbb R}^d\times{\mathbb R}^d)}:=
\int \hbox{Tr}[\tilde W_\eps(t,\vx,\bk;\zeta)
\tilde W_\eps^*(t,\vx,\bk;\zeta)]d\vx d\bk=
(2\pi\eps)^{-d/2}\|Q^\eps(t,\bq)\|_{L^2({\mathbb R}^d)}^2\nonumber\\
&&=(2\pi\eps)^{-d/2}\|\chi\|_{L^2({\mathbb R}^d)}.\label{L2-pure}
\end{eqnarray}
Therefore (\ref{assump-conv-intro}) may not hold for a pure state. Two
examples when assumption (\ref{assump-conv-intro}) holds are given by
the following lemma, which may be verified by a straightforward
calculation.  The first one arises when the initial data is random,
and the second comes from the time-reversal application.
\begin{lemma}\label{lemma-verify-L2} Assumption (\ref{assump-conv-intro}) is
satisfied in the following two cases:\newline (1) Statistical
averaging: the initial data is
$\bv_0^\eps(\vx;\zeta)=\psi(\vx)V(\vx/\eps;\zeta)$, where
$V(\vy;\zeta)$ is a mean zero, scalar spatially homogeneous random
process with a rapidly decaying two-point correlation function
$R(\vz)$: $E\left\{V(\vy)V(\vy+\vz)\right\}= \int
V(\vy;\zeta)V(\vy+\vz;\zeta)d\mu(\zeta)= R(\vz)\in L^2({\mathbb
R}^d)$, and $\psi(\vx)\in C_c({\mathbb R}^d)$. The limit Wigner
distribution is given by $W_0(\vx,\bk)=|\psi(\vx)|^2\tilde R(\bk)$,
where $\tilde R(\bk)$ is the inverse Fourier transform of
$R(\vy)$.\newline (2) Smoothing of oscillations: the initial data is
$\bv_0^\eps(\vx;\zeta)=\psi(\vx) e^{i\zeta\cdot\vx/\eps}$, where
$\psi(\vx)\in C_c({\mathbb R}^d)$. The measure $\mu$ is
$d\mu(\zeta)=g(\zeta)d\zeta$, $\zeta\in{\mathbb R}^d$, and $g\in
L^2({\mathbb R}^d)$. The limit Wigner distribution is
$W_0(\vx,\bk)=|\psi(\vx)|^2g(\bk)$.
\end{lemma}
\begin{proof}
We only verify case (2), the other case being similar:
\[
W_\eps^0(\vx,\bk)=\int e^{i\bk\cdot\vy}
\psi(\vx-\frac{\eps\vy}{2})\psi^*(\vx+\frac{\eps\vy}{2})
\hat g(\vy)\frac{d\vy}{(2\pi)^d}
\]
so that
\begin{equation}\label{in-error}
\|W_\eps^0-W_0\|_2^2=\int
\left(\psi(\vx-\frac{\eps\vy}{2})\psi^*(\vx+\frac{\eps\vy}{2})-
|\psi(\vx)|^2\right)^2|\hat g(\vy)|^2 \frac{d\vx d\vy}{(2\pi)^d}=
\int I_\eps(\vy)|\hat g(\vy)|^2 \frac{d\vy}{(2\pi)^d}.
\end{equation}
However, we have $\int|I_\eps(\vy)|\le 4\|\psi\|_{L^4}^4$ and
\[
I_\eps(\vy)=\int
\left(\psi(\vx-\frac{\eps\vy}{2})\psi^*(\vx+\frac{\eps\vy}{2})-
|\psi(\vx)|^2\right)^2d\vx\to 0
\]
as $\eps\to 0$ since $\psi\in C_c({\mathbb R}^d)$, pointwise in $\vy$.
Therefore $\|W_\eps^0-W_0\|_2\to 0$ by the Lebesgue dominated
convergence theorem.
\end{proof}

Note that if $g(\zeta)$ and $\psi$ in part (2) of Lemma
\ref{lemma-verify-L2} are sufficiently regular, then
$\|W_0^\eps-W_0\|_2=O(\eps)$ so that one may get the order of
convergence in (\ref{assump-conv-intro}).

\subsection{Approximation by the Liouville equations}

We now estimate directly the error between the mixed Wigner transform
and its semi-classical approximation. The dispersion matrix
$P_0^\delta(\vx,\bk)=ic_\delta(\vx)k_jD^j$ may be diagonalized as
\begin{equation}\label{P0-Piq}
  -iP_0^\delta(\bx,\bk)=\dsum_{q=0}^2 \lambda_q^\delta(\bx,\bk) \Pi_q(\bk),~~
  \quad \dsum_{q=0}^2 \Pi_q(\bk) = I.
\end{equation}
Here $\Pi_q$ is the projection matrix onto the eigenspace
corresponding to the eigenvalue $\lambda_q^\delta$. Notice that the
eigenspaces are independent of the spatial position $\bx$, hence of
the parameter $\delta$; see (\ref{P0-intro})-(\ref{eq:eigbs}).

As we have mentioned before, for a fixed $\delta>0$ the Wigner
transform $W_\eps^\delta(t,\vx,\bk)$ converges weakly as $\eps\to
0$ to its semi-classical limit $U^\delta(t,\vx,\bk)$ given by
\begin{equation}\label{U}
U^\delta(t,\vx,\bk)=\sum_q u_q^\delta(t,\vx,\bk)\Pi_q(\bk).
\end{equation}
The functions $u_q^\delta$ satisfy the Liouville equations
\begin{equation}\label{liouv}
\pdr{u_q^\delta}{t}+\nabla_\bk\lambda_q^\delta\cdot\nabla_\vx{u_q^\delta}-
\nabla_\vx\lambda_q^\delta\cdot\nabla_\bk{u_q^\delta}=0
\end{equation}
with initial data
$u_q^\delta(0,\vx,\bk)=\hbox{Tr}\Pi_qW_0(\vx,\bk)\Pi_q$.  Our goal is
to estimate the difference between $W_\eps^\delta$ and $U^\delta$ in
$L^2({\mathbb R}^d\times{\mathbb R}^d)$. Let us denote by
$\gamma_q^\delta(\vx,\bk)$ the largest eigenvalue of the matrix
$(F_q^\delta F_q^{\delta*})^{1/2}$, where
\[
F_q^\delta=\left(\begin{matrix}
-\dfrac{\partial^2\lambda_q^\delta}{\partial k_i\partial x_j}&
-\dfrac{\partial^2\lambda_q^\delta}{\partial k_i\partial k_j} \cr
\dfrac{\partial^2\lambda_q^\delta}{\partial x_i\partial x_j}&
\dfrac{\partial^2\lambda_q^\delta}{\partial x_i\partial k_j}\cr
                            \end{matrix}\right).
\]
Note that (\ref{d-i}) implies that
$\gamma_1^\delta(\vx,\bk)=\gamma_2^\delta(\vx,\bk)=\gamma^\delta(\vx,\bk)$,
while $\gamma_0=0$. The initial data $u_q^0$ is supported on a
compact set $S$ because $W_0$ is (see
(\ref{assump-conv-2-intro})). Then the set
\[
{\cal S}=\bigcup_{t\ge
0,~\delta\in(0,1]}\hbox{supp}~u_q^\delta(t,\vx,\bk)
\]
is bounded because the speed $c_\delta(\vx)$ is uniformly bounded
from above and below for $\delta$ sufficiently small (\ref{d-i}).
Therefore we have $\gamma^\delta(\vx,\bk)\le C/\delta^{3/2}$ with
a deterministic constant $C>0$.
We denote
$\bar\gamma_\delta=\sup_{{\cal S}}\gamma^\delta(\vx,\bk).$
We have the following approximation theorem.
\begin{theorem}\label{thm1}
  Let the acoustic speed $c_\delta(\vx)$ be of the form
  (\ref{c-fluct-delta}) and satisfy assumptions (\ref{d-i}). We assume
  that the Wigner transform $W_\eps^\delta$ satisfies
  (\ref{assump-conv-intro}) and that (\ref{assump-conv-2-intro})
  holds.  Moreover, we assume that the initial limit Wigner transform
  $W_0$ is of the form
\begin{equation}\label{no-offdiagonal}
W_0(\vx,\bk)=\sum_q u_q^0(\vx,\bk)\Pi_q(\bk).
\end{equation}
Let $U^\delta(t,\vx,\bk)=\sum_p u_p^\delta(t,\vx,\bk)\Pi_p(\bk)$,
where the functions $u_p^\delta$ satisfy the Liouville equations
(\ref{liouv}) with initial data $u_q^0(\vx,\bk)$. Then we have
\begin{equation}\label{eq-thm1}
\|W_\eps^\delta(t,\vx,\bk)-U^\delta(t,\vx,\bk)\|_2\le
C(\delta)
[ \eps \|W_0\|_{H^2}e^{2\bar\gamma_\delta t}+
\eps^2\|W_0\|_{H^3}e^{3\bar\gamma_\delta t}]+\|W_\eps^\delta(0)-W_0\|_2,
\end{equation}
where $C(\delta)$ is a rational function of $\delta$ with
deterministic coefficients
that may depend on the constant $C>0$ in the bound
(\ref{assump-conv-2-intro}) on the support of $W_0$.
\end{theorem}
Theorem \ref{thm1} shows that the semi-classical approximation is
valid for times $T\ll |\ln\eps|/\bar\gamma_\delta$. This is
reminiscent of the Ehrenfest time of validity of the semi-classical
approximation in quantum mechanics, see \cite{Paul,Robert} for recent
mathematical results in this direction for the Schr\"odinger
operators. The pre-factor constants on the right side of
(\ref{eq-thm1}) are not optimal but sufficient for the purposes of our
analysis.

The assumption that initially $W_0$ has no terms of the form
$\Pi_p\Pi_q$ with $p\neq q$ is necessary in general for the Liouville
equation to provide an approximation to $W_\eps^\delta$ in the strong sense.
This may be seen on the simple example of the solution
\[
u_\eps(t,x)=ae^{i(\bq\cdot\vx-ct)/\eps}+be^{i(\bq\cdot\vx+ct)/\eps}
\]
of the wave equation
\[
u_{tt}-c^2u_{xx}=0
\]
with a constant speed $c$. The cross-terms in the Wigner distribution
$W_\eps=[|a|^2+|b|^2+ab^*e^{-2ict/\eps}+a^*be^{2ict/\eps}]\delta(\bk-\bq)$
vanish only in the weak sense as a function of $t$ but not
strongly.

The Wigner distribution that arises in the time-reversal application
has an initial data that is described by part (2) of Lemma
\ref{lemma-verify-L2}:
\[
W_0(\vx,\bk)=|\chi(\vx)|^2\hat f(\bk)I
\]
and satisfies the assumption (\ref{no-offdiagonal}) with
$u_q^0(\vx,\bk)=|\chi(\vx)|^2\hat f(\bk)$ for all eigenspaces because
of the second equation in (\ref{P0-Piq}).  The error introduced by the
replacement of the initial data in (\ref{eq-thm1}) in that case is
given by (\ref{in-error}) and is $O(\eps)$ provided that $\chi$ and
$f$ are sufficiently regular.

The proof of Theorem \ref{thm1} is quite straightforward though
tedious.  We first obtain the evolution equation for $W_\eps^\delta$ in
Section \ref{sec:evol}, and show that it preserves the
$L^2$-norm. This allows us to replace the initial data in the equation
for $W_\eps^\delta$ by $W_0$ at the expense of the last term in
(\ref{eq-thm1}). We obtain the Liouville equations (\ref{liouv}) in Section
\ref{sec:liouv} and estimate the right side of (\ref{eq-thm1}) in terms
of the $H^3$-norm of its solution. Finally, in Appendix
\ref{sec:reg-liouv} we obtain the necessary estimates for the solution
of the Liouville equation.

\subsection{The evolution equation for the Wigner transform}
\label{sec:evol}

The $L^2$-norm of the Wigner transform $\tilde W(t,\vx,\bk;\zeta)$ of
a pure state, or a fixed $\zeta$, is preserved in time as follows from
the preservation of the $L^2$-norm of solutions of
(\ref{symm-version-intro}). We obtain now an evolution equation for
the Wigner transform $W_\eps$ of mixed states and show that its
$L^2$-norm is also preserved.  It is convenient to define the
skew-symmetric matrix symbol
\begin{equation}\label{P-eps-delta}
P_\eps^\delta(\vx,\bk)=P_0^\delta(\vx,\bk)+\eps
P_1^\delta(\vx),
\end{equation}
where $P_0^\delta$ is defined by (\ref{P0-intro})
and the symbol $P_1^\delta$ depends only on $\vx$:
\begin{equation}\label{P1}
P_1^\delta(\vx)=A_\delta(\vx)D^j\frac12\pdr{A_\delta}{x_j}(\vx)-\frac
12\pdr{A_\delta}{x_j}(\vx) D^j A_\delta(\vx)=
\frac 12\pdr{c_\delta}{x_j}(\bx)
\left[\ve_j\otimes\ve_{d+1}-\ve_{d+1}\otimes\ve_j\right].
\end{equation}
The latter equality follows from (\ref{eq:symmatADj}) and calculations
of the form
\begin{equation}
  \label{eq:diffadelta}
  D^j \pdr{A_\delta}{x_j}(\vx) = \pdr{c_\delta}{x_j}(\bx)
  \be_j\otimes \be_{d+1}.
\end{equation}
The following lemma describes the evolution of the Wigner transform
$W_\eps^\delta$.
\begin{lemma}\label{evol-weps}
The Wigner transform $W_\eps^\delta(t,\vx,\bk)$ satisfies the
evolution equation
\begin{equation}\label{wigeps-evol}
\eps\pdr{W_\eps^\delta}{t}+{\cal L}_\eps^\delta W_\eps^\delta=0
\end{equation}
with initial data $W_\eps^\delta(0,\vx,\bk)$. The operator ${\cal
  L}_\eps^\delta$ is given by
\begin{equation}\label{eq:Leps}
{\cal L}_\eps^\delta f(\bx,\bk) =
\dint \Big(P_\eps^\delta(\vy,\bq)e^{i\phi} f(\bz,\bp)
-f(\bz,\bp) e^{-i\phi} P_\eps^\delta(\vy,\bq)\Big)
\dfrac{d\bz d\bp d\vy d\bq}{(\pi\eps)^{2d}},
\end{equation}
where $\phi(\vx,\vz,\bk,\bp,\vy,\bq)=\frac{2}{\eps}((\bp-\bk)\cdot\by+
(\bq-\bp)\cdot\bx+(\bk-\bq)\cdot\vz)$. The integral of the trace and
the $L^2$-norm of the Wigner transform $W_\eps$ are preserved:
\begin{equation}\label{trace-preserve}
\int \rm{Tr}W_\eps^\delta(t,\vx,\bk)d\vx d\bk=
\int \rm{Tr}W_\eps^\delta(0,\vx,\bk)d\vx d\bk
\end{equation}
and
\begin{equation}\label{L2-preserve}
\int \rm{Tr}[W_\eps^\delta(t,\vx,\bk)W_\eps^{\delta*}(t,\vx,\bk)]d\vx d\bk=
\int \rm{Tr}[W_\eps^\delta(0,\vx,\bk)W_{\eps}^{\delta*}(0,\vx,\bk)]d\vx d\bk.
\end{equation}
\end{lemma}
This lemma is verified by a direct calculation that we present for the
convenience of the reader in Appendix \ref{app:lem-evol-weps}.

Note that the solution of (\ref{wigeps-evol}) with self-adjoint
initial data remains self-adjoint and the $L^2$-norm is preserved.
Therefore, we have the following corollary.
\begin{corollary}\label{cor:bd}
  Let $W_{0}(\bx,\bk)$ be a strong limit of $W_\eps(0)$ in $L^2$,
  which exists by assumption (\ref{assump-conv-intro}). Then the
  solutions $W_\eps^\delta(t,\bx,\bk)$ and $\bar
  W_\eps^\delta(t,\bx,\bk)$ of (\ref{wigeps-evol}) with initial
  conditions $W_{\eps}^0(\bx,\bk)$ and $W_0(\bx,\bk)$, respectively,
  satisfy
\begin{displaymath} \|\bar W_\eps^\delta(t)-W_\eps^\delta(t)\|_2 =
\|W_{\eps}^0-W_0\|_2 \to 0 \hbox{ as $\eps\to0$.}
\end{displaymath}
\end{corollary}
This shows that in the analysis of (\ref{wigeps-evol}), we can replace
strongly converging initial conditions by their limit, and consider
then the limit of $\bar W_\eps(t,\bx,\bk)$ as $\eps\to0$ with fixed
initial conditions. This is done in the following section.

\subsection{Derivation of the Liouville equations}
\label{sec:liouv}

We consider in this section the solution $\bar
W_\eps^\delta(t,\vx,\bk)$ of the evolution equation
(\ref{wigeps-evol}) with fixed initial data $W_0(\vx,\bk)$ and show
that it may be approximated by the solution of the Liouville equation.
We split the operator ${\cal L}_\eps^\delta={\cal L}_\eps^{\delta,0}+
\eps{\cal L}_\eps^{\delta,1}$, where
\[
{\cal L}_\eps^{\delta,j}f(\vx,\bk)= \dint
\Big(P_j^\delta(\vy,\bq)e^{i\phi} f(\bz,\bp)
-f(\bz,\bp) e^{-i\phi} P_j^\delta(\vy,\bq)\Big)
\dfrac{d\bz d\bp d\vy d\bq}{(\pi\eps)^{2d}},~~j=0,1,
\]
and the symbols $P_j^\delta$ are given by (\ref{P0-intro}) and (\ref{P1}). The
operator ${\cal L}_\eps^{\delta,0}$ is given explicitly by
\begin{eqnarray*}
&&{\cal L}_\eps^{\delta,0}f(\vx,\bk)=\int e^{i(\bk-\bp)\cdot\vy}
\left[c_\delta(\vx-\frac{\eps\vy}{2})ip_jD^jf(\vx,\bp)
-c_\delta(\vx+\frac{\eps\vy}{2})f(\vx,\bp)ip_jD^j\right]
\frac{d\bp d\vy}{(2\pi)^d}\\
&&\!\!\!\!\!\!+\frac{\eps}{2}\int  e^{i(\bk-\bp)\cdot\vy}\left(D^j\pdr{}{x_j}
\left[c_\delta(\vx-\frac{\eps\vy}{2})f(\vx,\bp)\right]
+\pdr{}{x_j}\left[c_\delta(\vx+\frac{\eps\vy}{2})f(\vx,\bp)\right]D^j\right)
\frac{d\bp d\vy}{(2\pi)^d}={\cal L}_{\eps,\delta}^{01}+
{\cal L}_{\eps,\delta}^{02}.
\end{eqnarray*}
We recast the operator ${\cal L}_{\eps,\delta}^{01}$ as
\begin{eqnarray*}
&&{\cal L}_{\eps,\delta}^{01}f(\vx,\bk)\!=\!
c_\delta(\vx)\!\left[ik_jD^jf(\vx,\bk)-\! f(\vx,\bk)ik_jD^j\right]\!
-\frac{\eps}{2}\pdr{c_\delta(\vx)}{x_m}
\pdr{}{k_m}\!\left(k_jD^jf(\vx,\bk)+f(\vx,\bk)k_jD^j\right)\\
&&+
\eps{\cal R}_{\eps,\delta}^{01}f
\end{eqnarray*}
with the correction term
\begin{eqnarray*}
&&{\cal R}_{\eps,\delta}^{01}f(\vx,\bk)=
\frac{1}{\eps}\int e^{i(\bk-\bp)\cdot\vy}
\left[\left(c_\delta(\vx-\frac{\eps\vy}{2})-c_\delta(\vx)+
\frac{\eps}{2}\vy\cdot\nabla c_\delta(\vx)\right)ip_jD^jf(\vx,\bp)\right.\\
&&~~~~~~~~~~~~~~~~~~~~~~~~~~~~~~~
\left.-\left(c_\delta(\vx+\frac{\eps\vy}{2})-c_\delta(\vx)-
\frac{\eps}{2}\vy\cdot\nabla c_\delta(\vx)\right)f(\vx,\bp)ip_jD^j\right]
\frac{d\bp d\vy}{(2\pi)^d}.
\end{eqnarray*}
Similarly, we have
\[
{\cal L}_{\eps,\delta}^{02}f(\vx,\bk)=\frac{\eps}{2}D^j
\pdr{}{x_j}\left(c_\delta(\vx)f(\vx,\bk)\right)
+\frac{\eps}{2}
\pdr{}{x_j}\left(c_\delta(\vx)f(\vx,\bk)\right)D^j+
\eps{\cal R}_{\eps,\delta}^{02}f
\]
with
\begin{eqnarray*}
&&{\cal R}_{\eps,\delta}^{02}f(\vx,\bk)=\frac{1}{2}\int e^{i(\bk-\bp)\cdot\vy}
\left(D^j\pdr{}{x_j}
\left[\left\{c_\delta(\vx-\frac{\eps\vy}{2})-
c_\delta(\vx)\right\}f(\vx,\bp)\right]
\right.\\
&&\left.~~~~~~~~~~~~~~~~~~~~~~~~~~~~~~~
+\pdr{}{x_j}\left[\left\{c_\delta(\vx+\frac{\eps\vy}{2})-c_\delta(\vx)\right\}
f(\vx,\bp)\right]D^j\right)
\frac{d\bp d\vy}{(2\pi)^d}.
\end{eqnarray*}
The operator ${\cal L}_\eps^{\delta,1}$ is given explicitly by
\begin{eqnarray}
&&{\cal L}_\eps^{\delta,1}f(\vx,\bk)=\int e^{i(\bk-\bp)\cdot\vy}
\left[P_1^\delta(\vx-\frac{\eps\vy}{2})f(\vx,\bp)-
f(\vx,\bp)P_1^\delta(\vx+\frac{\eps\vy}{2})\right]\frac{d\bp d\vy}{(2\pi)^d}
\nonumber\\
&&=P_1^\delta(\vx)f(\vx,\bk)-f(\vx,\bk)P_1^\delta(\vx)+
{\cal R}_{\eps,\delta}^1f(\vx,\bk)\label{L1eps1}
\end{eqnarray}
with the correction ${\cal R}_{\eps,\delta}^1$ defined by
\begin{equation}\label{Reps1}
{\cal R}_{\eps,\delta}^1f(\vx,\bk)=\!\int\! e^{i(\bk-\bp)\cdot\vy}
\left[\left(P_1^\delta(\vx-\frac{\eps\vy}{2})-P_1^\delta(\vx)\right)f(\vx,\bp)-
f(\vx,\bp)\left(P_1^\delta(\vx+\frac{\eps\vy}{2})-P_1^\delta(\vx)\right)\right]
\frac{d\bp d\vy}{(2\pi)^d}.
\end{equation}
Putting together the above expressions, we obtain the following
equation for $\bar W_\eps^\delta$
\begin{equation}
  \label{eq:app}
  \pdr{\bar W_\eps^\delta}{t}= \frac{1}{\eps}{\cal L}_\eps^\delta
\bar W_\eps^\delta=
\dfrac{[\bar W_\eps^\delta,P_0^\delta]}{\eps} + [\bar W_\eps^\delta,P_1^\delta]
   + \dfrac{1}{2i} ( \{\bar W_\eps^\delta,P_0^\delta\}-
\{P_0^\delta,\bar W_\eps^\delta\})
  - {\cal R}_\eps^\delta \bar W_\eps^\delta
\end{equation}
with ${\cal R}_\eps^\delta={\cal R}_{\eps,\delta}^{01}+
{\cal R}_{\eps,\delta}^{02}+
{\cal R}_{\eps,\delta}^{1}$.
Here $\{f,g\}$ is the standard Poisson bracket
\[
\{f,g\}=\nabla_\bk f\cdot\nabla_\vx g-\nabla_\vx f\cdot\nabla_\bk g
\]
and $[A,B]=AB-BA$ is the commutator.  We now introduce the expansion
\begin{equation}
  \label{eq:expW}
  \bar W_\eps^\delta=U^\delta+\eps U_{1}^\delta+U_{2,\eps}^\delta.
\end{equation}
We insert this ansatz into (\ref{eq:app}) and equating like powers
of $\eps$ obtain at the order $\eps^{-1}$
\begin{equation}
  \label{eq:order-1}
  [P_0^\delta,U^\delta]=0,
\end{equation}
which is equivalent to
\begin{equation}
  \label{eq:formW0}
  U^\delta=\dsum_{q=0}^2 \Pi_q U^\delta \Pi_q = \dsum_{q=0}^2 U_{q}^\delta,
\end{equation}
where $ U_{q}^\delta=\Pi_q U^\delta \Pi_q$, and for $q=1,2$ one has
$U_q^\delta=u_q^\delta\Pi_q$ with $u_q=\hbox{Tr}U_q^\delta$. The matrices
$\Pi_q$ are projections on the eigenspaces of $P_0^\delta$, as in
(\ref{P0-Piq}). This means that the matrix $U^\delta$ does not have
off-diagonal contributions in the eigenbasis of $P_0^\delta$.  The equation
of order $O(\eps^0)$ is given by
\begin{equation}
  \label{eq:order0}
   \pdr{U^\delta}{t}= [U_1^\delta,P_0^\delta] + [U^\delta,P_1^\delta]
   + \dfrac{1}{2i} ( \{U^\delta,P_0^\delta\}-\{P_0^\delta,U^\delta\}).
\end{equation}
Multiplying the above equation on both sides by $\Pi_q$ yields
\begin{equation}
  \label{eq:W0}
   \pdr{\Pi_q U^\delta \Pi_q}{t} = \Pi_q [U^\delta,P_1^\delta]\Pi_q+
\dfrac{1}{2i} \Pi_q\big(\{U^\eps,P_0^\delta\}-
\{P_0^\delta,U^\delta\}\big)\Pi_q.
\end{equation}
This is nothing but equation (6.16) of reference \cite{GMMP} for the Wigner
matrix without consideration of mixtures of states. The only
difference is that the leading order term $P_0$ depends on the
parameter $\delta$. This, of course, does not change the algebra, and following
\cite{GMMP} one obtains a system of decoupled Liouville equations for
$u_q^\delta=\hbox{Tr}U_q^\delta$, $q=1,2$,
\begin{equation}\label{liouville-eq}
\pdr{u_q^\delta}{t}+\left\{\lambda_q^\delta,u_q^\delta\right\}=0
\end{equation}
with initial data
$u_q^\delta(0,\vx,\bk)=\hbox{Tr}[\Pi_q(\bk)W_0(\vx,\bk)\Pi_q(\bk)]$.
The zero eigenvalue component of the matrix $U^\delta$, that is,
$U_0(t,\vx,\bk)=\Pi_0(\bk)W_0(\vx,\bk)\Pi_0(\bk)$, does not change in
time.

We have to show that the terms $U_1^\delta$ and $U_{2,\eps}^\delta$ in
(\ref{eq:expW}) are small.  In order to uniquely characterize
$U_1^\delta$, we assume that it is orthogonal to the terms of the form
(\ref{eq:formW0}), that is,
\begin{equation}
  \label{eq:W1form}
  U_1^\delta=\dsum_{p\not=q} \Pi_p U_1^\delta\Pi_q.
\end{equation}
Then, (\ref{eq:formW0}) and (\ref{eq:order0}) imply that
\begin{equation}
  \label{eq:W1result}
  \Pi_m U_1^\delta \Pi_p = \dfrac{1}{i(\lambda_m^\delta-\lambda_p^\delta)}
   \Pi_m {\cal B}(U^\delta) \Pi_p,
\end{equation}
where
\begin{displaymath}
{\cal B}(U^\delta)= [U^\delta,P_1^\delta]+
\dfrac{1}{2i}(\{U^\delta,P_0^\delta\}-\{P_0^\delta,U^\delta\}).
\end{displaymath}
We now analyze the term $U_{2,\eps}^\delta$ in (\ref{eq:expW}) and
show that it vanishes in the limit $\eps\to0$. The equation for the
$U_{2,\eps}^\delta$ is
\begin{equation}
  \label{eq:W1eps}
  \pdr{U_{2,\eps}^\delta}{t} =
\frac{1}{\eps}{\cal L}_\eps^\delta U_{2,\eps}^\delta + S_\eps,
\end{equation}
where
\begin{equation}
  \label{eq:Seps}
  S_\eps = \eps\Big([U_1^\delta,P_1^\delta]
 +\dfrac{1}{2i} ( \{U_1^\delta,P_0^\delta\}-\{P_0^\delta,U_1^\delta\})\Big)-
\eps\pdr{U_1^\delta}{t}
 - {\cal R}_\eps^\delta (U^\delta+\eps U_1^\delta).
\end{equation}
The initial condition for (\ref{eq:W1eps}) is
$U_{2,\eps}^\delta(0,\vx,\bk)=-\eps U_1^\delta(0,\vx,\bk)$ because of
(\ref{no-offdiagonal}), which implies that
$W_0(0,\vx,\bk)=U^\delta(0,\vx,\bk)$. We now use the fact that ${\cal
  L}_\eps^\delta$ is skew-symmetric to obtain the bound
\begin{equation}
  \label{eq:bderror2}
  \|U_{2,\eps}^\delta(t)\|_2\leq \eps\|U_1^\delta(0)\|_2+
\dint_0^t \|S_\eps(s)\|_2ds.
\end{equation}
The analysis of the convergence of the difference of $\bar W_\eps^\delta$ and
$U^\delta$ to zero thus relies on estimating the error term
$S_\eps$. The relevant bounds are provided by the following two lemmas.
Here we denote $\|f\|_{\dot H^s}=\|D^sf\|_{L^2}$.
\begin{lemma}\label{lemma-Seps-bound}
  There exists a constant $C>0$ that depends on the constant in
  the bound (\ref{assump-conv-2-intro}) on the support of $W_0$, and
on the constants $D_i$ in (\ref{d-i}) so that
\begin{equation}\label{eq-Seps-bound}
\|S_\eps\|_2\le
C\left[\frac{\eps}{\delta^{3}}\|U^\delta\|_{H^2}+
\frac{\eps^2}{\delta^{11/2}}\|U^\delta\|_{H^3}\right].
\end{equation}
\end{lemma}
\begin{lemma}\label{lemma-liov-regular}
The $\dot H^s({\mathbb R}^d\times{\mathbb R}^d)$-norm, $s=1,2,3$, of
$U^\delta(t)$ is bounded by
\begin{equation}\label{liouv-hsbd}
\|u_q^\delta(t)\|_{\dot H^s}\le C_s\|u_q^0\|_{\dot H^s}
\exp(s\bar\gamma_\delta t).
\end{equation}
Here $u_q^0=\rm{Tr}[\Pi_qW_0\Pi_q]$, the initial data for the
Liouville equation (\ref{liouville-eq}), the constant $C_s$ is a
deterministic rational function of $\delta$.
\end{lemma}
Note that the prefactors of the type $\delta^{-m}$ in Lemma
\ref{lemma-Seps-bound} are not as important as the terms
$\|U^\delta\|_{H^s}$ since the latter grow exponentially in
$\bar\gamma_\delta\sim\delta^{-3/2}$ according to Lemma
\ref{lemma-liov-regular}. 

{\bf Proof of Lemma \ref{lemma-Seps-bound}}.  Observe that thanks to
(\ref{eq:Seps}), we have
\begin{equation}\label{seps-interm}
\|S_\eps\|_2\le C\left[\frac{\eps}{\sqrt{\delta}}\|U_1^\delta\|_{H^1}+
\eps\left\|\pdr{U_1^\delta}{t}\right\|_2+
\|{\cal R}_\eps^\delta(U^\delta+\eps U_1^\delta)\|
\right].
\end{equation}
We have the following bound for $U_1^\delta$:
\begin{equation}\label{w1-bd}
\|U_1^\delta\|_{H^s}\le \frac{C}{\delta^{2s}}\|U^\delta\|_{H^{s+1}}
\end{equation}
with a constant $C>0$ that depends only on the constant in the bound
(\ref{assump-conv-2-intro}) on the support of $W_0$ and on the
constants $D_i$ in (\ref{d-i}).  Indeed, expression (\ref{eq:W1form})
implies that $\|U_1^\delta\|_{H^s}\le C\delta^{\frac 12-s}\|{\cal
  B}(U^\delta)\|_{H^{s}}$, while we have $\|{\cal
  B}(U^\delta)\|_{H^s}\le C\delta^{-s-\frac{1}{2}}
\|U^\delta\|_{H^{s+1}}$ so that (\ref{w1-bd}) follows.  This bound is
by no means optimal but will be sufficient for our purposes.
Furthermore, we have
\begin{equation}\label{u1t-bd}
\left\|\pdr{U_1^\delta}{t}\right\|_2\le
C\left\|{\cal B}\left(\pdr{U^\delta}{t}\right)\right\|_{2}\le
\frac{C}{\delta^{3}}\|U^\delta\|_{H^2}.
\end{equation}
In order to complete the bound (\ref{eq-Seps-bound}) for $S_\eps$ we
show that
\begin{equation}\label{Reps-bd}
\|{\cal R}_\eps^\delta f\|_{2}\le \frac{C\eps}{\delta^{3/2}}
\sum_j[\|k_jf\|_{H^2}+\|f\|_{H^2}].
\end{equation}
We only consider ${\cal R}_{\eps,\delta}^{01}$ as the corresponding
bounds for the operators ${\cal R}_{\eps,\delta}^{02}$ and ${\cal
  R}_{\eps,\delta}^1$ are obtained similarly.  We split ${\cal
  R}_{\eps,\delta}^{01}$ as ${\cal
  R}_{\eps,\delta}^{01}=I_{01}-I\!I_{01}$. We have
\begin{eqnarray*}
&&I_{01}f=\frac{1}{\eps}\int e^{i(\bk-\bp)\cdot\vy}
\left(c_\delta(\vx-\frac{\eps\vy}{2})-c_\delta(\vx)+
\frac{\eps}{2}\vy\cdot\nabla c_\delta(\vx)\right)ip_jD^j
f(\vx,\bp)\frac{d\bp d\vy}{(2\pi)^d}\\
&&=\frac{1}{4\eps}\int_0^\eps({\eps}-s)\int e^{i(\bk-\bp)\cdot\vy}y_ly_m
\frac{\partial^2c_\delta(\vx-\dfrac{s\vy}{2})}{\partial x_l\partial x_m}
ip_jD^jf(\vx,\bp)\frac{d\bp d\vy}{(2\pi)^d}ds=
\frac{1}{4\eps}\int_0^\eps({\eps}-s)\tilde I_{01}(s)fds.
\end{eqnarray*}
Moreover, we obtain that
\begin{eqnarray*}
&&\int|\tilde I_{01}(s)f(\vx,\bk)|^2d\vx d\bk=\hbox{Tr}
\int e^{i(\bk-\bp)\cdot\vy-
i(\bk-\bq)\cdot\vz}y_ly_m z_{l'}z_{m'}
\frac{\partial^2c_\delta(\vx-\dfrac{s\vy}{2})}{\partial x_l\partial x_m}
\frac{\partial^2c_\delta(\vx-\dfrac{s\vz}{2})}{\partial x_{l'}\partial x_{m'}}\\
&&\times
p_jq_rD^jf(\vx,\bp)f^*(\vx,\bq)D^r
\frac{d\bp d\vy d\bq d\vz d\vx d\bk}{(2\pi)^{2d}}\\
&&=\hbox{Tr}
\int e^{i(\bq-\bp)\cdot\vy}y_ly_m y_{l'}y_{m'}
\frac{\partial^2c_\delta(\vx-\dfrac{s\vy}{2})}{\partial x_l\partial x_m}
\frac{\partial^2c_\delta(\vx-\dfrac{s\vy}{2})}{\partial x_{l'}\partial x_{m'}}
p_jq_rD^jf(\vx,\bp)f^*(\vx,\bq)D^r
\frac{d\bp d\vy d\bq d\vx}{(2\pi)^{d}}\\
&&\le \frac{C}{\delta^3}\sum_j\|k_jf\|_{H^2}^2.
\end{eqnarray*}
Therefore the Minkowski inequality implies that
$\|I_{01}f\|_2\le C\eps \delta^{-3/2}\sum_j\|k_jf\|_{H^2}$,
and the same bound holds for $I\!I_{01}$. The operators ${\cal
R}_{\eps}^{02}$ and ${\cal R}_\eps^1$ may be bounded in a similar way as
$\|{\cal R}_\eps^{02}f\|_{L^2}+\|{\cal R}_\eps^{1}f\|_{L^2}
\le C\eps\delta^{-3/2}\|f\|_{H^2}$.
Therefore we have the bound (\ref{Reps-bd}) and then
(\ref{eq-Seps-bound}) follows from (\ref{seps-interm})-(\ref{Reps-bd}). 
\endproof

Theorem \ref{thm1} now follows from the bound (\ref{w1-bd}) for
$U_1^\delta$, the bound (\ref{eq:bderror2}) for $U_{2,\eps}^\delta$,
and Lemmas \ref{lemma-liov-regular} and \ref{lemma-Seps-bound}.  It
only remains to prove Lemma \ref{lemma-liov-regular}, which is done in
Appendix \ref{sec:reg-liouv}.

\section{The Liouville equations in a random medium}\label{sec:random}

We formulate in this section the main result concerning the
convergence of the expectation of the solution of the Liouville
equation (\ref{liouv-intro}) to the solution of the phase space
diffusion equation (\ref{diff-intro}) in the limit $\delta\to 0$. We
also show that the values of the solution of the Liouville equation at
different points in the phase space become independent in this limit.
This allows us to establish the self-averaging property in Theorem
\ref{theorem-main}.

\subsection{Preliminaries}
\label{prelim}

We let ${\cal C}_m:=C([0,+\infty);(\R^d)^m)$, and for any $R_1,\cdots,
R_m>0$ we denote by  ${\cal
C}_m(R_1,\cdots,R_m):=C\left([0,+\infty); S^{d-1}_{R_1}
\times\cdots\times S^{d-1}_{R_m}\right)$, where $S^{d-1}_R$ is
the sphere in $\R^d$ of radius $R>0$ centered at $\bze$.
We also let $\pi_t:{\cal C}_m\rightarrow (\R^d)^m$, $t>0$, be the
canonical mapping $\pi_t(K)=(K_1(t),\cdots,K_m(t))$, $
K=(K_1,\cdots,K_m)\in{\cal C}_m$.  For any $u\leq v$ we denote by
${\cal M}^{u,v}_m$ be the $\si$-algebra of subsets of ${\cal C}_m$
generated by $\pi_t$, $t\in[u,v]$, and let ${\cal M}_m :={\cal
M}^{0,+\infty}_m$ and ${\cal T}_m$ be the filtered measurable
space $\left({\cal C}_m,{\cal M}_m , \left({\cal
M}_m^{0,t}\right)_{t\geq0}\right)$. For any set $A\in{\cal
B}(\R^d)$ we denote ${\cal C}(A):=\si\{c_1(\bbx):\,\bbx\in A\}$.

We suppose further that
$ c_1:\R^d\times\Om\rightarrow\R$
is a scalar , measurable, strictly stationary, zero mean random field
that satisfies assumptions presented in Section \ref{sec-rand-assump},
that is, it satisfies the almost sure bounds (\ref{d-i}), is
exponentially $\phi$-mixing (\ref{DR}), and has a
$C^\infty$-correlation function $R(\vx)$.

We define the differential operator
\begin{equation}
\label{61102} {\cal L}F(\bbk)= \sum\limits_{p,q=1}^d
|\bbk|^2D_{p,q}(\hat{\bbk})\partial_{k_p,k_q}^2F(\bbk) +
\sum\limits_{p=1}^d |\bbk|E_{p}(\hat{\bbk})\partial_{k_p}F(\bbk),
\quad F\in C^\infty_0(\R^d\setminus\{\bze\})
\end{equation}
with the diffusion matrix $D$ given by (\ref{diff-matrix}) and the
drift ${\bf E}$ defined by
\[
E_{p}(\hat{\bbk})= -c_0\sum\limits_{q=1}^d
\int\limits_{0}^{+\infty}
s\,\partial_{x_p,x_q,x_q}^3\,R(c_0s\hat{\bbk})\,ds,\quad\forall\,p=1,\cdots,d.
\]
A simple calculation shows that ${\cal L}$ is a
generator of a diffusion on $S_{k_0}^{d-1}$ given by It\^{o} S.D.E.
\begin{equation}
\label{80203} \left\{
\begin{array}{l}
d\bmk(t)=|\bmk(t)|\left(\bcE(\hat{\bmk}(t))\,dt+\sqrt{2}\,\bD^{1/2}(\hat{\bmk}(t))\,d\bB(t)\right)\\
\bmk(0)=\bbk_0\not=\bze.
\end{array}
\right.
\end{equation}
Here  $\bcE=(E_1,\cdots,E_d)$ and $\bB(\cdot)$ is a
$d$-dimensional standard Brownian motion.
\begin{remark}\label{rm10}{\em
A simple calculation shows that the diffusion $\bmk(\cdot)$ given
by (\ref{80203}) is symmetric. Indeed the generator can be written
in the form
\[
 {\cal L}F(\bbk)=  \sum\limits_{p,q=1}^d\partial_{k_p}\left(
|\bbk|^2D_{p,q}(\hat{\bbk})\partial_{k_q}F(\bbk)\right), \quad F\in
C^\infty_0(\R^d\setminus\{\bze\}).
\]
For any $\bbk\not=\bze$ we denote by $\mathfrak Q_{\bbk}$ the law of
such a diffusion starting at $\bbk$, which
 is supported in ${\cal C}_1(k)$, $k=|\bk|$.}
\end{remark}

\begin{remark}
\label{rm10b}{\em
The matrix $\bD:=[D_{p,q}]$ is degenerate since
$\bD(\hat\bbk)\bbk=\bze$ for all $\bbk\in\R^d\setminus\{\bze\}$. It
can be shown however that under fairly general assumptions its
rank equals $d-1$.
\begin{proposition}
\label{lm72601} Suppose that $\hat{R}(\bze)>0$. Then, the rank of
$\bD$ equals  $d-1$.
\end{proposition}
\proof Suppose that $c_0=1$ and let
$H_\bbk:=[\bp\in\R^d:\,\bp\cdot\hat\bbk=0]$ be the hyperplane
orthogonal to $\bbk$. Then,
\begin{eqnarray*}
&&D_{ml}(\hat \bbk)=-\frac 12\int_{-\infty}^\infty
\partial^2_{x_m, x_l} R(s\hat \bbk)\,ds=
\frac 12\int_{-\infty}^\infty \left(\int_{{\mathbb R}^d}e^{is\hat
\bbk\cdot\bp} p_mp_l\hat R(\bp)d\bp\right)ds\\ &&=
\frac{1}{2^d\pi^{d-1}}\int_{H_\bbk}p_mp_l\hat R(\bp) d\bp
\end{eqnarray*}
and hence for any $\vxi\in{\mathbb R}^d$ we have
\begin{equation}
\label{80202} (\bD(\hat
\bbk)\vxi,\vxi)=D_{ml}(\hat\bbk)\xi_m\xi_l=
\frac{1}{2^d\pi^{d-1}}\int_{H_\bbk}(\bp\cdot\vxi)^2
\hat R(\bp) d\bp.
\end{equation}
Suppose that $\vxi\in H_\bbk$.
Then, since $\hat R(\bp)\ge 0$ the left hand side of (\ref{80202}) is
nonnegative.  We claim that in fact $(\bD(\hat \bbk)\vxi,\vxi)> 0$.
Indeed, if otherwise then, since $\hat R$ is continuous, we would have
$\hat R(\bp)(\bp\cdot\vxi)^2=0$ for all $\bp\in H_\bbk$, which is
impossible due to the fact that $\hat R(\bze)>0$ and the set
$H_{\vxi}\cap H_\bbk$ has the linear dimension $d-2$.  \endproof The
above argument shows that $\bD(\hat\bbk)$ is of rank $d-1$ if there
exists $\bp_0\in H_\bbk$ such that $\hat R(\bp_0)>0$. On the other
hand, if $\hat R(\bp)=0$ for all $\bp$ in the plane $H_\bbk$ then
$\bD(\hat\bbk)\vxi=0$ for all $\vxi\in{\mathbb R}^d$. Therefore the matrix
$\bD(\hat\bbk)$ either has rank $d-1$, or vanishes identically.
Another condition ensuring the latter does not happen is the radial
symmetry of $ R(\cdot)$. }
\end{remark}

\subsection{Two particle model}
We would like to show that solution $u^\delta(t,\vx,\bk)$ of
(\ref{liouv-intro}) decorrelates in the limit $\delta\to 0$ at two
different points, that is, that
\begin{equation}\label{decorrel}
\E\left\{u^\delta(t,\vx_1,\bk_1)
u^\delta(t,\vx_2,\bk_2)\right\}-\E\left\{u^\delta(t,\vx_1,\bk_1)\right\}
\E\left\{u^\delta(t,\vx_2,\bk_2)\right\}\to 0~~\hbox{as $\delta\to 0$}
\end{equation}
provided that $\bk_1\neq \bk_2$.  Recall that $u^\delta(t,\vx,\bk)$
may be represented as
\[
u_{q}^\delta(T,\bx,\bk)=
u_q^0(\vX^\delta(T,\vx,\bk),-\bK^\delta(T,\vx,\bk)),
\]
where $u_q^0$ is the initial data for (\ref{liouv-intro}), and
\begin{eqnarray}
      \dr{\vX^\delta(t)}{t} &=& \pdr{\lambda_q^\delta}{\bk}(\bX^\delta(t),\bK^\delta(t)) ,\qquad
    \vX^\delta(0)=\bx \nonumber\\
    \dr{\bK^\delta(t)}{t} &=& -\pdr{\lambda_q^\delta}{\bx}(\bX^\delta(t),\bK^\delta(t)),\qquad
    \bK^\delta(0)=-\bk.  \label{eq:odes}
\end{eqnarray}
In order to establish (\ref{decorrel}) we have to consider motion of two
particles that may start at the same physical point but are moving in
different directions.  The equations of motion for two
particles are governed by the Hamiltonian system
\begin{equation}\label{eq1}
\left\{
  \begin{array}{l}
\frac{d\bmx_{m}^{(\delta)}(t;\bbx_m,\bbk_m)}{dt}=\nabla_\bbk
\lambda_q^{\delta} \left(\bmx_{m}^{(\delta)}(t;\bbx_m,\bbk_m),
\bmk_{m}^{(\delta)}(t;\bbx_m,\bbk_m)\right)\\
\frac{d\bmk_{m}^{(\delta)}(t;\bbx_m,\bbk_m)}{dt} =-\nabla_\bbx
\lambda_q^{\delta} \left(\bmx_{m}^{(\delta)}(t;\bbx_m,\bbk_m),
\bmk_{m}^{(\delta)}(t;\bbx_m,\bbk_m)\right)\\
\bmx_{m}^{(\delta)}(0;\bbx_m,\bbk_m)=\bbx_m,
\quad\bmk_{m}^{(\delta)}(0;\bbx_m,\bbk_m)=\bbk_m,~~m=1,2.
  \end{array}
\right.
\end{equation}
We will assume that $\bbx_1=\bbx_2=\bze$, and
\begin{equation}\label{81001}
  \bbk_1\not=\bze,\quad
 \bbk_2\not=\bze\mbox{ and }\hat\bbk_1\not=\hat\bbk_2.
\end{equation}
The above system can be rewritten in the form
\begin{equation}\label{eq1b}
\left\{
  \begin{array}{l}
\frac{d\bmx_{m}^{(\delta)}(t;\bbx_m,\bbk_m)}{dt}=
\left[c_0+\sqrt{\delta}
c_1\left(\frac{\bmx_{m}^{(\delta)}(t;\bbx_m,\bbk_m)}{\delta}\right)
\right]\hat{\bmk}_{m}^{(\delta)}(t;\bbx_m,\bbk_m)\\
\frac{d\bmk_{m}^{(\delta)}(t;\bbx_m,\bbk_m)}{dt}
=-\frac{1}{\sqrt{\delta}}\nabla_\bbx
c_1\left(\frac{\bmx_{m}^{(\delta)}(t;\bbx_m,\bbk_m)}{\delta}\right)
|\bmk_{m}^{(\delta)}(t;\bbx_m,\bbk_m)| \\
\bmx_{m}^{(\delta)}(0;\bbx_m,\bbk_m)=\bze,
\quad\bmk_{m}^{(\delta)}(0;\bbx_m,\bbk_m)=\bbk_m,~~m=1,2.
  \end{array}
\right.
\end{equation}
The main result of this section is the following.
\begin{theorem}
\label{thm2-main} Suppose that the random field $c_1(\cdot)$ satisfies the
assumptions in Section \ref{sec-rand-assump} and that $d\geq3$.
Then, the laws of processes
$(\bmk_{1}^{(\delta)}(\cdot),\bmx^{(\delta)}_{1}(\cdot),
\bmk_{2}^{(\delta)}(\cdot),\bmx^{(\delta)}_{2}(\cdot))$ determined
by $(\ref{eq1})$,
  converge weakly in ${\cal C}_4$, as $\delta\rightarrow0$, to the law
of
$(\bmk_1(\cdot),\bmx_1(\cdot),\bmk_2(\cdot),\bmx_2(\cdot))$, where
$\bmk_j(\cdot)$, $j=1,2$ are independent symmetric diffusions given
by $(\ref{80203})$ starting at $\bbk_j$, $j=1,2$ respectively
and
\[
\bmx_j(t)=-c_0\int\limits_0^t\hat{\bmk}_j(s)ds, \quad j=1,2.
\]
\end{theorem}
Theorem \ref{theorem-main} is a simple corollary of Theorems \ref{thm1} and
\ref{thm2-main}.\\
{\bf Proof of Theorem \ref{theorem-main}}. First we observe that
\begin{eqnarray*}
\int \left|\int (W_\eps^\delta(t,\vx,\bk)-U^\delta(t,\vx,\bk)S(\bk)d\bk
\right|^2d\vx\le\|S\|_{L^2}^2
\int|W_\eps^\delta(t,\vx,\bk)-U^\delta(t,\vx,\bk)|^2 d\bk d\vx\to 0
\end{eqnarray*}
as $(\eps,\delta)\to 0$ in ${\cal K}_\mu$ and this convergence is
uniform in realizations of the random medium provided that the bounds
(\ref{d-i}) are satisfied. Therefore it suffices to study $\tilde
s^\delta(\vx)=\int U^\delta(t,\vx,\bk)S(\bk)d\bk$.
We observe that
\begin{eqnarray*}
&&\E\left\{\int \|\tilde s^\delta(\vx)-\bar s(\vx)\|^2d\vx\right\}=
\E\left\{\int\left\|\int(U^\delta(t,\vx,\bk)-\bar W(t,\vx,\bk))
S(\bk)d\bk\right\|^2d\vx\right\}\\
&&=\E\left\{S^*(\bk)_1\int(U^{\delta*}(t,\vx,\bk_1)-\bar W^*(t,\vx,\bk_1))
(U^\delta(t,\vx,\bk_2)-\bar W(t,\vx,\bk_2))S(\bk_2)d\bk_1d\bk_2d\vx
\right\}
\end{eqnarray*}
with $\bar s(\vx)$ and $\bar W(t,\vx,\bk)$ as in the formulation of
Theorem \ref{theorem-main}.  Theorem
\ref{thm2-main} implies that
\[
\E\left\{U^\delta(t,\vx,\bk)\right\}\to\bar W(t,\vx,\bk),~~
\E\left\{U^\delta(t,\vx,\bk_1)U^\delta(t,\vx,\bk_2)\right\}\to
\bar W(t,\vx,\bk_1)\bar W(t,\vx,\bk_2)
\]
pointwise in $\vx$ and $\bk$.  Recall that the functions
$U^\delta(t,\vx,\bk)$ and $\bar W(t,\vx,\bk)$ are uniformly compactly
supported and bounded in $L^\infty$. Therefore the Lebesgue dominated
convergence implies that
\[
\E\left\{\int \|\tilde s^\delta(\vx)-\bar s(\vx)\|^2d\vx\right\}\to 0
\]
and the proof of Theorem \ref{theorem-main} is complete. \endproof

\section{Proof of Theorem \ref{thm2-main}}
\label{asec1}

Before we present the proof of this result we wish to spend a few
words to lay out its main ideas. They are based in large part on
the ideas of \cite{kp} where the phase space diffusion equation
for the limit of the expectation of the solution of the Liouville
equation with the Hamiltonian
$H^\delta(\vx,\bk)=k^2/2+\sqrt{\delta}V(\vx/\delta)$ has been
obtained. The two-particle case introduces some additional
difficulties into the problem. Our first step in the proof, in
Section \ref{sec-cutoff} below, is to replace the processes
$(\bmk_1^{\delta}(\cdot),\bmk_2^{\delta}(\cdot))$ by
$(\bl_1^{\delta}(\cdot),\bl_2^{\delta}(\cdot))$ that agree with
$(\bmk_1^{\delta}(\cdot),\bmk_2^{\delta}(\cdot))$ up to certain
stopping times. These times are  determined by the stopping rules,
introduced by multiplying the Hamiltonian
$\lambda^{\delta}(\vx,\bk)$ by several cut-off functions.
 Their role is to prevent the trajectory of each particle to self-intersect
and also  not to allow the particles to get too close to each
other. We shall prove tightness of such modified processes by
showing that for any bounded, positive and continuous function $F$
one can find a constant $C>0$ such that
$F(\bl_1^{\delta}(t),\bl_2^{\delta}(t))+Ct$, $t\geq0$ are
sub-martingales (see e.g. \cite{stvar} Theorem 1.4.6), cf
(\ref{73101}). This fact will be established thanks to the
decorrelation properties of the random field $\nabla_\bbx
c_1(\cdot)$. More precisely, the latter imply mixing lemmas
contained in Section \ref{secmix}. The second ingredient of the
proof is a perturbative argument that allows us to replace the
trajectory $\bmx_i^{(\delta)}(\cdot)$ (in fact its modification
$\by^{(\delta)}_i(\cdot)$ that arises from the replacement of
$\bmk^\delta$ by $\bl^\delta$) by a linear approximation over the
time interval that is much longer than the correlation time (that
we recall is of order $O(\delta)$) yet is sufficiently short so we
can control the accuracy of the approximation, cf. Lemma
\ref{lm1}. In  order to ensure that the approximate motion (under
linear approximation) is not transverse to the direction of the
field at a given time, which could prevent us from using the
decorrelation properties of the field, but is rather propelled
forward, we have to introduce another stopping time rule, cf. the
condition on the scalar product of wave number directions
contained in (\ref{def1}).

Conducting the proof of tightness we also identify a certain
martingale property of any limiting law of
$(\bl_1^{\delta}(\cdot),\bl_2^{\delta}(\cdot))$, as
$\delta\rightarrow0$ that holds up to the aforementioned stopping
time. By proving that this time goes to infinity with the removal
of the cut-offs we are able to prove both the weak convergence of
the laws of $(\bmk_1^{\delta}(\cdot),\bmk_2^{\delta}(\cdot))$  and
identify a well-posed martingale problem associated with the
limiting measure. This step is done in Section \ref{cutoff}.

With no loss of generality we shall assume throughout this section
that $c_0=1$.

\subsection{The cut-off functions}\label{sec-cutoff}

\label{3.1.b}
Let $p,q>0$ and $  k\geq0$ be integers. Let $M$ be chosen in such a way that
\begin{equation}\label{80401}
 M\geq |\bbk_1|\vee|\bbk_2|\quad\mbox{ and }\quad|\bbk_1|\wedge|\bbk_2|\geq
 M^{-1}.
\end{equation}
Let $\hat\bbk_1\not=\hat\bbk_2$ be such as in the statement of
Theorem \ref{thm2-main}. Denote
\begin{equation}\label{81102}
{{\cal\bf K}}_N:= \left\{(\hat\bbk,\hat\bbk'):\, \,(\hat\bbk,\hat
\bbk_1)_{\R^d}\geq 1-\frac{1}{N+1}, \,(\hat\bbk',\hat
\bbk_2)_{\R^d}\geq 1-\frac{1}{N+1}\right\}
\end{equation}
and choose $N$ a positive integer such that
\begin{equation}\label{81101}
  \ga_N:=\inf\left\{|\hat\bbk-\hat\bbk'|:(\hat\bbk,\hat\bbk')\in {\cal\bf
  K}_N\right\}>0,
\end{equation}
that is, the cones of aperture $1/(N+1)$ centered at $\hat\bk_1$
and $\bbk_2$ are separated. As a consequence of (\ref{81101}) we
may choose a positive integer $q$ so that
\begin{equation}\label{81101b}
  \la_N(p):=
  \inf\left\{\left|\frac{1}{p}\hat\bbk-\rho\hat\bbk'\right|\wedge
  \left|\frac{1}{p}\hat\bbk'-\rho\hat\bbk\right|:\,\rho\in\left[0,\frac{1}{p}\right],
  \quad(\hat\bbk,\hat\bbk')\in {\cal \bf K}_N\right\}\geq\frac{4}{q}.
\end{equation}
We define now several auxiliary functions that will be used to
introduce the cut-offs in the dynamics. The function
$\psi:\R^d\times (S_1^{d-1})^2\rightarrow [0,1]$ is $C^\infty$ and
has the property that
\begin{equation} \label{def1}
\psi(\bbk,\bbl_1,\bbl_2)=\left\{
\begin{array}{l}
1, \phantom{aaaaaa}\mbox{ if    }~~\hat{\bbk}\cdot \bbl_1\geq
1-\frac{1}{N+1}~~\mbox{ and }~~\hat{\bbk}\cdot \bbl_2\geq
1-\frac{1}{N+1}\vphantom{\lim\limits_{s^{(p_1)}_{k}}}\\\phantom{aaaaaa}
\vphantom{\lim\limits_{s^{(p_1)}_{k}}}\phantom{aaaaaa}\mbox{ and
}\phantom{aaaaaa}M^{-1}\leq |\bbk|\leq M\\
0,\phantom{aaaaaa}\vphantom{\int\limits_{s^{(p_1)}_{k}}} \mbox{ if
}~~\hat{\bbk}\cdot \bbl_1\leq 1-\frac{2}{N+1}~~\mbox{ or
}~~\hat{\bbk}\cdot \bbl_2\leq
1-\frac{2}{N+1}\\\phantom{aaaaaa}\phantom{aaaaaa}~~\mbox{ or
}~~|\bbk|\leq
(2M)^{-1}\vphantom{\int\limits_{s^{(p_1)}_{k}}}\phantom{aaa}
\mbox{ or }\phantom{aaa} |\bbk|\geq 2M.
\end{array}
\right.
\end{equation}
The function $\phi_k: \R^d\times{\cal C}_1\rightarrow [0,1]$ is
$C^\infty$ for a fixed path $K(t)$ and satisfies
\begin{equation}
\label{def11} \phi_k(\bby;K)=\left\{
\begin{array}{ll}
1,&  \mbox{ if    }~~\inf\limits_{0\leq t\leq
t^{(p)}_{k-1}}\left|\bby-\int\limits_0^tK(s)ds\right|\geq
\frac{2}{q}\\ 0,& \mbox{ if    }~~\inf\limits_{0\leq t\leq
t^{(p)}_{k-1}}\left|\bby-\int\limits_0^tK(s)ds\right|\leq
\frac{1}{q}.
\end{array}
\right.
\end{equation}
Here $t^{(p)}_k:=kp^{-1}$ and, by convention, $K(s):=K(0)$,
$s\leq0$. The function $\xi_k: \R^d\times \R^d\times{\cal
C}_2\rightarrow [0,1]$
is smooth when the paths $K_1(\cdot),K_2(\cdot)\in {\cal C}_1 $
are fixed. We let
\begin{equation}\label{70501}
p_1:=  2^q[8(1+D_0)] p
\end{equation}
and $s^{(p_1)}_k:=kp_1^{-1}$ be a sub-partition of $t_k$, 
and define
\begin{equation}
\label{def111} \xi_k(\bby_1,\bby_2;K_1(\cdot),K_2(\cdot))=\left\{
\begin{array}{lr}
1,& \mbox{ if    }~~\inf\limits_{0\leq t\leq
s^{(p_1)}_{k}}\left|\bby_1-\int\limits_0^t K_2(s)ds\right|\geq
\frac{2}{q}\\ \phantom{aaaaaa}\mbox{ and
}&\phantom{aaaa}\inf\limits_{0\leq t\leq
s^{(p_1)}_{k}}\left|\bby_2-\int\limits_0^t K_1(s)ds\right|\geq
\frac{2}{q}\\ 0,& \mbox{ if    }~~\inf\limits_{0\leq t\leq
s^{(p_1)}_{k}}\left|\bby_1-\int\limits_0^t K_2(s)ds\right|\leq
\frac{1}{q}\\ \phantom{aaaaa}\mbox{or}&~~\inf\limits_{0\leq t\leq
s^{(p_1)}_{k}}\left|\bby_2-\int\limits_0^t K_1(s)ds\right|\leq
\frac{1}{q}.
\end{array}
\right.
\end{equation}
For $j=1,2$ we set
\begin{equation}
\label{def2} \Phi_j(t,\bby;K(\cdot)):=\left\{
\begin{array}{ll}
1,& \mbox{ if    }0\leq t< t^{(p)}_1\\
\phi_k(\bby;K(\cdot)),& \mbox{ if }t^{(p)}_k\leq t< t^{(p)}_{k+1}.
\end{array}
\right.
\end{equation}
Each $\Phi_j(\cdot)$ shall be used to modify the dynamics of the
corresponding particle in order to avoid a possibility of
self-intersections of its trajectory. The cut-off
function
\begin{equation} \label{def21}
\Psi(t,\bbk;K(\cdot)):=\left\{
\begin{array}{ll}
\psi\left(\bbk,\hat{K}\left(t^{(p)}_{k-1}\right),
\hat{K}\left(t^{(p)}_{k}\right)\right)&\mbox{ for }t\in[t_k^{(p)},t_{k+1}^{(p)})\mbox{ and }k\geq1\\
\psi(\bbk,\hat{K}(0), \hat{K}(0))&\mbox{ for }t\in[0,t_{1}^{(p)})
\end{array}
\right.
\end{equation}
will allow us to control the direction of the particle motion over
each interval of the partition as well as not to allow the
trajectory to escape to the regions where the change of velocity
can be uncontrollable. The cut-off
\begin{equation} \label{def211}
\Xi(t,\bby_1,\bby_2;K_1(\cdot),K_2(\cdot))=\left\{
\begin{array}{ll}
1,& \mbox{ if    }0\leq t< t^{(p)}_{1}\\
\xi_k(\bby_1,\bby_2;K_1(\cdot),K_2(\cdot)),& \mbox{ if
}s^{(p_1)}_{k}\leq t< s^{(p_1)}_{k+1}\mbox{ and }t^{(p)}_{1}\leq
s^{(p_1)}_{k}
\end{array}
\right.
\end{equation}
is introduced in order not to allow the two trajectories to come
too close to each other. Note that this cut-off is "switched on"
only after time $t=t_1^{(p)}$ so as to allow the two particles to
separate initially. After this time it is updated every $1/p_1$
time step, that is, more frequently that the cut-offs that control
the self-intersections of each trajectory that are updated only at
each $1/p$ time step.

The following lemma can be checked by a direct calculation.  Both here
and in what follows we denote by $D_{\bullet,\bt}$ the partial with
respect to the $\bt$ component of the given vector variable.
\begin{lemma}
\label{lm3} Let ${\bf m}=(m_1,\cdots,m_d)$ be a multi-index with
nonnegative integer valued components, $m=\sum\limits_{p=1}^dm_p$.
There exist constants $C_3$, $C_4>0$ depending
only on $M,N,p,q$, $m$ such that
\[
|D_\bby^{\bf m}\Phi_j(t,\bby)|\leq C_3,\quad |D_{\bby_j}^{\bf
m}\Xi(t,\bby_1,\bby_2)|\leq C_4,\quad j=1,2.
\]
\end{lemma}

Let $K=(K_1,K_2)\in{\cal C}_2$ and denote
\begin{equation}
\label{70605}
\Theta_j(s,\by_1,\by_2,\bbl;K):=\Psi(s,\bbl;K_j)\Phi_j\left(s,\by_j;K_j\right)
\Xi\left(s,\by_1,\by_2;K\right),
\end{equation}
\begin{equation}
\label{70605b}
\Lambda_j(s,\bby_1,\bby_2,\bby_1',\bby_2',\bbl;K)
:=\Theta_j(s,\bby_1,\bby_2,\bbl;K)\Theta_j(s,\bby_1',\bby_2',\bbl;K).
\end{equation}
We also introduce a random transformation of paths
 $ \tilde K(\cdot)=(\tilde K_1(\cdot),\tilde K_2(\cdot))$ for
any $K\in{\cal C}_2$  given by
\begin{equation}
\label{80405} \tilde{K}_j(t)=\left[1+\sqrt{\delta}
c_1\left(\frac{K_j(t)}{\delta}\right) \right]\hat{K}_{j}(t),\quad
t\geq0 .
\end{equation}
Finally, let us set
\begin{equation}
\label{70613} F_j(t,\bby_1,\bby_2,\bbl;K)
=\Theta_j(t,\delta\bby_1,\delta\bby_2,\bbl;K) \nabla_{\bby_j}
c_1\left(\bby_j\right)|\bbl|,\quad j=1,2.
\end{equation}

The modified two particle system with the cut-offs that we will
consider is given by
\begin{equation}\label{eq2}
\left\{
  \begin{array}{l}
\frac{d\by_j^{(\delta)}(t)}{dt}=\left[1+\sqrt{\delta}
c_1\left(\frac{\by_{j}^{(\delta)}(t;\bbx_j,\bbk_j)}{\delta}\right)
\right]\hat{\bl}_{j}^{(\delta)}(t;\bbx_j,\bbk_j)\\
\frac{d\bl_j^{(\delta)}(t)}{dt}=-\frac{1}{\sqrt{\delta}}\,F_j\left(t,
\frac{\by_1^{(\delta)}(t)}{\delta},\frac{\by_2^{(\delta)}(t)}{\delta},\bl_j^{(\delta)}(t);
\tilde{\bl}^{(\delta)}(\cdot)\right)\\
\by_j^{(\delta)}(0)=\bze,\quad\bl_j^{(\delta)}(0)=\bbk_j,\quad
j=1,2,
  \end{array}
\right.
\end{equation}
where the path
$\tilde{\bl}^{(\delta)}(\cdot)=(\tilde{\bl}^{(\delta)}_1(\cdot),\tilde{\bl}^{(\delta)}_2(\cdot))$
is obtained from $\bl(\cdot)$ by the transformation (\ref{80405}).  We
will denote by $Q^{\delta}(\cdot;M,N,p,q)$ the law of
$(\bl_1^{(\delta)}(\cdot),\by_1^{(\delta)}(\cdot),\bl_2^{(\delta)}(\cdot),\by_2^{(\delta)}(\cdot))$
for a given $\delta>0$ over ${\cal C}_4$.

\subsection{The Mixing Lemmas}
\label{secmix}
For any $t\geq0$ we denote by ${\cal F}_t$ the $\si$-algebra generated
by $(\bl_1^{(\delta)}(s),\bl_2^{(\delta)}(s))$, $s\leq t$.
Throughout this section we assume that $X_1,X_2:\R\times\R^d\times
\R^{d^2}\rightarrow\R$ are certain continuous functions, $Z$ is a
random variable and $g_1,g_2$ are $\R^d$-valued random vectors. We
suppose further that $Z,g_1,g_2$, are ${\cal F}_t$-measurable,
while $X_1,X_2$ are random fields of the form $
X_i(\bbx)=X_i\left(c_1(\bbx),\nabla_\bbx c_1(\bbx),\nabla_\bbx^2
c_1(\bbx)\right), $ satisfy $
\lim\limits_{|\bbx|\rightarrow0}\|X_i(\bbx)-X_i(\bze)\|_\infty=0,\quad
i=1,2. $ We also let
\begin{equation}\label{80101}
U(\theta_1,\theta_2):= \bbE\left[X_1(\theta_1)X_2(\theta_2)\right]
,\quad (\theta_1,\theta_2)\in(\R^d)^2.
\end{equation}
The following mixing lemmas will be of crucial importance for us
in the sequel.
\begin{lemma}
\label{mix1} Assume that $r,t\geq0$ and
\begin{equation} \label{70202}
 \inf\limits_{u\leq t}\left|g_i-\frac{\by_j^{(\delta)}(u)}{\delta}\right|\geq
\frac{r}{\delta}
\end{equation}
$\bbP$--a.s. on the set $Z\not=0$ for  $i,j=1,2$. Then, we have
\begin{equation}
\label{70201}
\left|\bbE\left[X_1(g_1)X_2(g_2)Z\right]-\bbE\left[
U(g_1,g_2) Z\right]\right| \leq
2\phi\left(\frac{r}{2\delta}\right)\|X_1\|_{\infty}\|X_2\|_{\infty}\|Z\|_1.
\end{equation}
\end{lemma}
\proof The proof is a modification of the proof of Lemma 2 of
\cite{kp} so we only highlight its main points.  Choose an arbitrary
$\eta>0$. By a suitable modification of $g_1,g_2$ on the event $Z=0$,
so that the modified r.v. remain ${\cal F}_t$--measurable, we can
guarantee that (\ref{70202}) holds $\bbP$--a.s. Let
\[
\bbi=(i_1,\cdots,i_d)\in \mathbb Z^d\mbox{ and
}C_\bbi:=[i_1/2^{M_1},(i_1+1)/2^{M_1})\times\cdots\times[i_d/2^{M_1},(i_d+1)/2^{M_1})
\]
and
\[
c_\bbi:=((2i_1+1)/2^{M_1+1},\cdots,(2i_d+1)/2^{M_1+1}).
\]
Here $M_1>0$ is a sufficiently large integer so that
\begin{equation}\label{7301}
\|X_i(\bbx)-X_i(c_\bbi)\|_\infty\leq
\eta,\quad\forall\,\bbi\in\mathbb Z^d,\,\bbx\in C_\bbi,\,i=1,2
\end{equation}
and $2^{-M_1}<r/(20\delta)$. We let
\[
D_{\bbi,\bbj}:=[\bbz:\,dist(\bbz,C_\bbi\cup
C_\bbj)>r(2\delta)^{-1}]
\]
and
\[
\mathfrak
Y^{(\delta)}_t:=
\left[\frac{1}{\delta}(\by_1^{(\delta)}(s),\by_2^{(\delta)}(s)):s\leq
t\right] .
\]
Let us denote by $I_{\bbi,\bbj}$ the indicator of the event
$[(g_1,g_2)\in C_\bbi\times C_\bbj]$ and the event $
A_{\bbi,\bbj}=[\om:\,\mathfrak Y^{(\delta)}_t(\om)\subseteq
D_{\bbi,\bbj}]. $ Note that
\begin{equation}
\label{7302}
\bbE\left[X_1(g_1)X_2(g_2)Z\right]=\sum\limits_{\bbi,\bbj}
\bbE\left[X_1(g_1)X_2(g_2)ZI_{\bbi,\bbj}\chi_{A_{\bbi,\bbj}}\right].
\end{equation}
Using precisely the same argument as in \cite{kp} we prove that
$ZI_{\bbi,\bbj}\chi_{A_{\bbi,\bbj}}$ is ${\cal
C}(D_{\bbi,\bbj})$--measurable for each $\bbi,\bbj\in\mathbb Z^d$.
Note however that the right hand side of (\ref{7302}) is equal, up
to a term of order $O(\eta)$, to
\begin{equation}\label{7303}
\sum\limits_{\bbi,\bbj}
\bbE\left[X_1(c_\bbi)X_2(c_\bbj)ZI_{\bbi,\bbj}\chi_{A_{\bbi,\bbj}}\right].
\end{equation}
The random variable $X_1(c_\bbi)X_2(c_\bbj)$ is however ${\cal
C}(C_\bbi\cup C_\bbj)$--measurable. Therefore we can write, see
e.g. \cite{bill} p.171, that
\begin{equation}\label{80103}
 \sum\limits_{\bbi,\bbj}
\left|\bbE\left[X_1(c_\bbi)X_2(c_\bbj)ZI_{\bbi,\bbj}\chi_{A_{\bbi,\bbj}}\right]-
U(c_\bbi,c_\bbj)\bbE\left[ZI_{\bbi,\bbj}\chi_{A_{\bbi,\bbj}}\right]\right|
\end{equation}
\[\leq
\sum\limits_{\bbi,\bbj}
\phi\left(\frac{r}{2\delta}\right)\left|\bbE\left[ZI_{\bbi,\bbj}\chi_{A_{\bbi,\bbj}}\right]\right|
\|X_1\|_{\infty}\|X_2\|_{\infty}.
\]
However, $U(c_\bbi,c_\bbj)$ equals, up to a term of order
$O(\eta)$, to $U(g_1,g_2)$ on the event corresponding to
$I_{\bbi,\bbj}$. The conclusion of Lemma \ref{mix1} follows upon
the passage to the limit $M_1\rightarrow+\infty$ and $\eta\da$.
\endproof

\begin{lemma}
\label{mix2} Assume that $r,t$ are as in the previous lemma.
Let $\bbE X_1=0$. Furthermore, we assume that $g_2$ satisfies
$(\ref{70202})$,
\begin{equation} \label{70202b}
 \inf\limits_{u\leq t}\left|g_1-\frac{\by_j^{(\delta)}(u)}{\delta}\right|\geq
\frac{r+r_1}{\delta},\quad j=1,2
\end{equation}
and
\begin{equation}\label{70601}
  |g_1-g_2|\geq \frac{r_1}{\delta},
  \end{equation}
for some $r_1\geq0$, $\bbP$-a.s. on the event  $Z\not=0$.
\commentout{

 Let also
\[
V(\theta_1,\theta_2):=
\bbE\left[X_1\circ\tau_{\theta_1}X_2\circ\tau_{\theta_2}\right]
,\quad (\theta_1,\theta_2)\in(\R^d)^2.
\]

 Let also
\[
V(\theta_1,\theta_2):=
\bbE\left[X_1\circ\tau_{\theta_1}X_2\circ\tau_{\theta_2}\right]
,
\]
\[
\tilde{V}(\theta_1,\theta_2):=
X_1\circ\tau_{\theta_1}X_2\circ\tau_{\theta_2}-V(\theta_1,\theta_2)
,\quad (\theta_1,\theta_2)\in(\R^d)^2,
\]

} Then we have
\begin{equation}
\label{70201b}
\left|\bbE\left[X_1(g_1)X_2(g_2)\,Z\right]-\bbE\left[
U(g_1,g_2) Z\right]\right|
\leq C_5\phi^{1/2}\left(\frac{r}{2\delta}\right)
\phi^{1/2}\left(\frac{r_1}{2\delta}\right)\|X_1\|_{\infty}\|X_2\|_{\infty}\|Z\|_1
\end{equation}
for some absolute constant $C_5>0$ 
Here the function $U$ is given by $(\ref{80101})$.
\end{lemma}
\proof We prove that the left hand side of (\ref{70201b}) is bounded
by
\begin{equation}\label{80102}
  C_6
\phi\left(\frac{r_1}{2\delta}\right)\|X_1\|_{\infty}\|X_2\|_{\infty}\|Z\|_1.
\end{equation}
This together with the result of the previous lemma imply
(\ref{70201b}).

Let $\eta>0$ and $M_1$ be as in the proof of Lemma \ref{mix1}, and
in addition $2^{-M_1}<r_1/(20\delta)$.
 Note that $X_2(c_\bbj)ZI_{\bbi,\bbj}\chi_{A_{\bbi,\bbj}}$ (in
the notation of the proof of Lemma \ref{mix1}) is ${\cal
C}(D_{\bbi,\bbj}\cup C_\bbj)$--measurable. In addition, we have
$dist(C_\bbi,D_{\bbi,\bbj}\cup C_\bbj)> r_1(2\delta)^{-1}$ thus,
using the mixing coefficient as in e.g. \cite{bill} p.171 we can
estimate
\[
\sum\limits_{\bbi,\bbj}
\left|\bbE\left[X_1(c_\bbi)X_2(c_\bbj)ZI_{\bbi,\bbj}\chi_{A_{\bbi,\bbj}}\right]\right|\leq
2\phi\left(\frac{r_1}{2\delta}\right)\|X_1\|_{\infty}\|X_2\|_{\infty}\|Z\|_1.
\]
On the other hand, we have $I_{\bbi,\bbj}\not=0$ only if
$|c_\bbi-c_\bbj|\geq r_1(2\delta)^{-1}$, which in turn implies
that
\[
|U(c_\bbi,c_\bbj)|\leq C_7
\phi\left(\frac{r_1}{2\delta}\right)\|X_1\|_{\infty}\|X_2\|_{\infty},
\]
with the constant $C_7$ independent of $\eta>0$.
Summarizing, we have shown that 
\[
\sum\limits_{\bbi,\bbj} \left|
U(c_\bbi,c_\bbj)\bbE\left[ZI_{\bbi,\bbj}\chi_{A_{\bbi,\bbj}}\right]\right|\leq
C_8\phi\left(\frac{r_1}{2\delta}\right)\|X_1\|_{\infty}\|X_2\|_{\infty}\|Z\|_1,
\]
with the constant $C_8$ independent of $\eta>0$. Letting
$\eta\rightarrow0$ and using (\ref{7301})
we conclude (\ref{70201b}).
\endproof

\subsection{Tightness and the martingale property of limiting measures}
In this section we prove tightness of the family
$Q^{\delta}(\cdot;M,N,p,q)$, $\delta\in(0,1]$ and show that any
weak limit point $Q(\cdot;M,N,p,q)$ of this family as
$\delta\rightarrow0$, has a certain martingale property.

Let ${\cal L}^{M,N,p,q}$ be a random partial differential operator
defined on $C_0^\infty((\R^d)^2)$ as follows. For any
$K=(K_1,K_2)\in{\cal C}_2$ and  $G\in C_0^\infty((\R^d)^2)$
we set $Y=(Y_1,Y_2)\in{\cal C}_2$,
\begin{equation}\label{80701}
 Y_i(t)=\int\limits_0^tK_i(s)ds,\quad i=1,2,
\end{equation}
\[
\Theta_i(t;K):= \Phi_{i,*}(t;K_i)\Psi_*(t;K_i)\Xi_*(t;K),\
\]
where
\[
\Phi_{i,*}(t;K_i):=\Phi_i\left(t,Y_i(t);K_1\right),
~~ \Psi_*(t;K_i):=\Psi(t,K_i(t);K_i),
~~\Xi_*(t;K):=\Xi\left(t,Y_1(t),Y_2(t);K\right).
\]
We let
\[
({\cal L}^{M,N,p,q}G)(\bbk_1,\bbk_2;K) := \Theta_1^2(t;K){\cal
L}_{\bbk_1}G(\bbk_1,\bbk_2) +\Theta_2^2(t;K) {\cal
L}_{\bbk_2}G(\bbk_1,\bbk_2),
\]
with ${\cal L}_{\bbk_i}$, $i=1,2$ given by (\ref{61102}).

Let  $\zeta\in C_b((\R^d)^{2n})$ be an arbitrary nonnegative
function, let
 $0\leq t_1<\cdots< t_n\leq t<u$ and define
$\zeta(K):=\zeta(K(t_1),\cdots,K(t_n))$. We will show that for any
 function $G\in C_0^\infty((\R^d)^2)$ there exists a
deterministic constant $C_9>0$ 
such that
\begin{eqnarray}
\label{73101}
&&\left|\bbE\left\{\left[G(\bl^{(\delta)}_1(u),\bl^{(\delta)}_2(u))-
G(\bl^{(\delta)}_1(t),\bl^{(\delta)}_2(t))\right]\zeta(\bl^{(\delta)}_1(\cdot),
\bl^{(\delta)}_2(\cdot))\right\}\right|\\
&&\leq C_9(u-t)
\bbE[\zeta(\bl^{(\delta)}_1(\cdot),
\bl^{(\delta)}_2(\cdot))],\quad\forall\,\zeta(\cdot),\delta\in(0,1].\nonumber
\end{eqnarray}
The choice of the constant $C_9$ may depend on a particular
function $G$ but should be the same for all the spatial translates
of $G$, and may not depend on the test function $\zeta$. This,
according to Theorem 1.4.6 of \cite{stvar}, implies tightness of
the laws of $(\bl^{(\delta)}_1(\cdot),\bl^{(\delta)}_2(\cdot))$,
$\delta\in(0,1]$ over ${\cal C}_2$.

Additionally, we prove that if $Q(\cdot;M,N,p,q)$ is any limiting
law of $Q^{\delta_n}(\cdot;M,N,p,q)$, as $\delta_n\rightarrow0$
then
\begin{eqnarray}\label{51401}
&&\lim\limits_{n\rightarrow+\infty}
\bbE\left\{\left[G(\bl^{(\delta_n)}_1(u),\bl^{(\delta_n)}_2(u))-
G(\bl^{(\delta_n)}_1(t),\bl^{(\delta_n)}_2(t))\right]
\zeta(\bl^{(\delta_n)}_1(\cdot),
\bl^{(\delta_n)}_2(\cdot))\right\}\\
&&
~~~~~~~=\int\left\{\left[\int\limits_t^u ({\cal
L}^{M,N,p,q}G)(K(s);K)ds\right]\zeta(K)\right\}Q(dK;M,N,p,q)\nonumber
\end{eqnarray}
for any $u>t$. This property will be used in the next section to
identify the limiting law of
$(\bmk_1^{(\delta)}(\cdot),\bmk_2^{(\delta)}(\cdot))$, as
$\delta\rightarrow0$.

Throughout the remainder of this section we suppress writing both
the superscript $\delta$ and  the cut-off parameters $M,N,p,q$ of
the respective measures.
 With no loss of generality we assume that there exists $k_1$ such
that $s_{k_1}^{(p_1)}\leq t<u\leq s_{k_1+1}^{(p_1)}$, cf.
(\ref{70501}). Given $s\geq\si>0$, we define the linear
approximation
\[
\bmL_j(\si,s):=\by_j(\si)+(s-\si)\hat{\bl}_j(\si),
\]
and
\[
\bmR_j(v,\si,s):=(1-v)\bmL_j(\si,s)+v\by_j(s),\quad j=1,2.
\]
The following simple lemma can be verified by a direct
calculation.
\begin{lemma}
\label{lm1}
Suppose that $s\geq\si$. Then,
\[
|\by_{j}(s)-L_{j}(\si,s)|\leq
\frac{D_1(s-\si)^2}{2\sqrt{\delta}}+D_0\sqrt{\delta}(s-\si),
\quad\forall\,\delta>0,\, j=1,2 .
\]
\end{lemma}
\begin{remark}
{\em Throughout this argument we use
\begin{equation}\label{80703}
  \si(s):=\max[t,s-\delta^{1-\ga_1}]\mbox{ for some }\ga_1\in(0,1/8).
\end{equation}
The above lemma proves that for this choice of $\si$ the linear
approximation $L_{j}(\si,s)$ of the particle position given by
$\by_{j}(s)$ is exact, up to a term of order
\begin{equation}\label{80704}
 O(\delta^{3/2-2\ga_1}).
\end{equation}
}
\end{remark}

We begin now the proof of (\ref{73101}). Our strategy is based on the
perturbation method: the trajectory is approximated by the iterated
linear approximation sufficiently many times so that the error becomes
deterministically small. The terms that involve the linear
approximation are potentially large but are handled with the help of
the mixing lemmas.  Note that
\begin{eqnarray}
\label{51402}
&&G(\bl^{(\delta)}_1(u),\bl^{(\delta)}_2(u))-
G(\bl^{(\delta)}_1(t),\bl^{(\delta)}_2(t))\\
&&
=-\frac{1}{\sqrt{\delta}}\sum\limits_{j,\al}
\int\limits_t^uD_{\bbl_j,\al}G(\bl_1(s),\bl_2(s))
F_{j,\al}\left(s,\frac{\by_1(s)}{\delta},
\frac{\by_2(s)}{\delta},\bl_j(s)\right)ds.\nonumber
\end{eqnarray}
We can rewrite (\ref{51402}) in the form
\begin{equation}
\label{51403} I^{(1)} + I^{(2)} + I^{(3)},
\end{equation}
where
\begin{eqnarray*}
&&I^{(1)}:=-\frac{1}{\sqrt{\delta}}
\sum\limits_{j,\al}\int\limits_t^uD_{\bbl_j,\al}G(\bl_1(\si),\bl_2(\si))
F_{j,\al}\left(s,\frac{\by_1(s)}
{\delta},\frac{\by_2(s)}{\delta},\bl_j(\si)\right)ds,\\
&&
I^{(2)}:=\frac{1}{\delta}\sum\limits_{j,\al}\sum\limits_{i,\bt}
\int\limits_t^u\int\limits_\si^s
D_{\bbl_j,\al}G(\bl_1(\rho),\bl_2(\rho))
D_{\bbl_j,\bt}F_{j,\al}
\left(s,\frac{\by_1(s)}{\delta},\frac{\by_2(s)}{\delta},\bl_j(\rho)\right)\\
&&~~~~~\times F_{j,\bt}\left(\rho,\frac{\by_1(\rho)}{\delta},
\frac{\by_2(\rho)}{\delta},\bl_j(\rho)\right)ds\,d\rho,\\
&&
I^{(3)}:=\frac{1}{\delta}\sum\limits_{j,\al}\sum\limits_{i,\bt}
\int\limits_t^u\int\limits_\si^s
D_{\bbl_i,\bt}D_{\bbl_j,\al}G(\bl_1(\rho),\bl_2(\rho))\\
&&~~~~~\times F_{j,\al}\left(s,\frac{\by_1(s)}{\delta},
\frac{\by_2(s)}{\delta},\bl_j(\rho)\right)
F_{i,\bt}\left(\rho,\frac{\by_1(\rho)}{\delta},
\frac{\by_2(\rho)}{\delta},\bl_i(\rho)\right)ds\,d\rho.
\end{eqnarray*}

\subsubsection{Term  $\bE[I^{(1)}\zeta]$.}\label{seca41}
The term $I^{(1)}$ can be rewritten in the form
\[
J^{(1)}+J^{(2)},
\]
where
\[
J^{(1)}:=-\frac{1}{\sqrt{\delta}}\sum\limits_{j,\al}
\int\limits_t^uD_{\bbl_j,\al}G(\bl_1(\si),\bl_2(\si))
F_{j,\al}\left(s,\frac{\bmL_1(\si,s)}{\delta},
\frac{\bmL_2(\si,s)}{\delta},\bl_j(\si)\right)ds,
\]
and
\begin{eqnarray}
&&J^{(2)}:=-\frac{1}{\delta^{3/2}}\sum\limits_{j,\al}\sum\limits_{i,\bt}
\int\limits_t^u\int\limits_0^1D_{\bbl_j,\al}G(\bl_1(\si),\bl_2(\si))
D_{\bby_i,\bt}F_{j,\al}
\left(s,\frac{\bf\bmR_1(v,\si,s)}{\delta},\frac{\bmR_2(v,\si,s)}{\delta},
\bl_j(\si)\right)\nonumber\\
&&~~~~~~~~~~~\times (y_{i,\bt}(s)-L_{i,\bt}(\si,s))\,ds\,dv.\label{53105}
\end{eqnarray}
Note that we have replaced $\by_j$ by its linearization
$\bmL_j$ in the term $J^{(1)}$. The linear approximation is always
``propelled forward'', which allows us to use Lemma \ref{mix1} to
handle the term $\bE[J^{(1)}\zeta]$.  Suppose that $k$ is such that
$s,t\in[t^{(p)}_k,t^{(p)}_{k+1})$, recall also that
$s,t,u\in[s_{k_1}^{(p_1)}, s_{k_1+1}^{(p_1)})$, and let us fix one
trajectory by setting, for instance, $j=1$. We will use Lemma \ref{mix1}
with
$X_1(\bbx)=-\nabla_\bbx c_1(\bbx)$, $X_2(\bbx)\equiv\bone$,
\[
Z=\Theta_1\left(s^{(p_1)}_{k_1},\frac{\bmL_1(\si,s)}{\delta},
\frac{\bmL_2(\si,s)}{\delta},\bl_1(\si)\right)
|\bl_1(\si)|D_{\bbl_1}G(\bl_1(\si),\bl_2(\si))\zeta
\]
and $g_1=\bmL_1(\si,s)\delta^{-1}$, cf. (\ref{70605}).  We
need to verify (\ref{70202}). Suppose therefore that $Z\not=0$. 
For $\rho\in[0, t^{(p)}_{k-1}]$ we have
$|\bmL_1(\si,s)-\by_1(\rho)|\geq (2q)^{-1}$, provided that
$0<\delta<\left(2q\right)^{-1/(1-\gamma_1)}$. For
$\rho\in[t^{(p)}_{k-1},\si]$ we have
\begin{eqnarray}
\label{80705}
&&(\bmL_1(\si,s)-\by_1(\rho))\cdot
\hat{\bl}_1\left(t^{(p)}_{k-1}\right)\\
&&\geq (s-\si)\hat{\bl}_1\left(\si\right)\cdot
\hat{\bl}_1\left(t^{(p)}_{k-1}\right)+
\int\limits_\rho^\si\left[1+\sqrt{\delta}
c_1\left(\frac{\by_1(\rho_1)}{\delta}\right)\right]
\hat{\bl}_1\left(\rho_1\right)\cdot
\hat{\bl}_1\left(t^{(p)}_{k-1}\right)d\rho_1\nonumber\\
&&
\geq(s-\si)\left(1-\frac{2}{N+1}\right)+(1-\sqrt{\delta}
D_0)(s-\rho)\left(1-\frac{2}{N+1}\right)\geq(s-\si)
\left(1-\frac{2}{N+1}\right),\nonumber
\end{eqnarray}
provided that $\delta<1/D_0^2$. We see from (\ref{80705}) that
(\ref{70202}) is satisfied with $r=\left(1-\frac{2}{N+1}\right)(s-\si)$
and $j=1$.

We verify next that $g_1$ is also separated from $\by_2(\rho)\delta^{-1}$,
$\rho\in[0,\si]$.  Consider two cases. First, when
$s,t\in[0,t^{(p)}_1)$, using condition (\ref{81101}) we obtain then
that there exists $\ga_N'>0$ depending only on $N$ such that
\[
\left|g_1-\frac{\by_2(\rho)}{\delta}\right|\geq
\frac{\ga_N'(s-\si)}{\delta}.
\]

Suppose then that $s,t\geq 1/p$ and
$s,t\in[s_{k_1}^{(p_1)}, s_{k_1+1}^{(p_1)})$.
Then we have for $\rho\in[0, s^{(p_1)}_{k_1}]$, with $p_1$ given by
(\ref{70501}), $|\bmL_1(\si,s)-\by_2(\rho)|\geq (2q)^{-1}$,
provided that $\delta$ is as above. For $\rho\in[s^{(p_1)}_{k_1},\si]$
we get, thanks to (\ref{70501}),
\begin{eqnarray*}
&&|\bmL_1(\si,s)-\by_2(\rho)|\geq
\left|\bmL_1(\si,s)-\by_2\left(s^{(p_1)}_{k_1}\right)\right|-
\left|\by_2\left(s^{(p_1)}_{k_1}\right)-\by_2(\rho)\right|\\
&&~~~~~~~~~~~~~~~~~~~~~~~
\geq\frac{1}{2q}-\frac{1+D_0}{p_1}\geq\frac{1}{4q}\geq(s-\si)
\left(1-\frac{2}{N+1}\right),
\end{eqnarray*}
provided that $\delta<(4q)^{-(1-\ga_1)}$.

Using Lemma \ref{mix1} we estimate
\begin{eqnarray}
\label{70502}
&&\left|\bbE[J^{(1)}\zeta]\right|\leq\frac{MD_0}{\sqrt{\delta}}
\|\nabla G\|_{L^\infty((\R^d)^2)}
\bbE[\zeta]\int\limits_t^u
\,\phi\left(C_{10}\frac{s-\si}{\delta}\right)ds\\
&&~~~~~~~~~~~~~\leq C_{11}(\delta)(u-t)\|\nabla G\|_{L^\infty((\R^d)^2)}
\bbE[\zeta],\nonumber
\end{eqnarray}
where $C_{10}:=\min[\ga_N',1/2(1-2/(N+1))]$, and $C_{11}(\delta)$
depends only on $\delta$ and vanishes as
$\delta\rightarrow0$.

On the other hand, the term $J^{(2)}$ defined by (\ref{53105}) may be
written as
\[
J^{(2)}=J^{(2)}_1+J^{(2)}_2,
\]
where
\begin{eqnarray*}
&&J^{(2)}_1:=-\frac{1}{\delta^{3/2}}\sum\limits_{j,\al}\sum\limits_{i,\bt}
\int\limits_t^u D_{\bbl_j,\al}G(\bl_1(\si),\bl_2(\si))\\
&&~~~~~~~~\times
D_{\bby_i,\bt}F_{j,\al}
\left(s,\frac{\bmL_1(\si,s)}{\delta},\frac{\bmL_2(\si,s)}{\delta},
\bl_j(\si)\right)(y_{i,\bt}(s)-L_{i,\bt}(\si,s))\,ds
\end{eqnarray*}
and
\begin{eqnarray}
\label{53106}
&&J^{(2)}_2:=-\frac{1}{\delta^{5/2}}
\sum\limits_{j,\al}\mathop{\sum\limits_{i,\bt}}\limits_{k,\ga}
\int\limits_t^u\int\limits_0^1\int\limits_0^1
D_{\bby_k,\ga}D_{\bby_i,\bt}F_{j,\al}\left(s,\frac{\bmR_1(\theta
v,\si,s)}{\delta}, \frac{\bmR_2(\theta
v,\si,s)}{\delta},\bl_j(\si)\right)\\
&&~~~~~~~~\times D_{\bbl_j,\al}G(\bl_1(\si),\bl_2(\si))v\,
(y_{i,\bt}(s)-L_{i,\bt}(\si,s))
(y_{k,\ga}(s)-L_{k,\ga}(\si,s))\,ds\,dv\,d\theta.\nonumber
\end{eqnarray}
The second term may be handled easily with the help of Lemma \ref{lm1}
and (\ref{80704}). We have
\begin{equation}
\label{53107}
|\bbE[J^{(2)}_2\zeta]|\leq C_{12}D_2\bbE[\zeta]
\|\nabla
G\|_{L^\infty((\R^d)^2)}(u-t)\delta^{-5/2}\delta^{3-4\ga_1}
\leq C_{13}\delta^{1/2-4\ga_1}(u-t)\bbE[\zeta] \|\nabla
G\|_{L^\infty(\R^{2d})}.
\end{equation}
In order to estimate $J^{(2)}_1$ we split it as
\begin{equation}
\label{60101}
J^{(2)}_1=J^{(2)}_{1,1}+J^{(2)}_{1,2}
\end{equation}
where
\begin{eqnarray}
&&J^{(2)}_{1,1}:=-\frac{1}{\delta^{3/2}}
\sum\limits_{j,\al}\sum\limits_{i,\bt}\int\limits_t^u\int\limits_\si^s
 D_{\bbl_j,\al}G(\bl_1(\si),\bl_2(\si))
D_{\bby_i,\bt}F_{j,\al}\left(s,\frac{\bmL_1(\si,s)}{\delta},
\frac{\bmL_2(\si,s)}{\delta},\bl_j(\si)\right)\nonumber\\
&&~~~~~~\times(s-\rho_1)\frac{d}{d\rho_1}\hat{l}_{i,\bt}(\rho_1)\,ds\,d\rho_1,
\label{J_11^2}
\end{eqnarray}
and
\begin{eqnarray}
\label{60102}
&&J^{(2)}_{1,2}:=-\frac{1}{\delta}\sum\limits_{j,\al}
\sum\limits_{i,\bt}\int\limits_t^u\int\limits_\si^s
D_{\bbl_j,\al}G(\bl_1(\si),\bl_2(\si))
D_{\bby_i,\bt}F_{j,\al}\left(s,\frac{\bmL_1(\si,s)}{\delta},
\frac{\bmL_2(\si,s)}{\delta},\bl_j(\si)\right)
\\
&&~~~~~~\times
c_1\left(\frac{\by_{i}(\rho)}{\delta}\right)\hat{l}_{i,\bt}(\rho)\,ds\,d\rho,
\nonumber
\end{eqnarray}
with
\begin{equation}
\label{x1}
\frac{d}{d\rho_1}\hat{l}_{i,\bt}(\rho_1)=|\bl(\rho_1)|^{-1}
\left[\frac{d}{d\rho_1}l_{i,\bt}(\rho_1)-
(\hat{\bl}_{i}(\rho_1),\frac{d}{d\rho_1}\bl_{i}
(\rho_1))_{\R^d}\hat{l}_{i,\bt}(\rho_1)\right].
\end{equation}

We deal with $J^{(2)}_{1,2}$ first. It may be split as
$J^{(2)}_{1,2}=J^{(2)}_{1,2,1}+J^{(2)}_{1,2,2}+J^{(2)}_{1,2,3}$, where
\begin{eqnarray}
\label{70612}
&&J^{(2)}_{1,2,1}:=-\frac{1}{\delta}\sum\limits_{j,\al}
\sum\limits_{i,\bt}\int\limits_t^u\int\limits_\si^s
D_{\bbl_j,\al}G(\bl_1(\si),\bl_2(\si))
D_{\bby_i,\bt}F_{j,\al}\left(s,\frac{\bmL_1(\si,s)}{\delta},
\frac{\bmL_2(\si,s)}{\delta},\bl_j(\si)\right)\\
&& ~~~~~~~~~~
\times
c_1\left(\frac{L_{i}(\si,\rho)}{\delta}\right)\hat{l}_{i,\bt}(\si)\,ds\,d\rho,
\nonumber
\end{eqnarray}
\begin{eqnarray*}
&&J^{(2)}_{1,2,2}:=
-\frac{1}{\delta^2}\sum\limits_{j,\al}\sum\limits_{i,\bt,\ga}
\int\limits_t^u\int\limits_\si^s\int\limits_0^1
D_{\bbl_j,\al}G(\bl_1(\si),\bl_2(\si))
D_{\bby_i,\bt}F_{j,\al}\left(s,\frac{\bmL_1(\si,s)}{\delta},
\frac{\bmL_2(\si,s)}{\delta},\bl_j(\si)\right)\\
&&~~~~~~~~~~\times
(D_{\bby_i,\ga}c_1)\left(\frac{R_{i}(v,\si,\rho)}{\delta}\right)
(y_{i,\ga}(\rho)-L_{i,\ga}(\si,\rho))\hat{l}_{i,\bt}(\rho)\,ds\,d\rho\,dv
\end{eqnarray*}
and
\begin{eqnarray*}
&&J^{(2)}_{1,2,3}:=-\frac{1}{\delta}\sum\limits_{j,\al}\sum\limits_{i,\bt}
\int\limits_t^u\int\limits_\si^s\int\limits_\si^{\rho}
D_{\bbl_j,\al}G(\bl_1(\si),\bl_2(\si))
D_{\bby_i,\bt}F_{j,\al}\left(s,\frac{\bmL_1(\si,s)}{\delta},
\frac{\bmL_2(\si,s)}{\delta},\bl_j(\si)\right)\\
&&~~~~~~~~~~\times c_1\left(\frac{\bmL_i(\si,s)}{\delta}\right)
\frac{d}{d\rho_1}\hat{l}_{i,\bt}(\rho_1)\,ds\,d\rho\,d\rho_1.
\end{eqnarray*}
By virtue of Lemma \ref{lm1}, (\ref{80704}) and the definition
(\ref{70613}), we obtain easily
\begin{equation}\label{80301}
 |\bbE[J^{(2)}_{1,2,2}\zeta]|= O(\delta^{1/2-2\ga_1})\|\nabla
G\|_{L^\infty((\R^d)^2)}(u-t)\bE\zeta,\quad\mbox{ as
}\delta\rightarrow0.
\end{equation}
The same argument also shows that $|\bbE[J^{(2)}_{1,2,3}\zeta]|$ is of
the order of magnitude of the right hand side of (\ref{80301}).

Using Lemma \ref{lm3} and the definition (\ref{70613}) we conclude
that
\[
D_{\bby_i}\Theta_i\left(s,\frac{\bmL_1(\si,s)}{\delta},
\frac{\bmL_2(\si,s)}{\delta},\bl_i(\si)\right)=O(\delta).
\]
Therefore, $|\bE[J^{(2)}_{1,2,1}\zeta]|$ is equal, up to a term of
order $O(\delta^{1-\ga_1})(u-t)\|\nabla
G\|_{L^\infty((\R^d)^2)}\bE\zeta $, to
\begin{eqnarray}
\label{60103a}
&&-\frac{1}{\delta}\sum\limits_{i,\al,\bt}\int\limits_t^u\int\limits_\si^s
\bE\left[D_{\bbl_i,\al}G(\bl_1(\si),\bl_2(\si))
\Theta_i\left(s^{(p_1)}_{k_1},\frac{\bmL_1(\si,s)}{\delta},
\frac{\bmL_2(\si,s)}{\delta},\bl_i(\si)\right)\right.\\
&&\times
\left.D_{\bby_i,\bt}D_{\bby_i,\al}c_1\left(\frac{\bmL_i(\si,s)}{\delta}\right)
c_1\left(\frac{\bmL_{i}(\si,\rho)}{\delta}\right)
|\bl_i(\si)|\hat{l}_{i,\bt}(\si)\,\zeta\right]\,ds\,d\rho.\nonumber
\end{eqnarray}
Let $\delta<(2p_1)^{1/(1-\ga_1)}$ and fix $i$. We may apply Lemma
\ref{mix2}, with
\begin{eqnarray*}
&&Z=D_{\bbl_i,\al}G(\bl_1(\si),\bl_2(\si))\Theta_i
\left(s^{(p_1)}_{k_1},\frac{\bmL_1(\si,s)}{\delta},
\frac{\bmL_2(\si,s)}{\delta},\bl_i(\si)\right)
|\bl_i(\si)|\hat{l}_{i,\bt}(\si)\zeta,\\
&& X_1:=D_{\bby_i,\bt}D_{\bby_i,\al}c_1(\bbx),\quad  X_2:=c_1(\bbx),\\
&&
 g_1:=\frac{\bmL_i(\si,s)}{\delta},\quad
 g_2:=\frac{\bmL_{i}(\si,\rho)}{\delta},\quad
 r=C_{13}(\rho-\si),\,r_1=C_{13}(s-\rho),
\end{eqnarray*}
where $C_{13}>0$ depends only on $N$.  We conclude that
\begin{eqnarray}
\label{60103}
&&\left|\bbE\left[J^{(2)}_{1,2,1}\zeta\right]+\frac{1}{\delta}
\sum\limits_{i,\al,\bt}\int\limits_t^u\int\limits_\si^s\bbE\left[
\Theta_i\left(\si,\frac{\bmL_1(\si,s)}{\delta},
\frac{\bmL_2(\si,s)}{\delta},\bl_i(\si)\right)
D_{\bbl_i,\al}G(\bl_1(\si),\bl_2(\si))\right.\right.\\
&&
\left.\left.\times |\bl_i(\si)|\hat{l}_{i,\bt}(\si)
\partial^2_{\al,\bt}
R\left(\frac{\bmL_i(\si,s)-\bmL_i(\si,\rho)}{\delta}\right)\,\zeta\right]
\,ds\,d\rho
\right|\nonumber\\
&&
\leq C_{14}\delta^{-1}\|\nabla G\|_{L^\infty((\R^d)^2)}\bE[\zeta]
\int\limits_t^u\int\limits_\si^s
\phi^{1/2}\left(\,\frac{C_{13}(\rho-\si)}{2\delta}\right)
\phi^{1/2}\left(\,\frac{C_{13}(s-\rho)}{2\delta}\right)\,ds\,d\rho.\nonumber
\end{eqnarray}
Here we used the fact that
\[
\Theta_i\left(\si,\frac{\bmL_1(\si,s)}{\delta},
\frac{\bmL_2(\si,s)}{\delta},\bl_i(\si)\right)=
\Theta_i\left(s^{(p_1)}_{k_1},\frac{\bmL_1(\si,s)}{\delta},
\frac{\bmL_2(\si,s)}{\delta},\bl_i(\si)\right).
\]
The  right hand side of (\ref{60103}) is of the form 
$C_{15}(\delta)(u-t)\|\nabla
G\|_{L^\infty((\R^d)^2)}\bE\zeta$, where $C_{15}(\delta)$ vanishes,
as $\delta\rightarrow0$. The second term appearing on the left
hand side  equals to
\begin{eqnarray}
\label{70206}
&&-\sum\limits_{i,\al,\bt}\int\limits_t^u\bbE\left\{
\Theta_i\left(\si,\frac{\bmL_1(\si,s)}{\delta},
\frac{\bmL_2(\si,s)}{\delta},\bl_i(\si)\right)
D_{\bbl_i,\al}G(\bl_1(\si),\bl_2(\si))\right.\\
&&\left. \times\left[\int\limits_\si^s\frac{d}{d\rho}
\partial_{\al}R\left(\frac{s-\rho}{\delta}\,\hat{\bl}_i(\si)\right)\,
d\rho\right]|\bl_i(\si)|\zeta\right\}\,ds\nonumber\\
&&
=\sum\limits_{i,\al,\bt}\int\limits_t^u\bbE\left[
\Theta_i\left(\si,\frac{\bmL_1(\si,s)}{\delta},
\frac{\bmL_2(\si,s)}{\delta},\bl_i(\si)\right)
D_{\bbl_i,\al}G(\bl_1(\si),\bl_2(\si))
\partial_{\al}R\left(\frac{s-\si}{\delta}\,\hat{\bl}_i(\si)\right)
|\bl_i(\si)|\zeta\right]\,ds\,\nonumber
\end{eqnarray}
thanks to the fact that $\nabla_\by R(\bze)=\bze$. The term appearing
on the right hand side of (\ref{70206}) vanishes as
$\delta\rightarrow0$ and, in consequence we have shown that
$|\bbE[J^{(2)}_{1,2,1}\zeta]|= C_{16}(\delta)(u-t)\|\nabla
G\|_{L^\infty((\R^d)^2)}\bE\zeta$, where $C_{16}(\delta)$ vanishes, as
$\delta\rightarrow0$.

We now estimate $J_{1,1}^{(2)}$ given by (\ref{J_11^2}). Note that
according to (\ref{x1}) and (\ref{eq2}) we have
\[
J^{(2)}_{1,1}=J^{(2)}_{1,1,1}+J^{(2)}_{1,1,2},
\]
where
\begin{eqnarray*}
&&J^{(2)}_{1,1,1}:= \frac{1}{\delta^{2}}\sum\limits_{j,\al}\sum\limits_{i,\bt}
\int\limits_t^u\int\limits_\si^s
D_{\bbl_j,\al}G(\bl_1(\si),\bl_2(\si))
D_{\bby_i,\bt}F_{j,\al}\left(s,\frac{\bmL_1(\si,s)}{\delta},
\frac{\bmL_2(\si,s)}{\delta},\bl_j(\si)\right)\\
&&~~~~~~~~\times (s-\rho_1)\Gamma_{i,\bt}\left(\rho_1,
\frac{\by_1(\rho_1)}{\delta},\frac{\by_2(\rho_1)}{\delta},\bl_i(\si)\right)
\,ds\,d\rho_1,
\end{eqnarray*}
with
\[
\Gamma_i\left(\rho,\by_1,\by_2,\bbl\right):=|\bbl|^{-1}\left[
F_{i}\left(\rho,\by_1,\by_2,\bbl\right)-
\left(\hat{\bbl},F_{i}\left(\rho,\by_1,\by_2,\bbl\right)\right)_{\R^d}
\hat{\bbl}\right],
\]
while
\begin{eqnarray}
&& J^{(2)}_{1,1,2}:= \frac{1}{\delta^{2}}\sum\limits_{j,\al}\sum\limits_{i,\bt}
\int\limits_t^u\int\limits_\si^s
\int\limits_\si^{\rho_1}
D_{\bbl_j,\al}G(\bl_1(\si),\bl_2(\si))
D_{\bby_i,\bt}F_{j,\al}\left(s,\frac{\bmL_1(\si,s)}{\delta},
\frac{\bmL_2(\si,s)}{\delta},\bl_j(\si)\right)\nonumber\\
&&~~~~~~~~\times(s-\rho_1)\frac{d}{d\rho_2}\Gamma_{i,\bt}
\left(\rho_1,\frac{\by_1(\rho_1)}{\delta},
\frac{\by_2(\rho_1)}{\delta},\bl_i(\rho_2)\right)\,
ds\,d\rho_1\,d\rho_2.\label{80801}
\end{eqnarray}
A straightforward computation, using Lemma \ref{lm1} (note that $\frac{d}{d\rho_2}\Gamma_{i,\bt}\sim \delta^{-1/2}$
in (\ref{80801})), shows that
\[
|\bbE[J^{(2)}_{1,1,2}\zeta]|\leq O(\delta^{1/2-3\ga_1})(u-t)\|\nabla
G\|_{L^\infty((\R^d)^2)}\bbE[\zeta].
\]

An application of Lemma  \ref{lm1}, in the same fashion as it was
done in the calculations concerning the terms $\bE[J^{(2)}_{1,2,2}\zeta]$ and
$\bE[J^{(2)}_{1,2,3}\zeta]$,
yields that $\bbE[J^{(2)}_{1,1,1}\zeta]$ is equal, up to a term of
the order 
$C_{17}(\delta)(u-t)\|\nabla
G\|_{L^\infty((\R^d)^2)}\bbE[\zeta]$, where
$\lim\limits_{\delta\rightarrow0}C_{17}(\delta)=0$, to
\begin{eqnarray}
\label{52701} && \frac{1}{\delta^{2}}
\sum\limits_{i,\al,\bt}\int\limits_t^u\int\limits_\si^s (s-\rho_1)
\bbE\left[D_{\bbl_i,\al}G(\bl_1(\si),\bl_2(\si))
D_{\bby_i,\bt}F_{i,\al}\left(s,\frac{\bmL_1(\si,s)}{\delta},
\frac{\bmL_2(\si,s)}{\delta},\bl_i(\si)\right) \right.\\
&&\times\left.\Gamma_{i,\bt}\left(\rho_1,\frac{\bmL_1(\si,\rho_1)}{\delta},
\frac{\bmL_2(\si,\rho_1)}{\delta},\bl_i(\si)\right)\zeta\right]
\,ds\,d\rho_1.\nonumber
\end{eqnarray}
We denote
\[
V_{i,\bt}(\by_1,\by_2,\by_1',\by_2',\bbl):=\left(\sum\limits_{\ga,q}
\partial_{\bt,\ga,q}^3
R(\by_i-\by_i')\hat{l}_{q}\hat{l}_{\ga}-\sum\limits_{\ga}
\partial_{\bt,\ga,\ga}^3
R(\by_i-\by_i')\right)|\bbl|.
\]
Applying Lemmas  \ref{lm3} and \ref{mix2}, as in (\ref{60103a})
(\ref{60103}), we conclude that (\ref{52701}) is equal, up to a
term of order 
$C_{18}(\delta)(u-t)\|\nabla
G\|_{L^\infty((\R^d)^2)}\bbE[\zeta]$, where
$\lim\limits_{\delta\rightarrow0}C_{18}(\delta)=0$, to
\begin{equation}
\label{53001}
 \frac{1}{\delta^{2}}\sum\limits_{i,\al}
\int\limits_t^u\int\limits_\si^s (s-\rho_1)
\bbE\left\{D_{\bbl_i,\al}G(\bl_1(\si),\bl_2(\si))
\Lambda_{i}(\si,P_{i};\tilde\bl(\cdot))V_{i,\al}(P_{i})\,\zeta\right\}
\,ds\,d\rho_1,
\end{equation}
with $\Lambda_i$ defined by (\ref{70605b}), and
$
P_{i}=\left(\bmL_1(\si,s),\bmL_2(\si,s),
\bmL_1(\si,\rho_1),\bmL_2(\si,\rho_1),\bl_i(\si)\right).
$
\commentout{

Using condition DR) we conclude that
\[
|W_{i,\ga,j,\bt}(\by_1,\by_2,\by_1',\by_2',\bbl,\bbl')|\leq
C_6\phi(|\by_j-\by_i'|).
\]
The above estimate and Lemma \ref{lm3} together with (\ref{70613})
allow us to conclude that the second term of (\ref{53001}) can be
estimated by
\[
 \frac{C_7}{\delta}\int\limits_t^u\int\limits_\si^s
 \int\limits_\si^\rho\phi\left(\frac{s-\rho_1}{2N\delta}\right)\,ds\,d\rho\,d\rho_1=o(1).
\]

}Note, however, that for $s\in[s^{(p_1)}_{k_1},s^{(p_1)}_{k_1+1}]$
\[
\Xi\left(s,\bmL_1(\si,s),\bmL_2(\si,s)
\right)=
\Xi\left(s^{(p_1)}_{k_1},\bmL_1(\si,s),\bmL_2(\si,s)\right)
\]
and
\[
\left|\Xi\left(\si,\bmL_1(\si,s),\bmL_2(\si,s)\right)-
\Xi\left(\si,\by_1(\si),\by_2(\si)\right)\right|
\leq C\sum\limits_{p=1}^2
\left|L_p(\si,s)-\by_p(\si)\right|
\leq C(s-\si)\leq C\delta^{1-\ga_1}.
\]
A similar estimate holds also for the terms containing
$\bmL_i(\si,\rho_1)$ and we conclude that the expression in (\ref{53001})
is equal, up to a term of order 
$C_{19}(\delta)(u-t)\|\nabla G\|_{L^\infty((\R^d)^2)}\bbE[\zeta]$,
where $\lim\limits_{\delta\rightarrow0}C_{19}(\delta)=0$, to
\begin{equation}
\label{53101} \frac{1}{\delta^{2}}\sum\limits_{i,\al}
\int\limits_{t+\delta^{1-\gamma_1}}^u
\bbE\left\{D_{\bbl_j,\al}G(\bl_1(\si),\bl_2(\si))\overline{\Theta}_i^2(\si)
\left[ \int\limits_\si^s
(s-\rho_1)V_{i,\al}(P_{i})\,d\rho_1\right]\,\zeta\right\}\,ds,
\end{equation}
\commentout{
 \frac{1}{\delta^{2}}\sum\limits_{i,\al}
\int\limits_t^u\int\limits_\si^s \int\limits_\si^\rho
\bbE\left\{D_{\bbl_i,\al}G(\bl_1(\si),\bl_2(\si))
E^{(1)}_i(\si,\si,Q_{i};\bl(\cdot),
\tilde\bl(\cdot))V_{i,\al}(P_{i})\,\zeta\right\}\,ds\,d\rho\,d\rho_1
\[
=
\]

}with
\[
\overline{\Theta}_i(\si):=
\Theta_{i}(\si,\by_1(\si),\by_2(\si),\bl(\si);\tilde\bl(\cdot)).
\]
Note that, for $s>t+\delta^{1-\ga_1}$ we have
\begin{eqnarray}
\label{70606}
&& \frac{1}{\delta^{2}}\sum\limits_{i,\al}\int\limits_{s-\delta^{1-\ga_1}}^s
 (s-\rho_1)V_{i,\al}(P_{i})\,d\rho_1= \frac{1}{\delta^{2}}\sum\limits_{i,\al}
|\bl_i(\si)|\int\limits_{s-\delta^{1-\ga_1}}^s
 (s-\rho_1)\\
&&\times \left[\sum\limits_{\ga,q}\partial_{\al,\ga,q}^3
R\left(\frac{s-\rho_1}{\delta}\,\hat{\bl}_i(\si)\right)
\hat{l}_{i,q}(\si)\hat{l}_{i,\ga}(\si)\right.
\left.-\sum\limits_\ga\partial_{\al,\ga,\ga}^3
 R\left(\frac{s-\rho_1}{\delta}\,\hat{\bl}_i(\si)\,\right)
\right]\,d\rho_1\nonumber\\
&&=\sum\limits_{i,\al}|\bl_i(\si)|\int\limits_0^{\delta^{-\ga_1}}
 \rho_1 \left[\sum\limits_{\ga,q}\partial_{\al,\ga,q}^3
R\left(\rho_1\,\hat{\bl}_i(\si)\right)
 \hat{l}_{i,q}(\si)\hat{l}_{i,\ga}(\si)-
\sum\limits_\ga
\partial_{\al,\ga,\ga}^3R\left(\rho_1\,\hat{\bl}_i(\si)\right)
\right]
 \,d\rho_1.\nonumber
\end{eqnarray}
Using the fact that
\[
\sum\limits_q\partial_{\al,\ga,q}^3R\left(\rho_1\,\hat{\bl}_i(\si)\right)
 \hat{l}_{i,q}(\si)=\frac{d}{d\rho_1}
\partial_{\al,\ga}^2R\left(\rho_1\,\hat{\bl}_i(\si)\right)
\]
we obtain, upon the integration by parts performed in the first
term on the utmost right hand side of (\ref{70606}), that this
term equals to
\begin{eqnarray}
\label{70607}
&&\sum\limits_{i,\al,\ga}|\bl_i(\si)|
\left[
\delta^{-\ga_1}\partial_{\al,\ga}^2
R\left(\delta^{-\ga_1}\,\hat{\bl}_i(\si)\right)
\hat{l}_{i,\ga}(\si)-\int\limits_0^{\delta^{-\ga_1}}
\partial_{\al,\ga}^2R\left(\rho_1\,\hat{\bl}_i(\si)\right)\hat{l}_{i,\ga}(\si)
\,d\rho_1\right.\\
&&-\left.
\int\limits_0^{\delta^{-\ga_1}}
\rho_1  \partial_{\al,\ga,\ga}^3R\left(\rho_1\,\hat{\bl}_i(\si)\right)\,d\rho_1
\right]\nonumber\\
&&\!\!\!=\sum\limits_{i,\al,\ga} |\bl_i(\si)|
\left[
\delta^{-\ga_1}\partial_{\al,\ga}^2
R\left(\delta^{-\ga_1}\,\hat{\bl}_i(\si)\right)
\hat{l}_{i,\ga}(\si)
-\partial_{\al}R\left(\delta^{-\ga_1}\,\hat{\bl}_i(\si)\right)
\right.\left.-\! \int\limits_0^{\delta^{-\ga_1}}
 \rho_1  \partial_{\al,\ga,\ga}^3R\left(\rho_1\,\hat{\bl}_i(\si)\right)
\,d\rho_1
 \right].\nonumber
\end{eqnarray}
We have used here the fact that $\nabla R(\bze)=\bze$ and
\[
\sum\limits_{\ga}\partial_{\al,\ga}R\left(\rho_1\,\hat{\bl}_i(\si)\right)
\hat{l}_{i,\ga}(\si)
=\frac{d}{d\rho_1}\partial_{\al}R\left(\rho_1\,\hat{\bl}_i(\si)\right).
\]
Summarizing the work done in this section, we have shown that
\begin{equation}\label{80305}
  |\bE[I^{(1)}\zeta]|\leq C_{20}(u-t)\|\nabla
  G\|_{L^\infty((\R^d)^2)}\bE\zeta,
\end{equation}
where the constant $C_{20}$ does not depend on $\delta$ and $G$.

\subsubsection{The terms $\bbE[I^{(2)}\zeta]$ and $\bbE[I^{(3)}\zeta]$}

 The calculations concerning these terms essentially follow the
respective steps performed in the previous section 
so we only highlight their main points. First, we note that
because $\bl_i(\rho)-\bl_i(\si)\sim \delta^{1/2-\ga_1}$ we have
that
$\bbE[I^{(2)}\zeta]$ is, up to a term 
$C_{21}(\delta)(u-t)\|\nabla G\|_{L^\infty((\R^d)^2)}\bbE[\zeta]$,
where $\lim\limits_{\delta\rightarrow0}C_{21}(\delta)=0$,  equal to
\begin{eqnarray}
\label{70611}
&&\frac{1}{\delta}\sum\limits_{j,\al,\bt}
\int\limits_t^u\int\limits_\si^s\bbE\left[D_{\bbl_j,\al}G(\bl_1(\si),\bl_2(\si))
D_{\bbl_j,\bt}F_{j,\al}\left(s,\frac{\by_1(s)}{\delta},
\frac{\by_2(s)}{\delta},\bl_j(\si)\right)\right.\\
&&\times\left. F_{j,\bt}\left(\rho,\frac{\by_1(\rho)}{\delta},
\frac{\by_2(\rho)}{\delta},\bl_j(\si)\right)\zeta\right]
ds\,d\rho.\nonumber
\end{eqnarray}
Replacing $\rho$ by $\si$ as the argument of $\bl_1(\cdot)$,
$\bl_2(\cdot)$ in (\ref{70611}) needs a correction that is of
order of magnitude $O(\delta^{1/2-2\ga_1})(u-t)\|\nabla
G\|_{L^\infty((\R^d)^2)}\bbE[\zeta]$, since $\ga_1\in(0,1/8]$.
Next we note that (\ref{70611}) equals
\begin{eqnarray}
\label{60104}
&&\frac{1}{\delta}
\sum\limits_{j,\al,\bt}\int\limits_t^u\int\limits_\si^s \bbE\left[
D_{\bbl_j,\al}G(\bl_1(\si),\bl_2(\si))
D_{\bbl_j,\bt}F_{j,\al}\left(s,\frac{\bmL_1(\si,s)}{\delta},
\frac{\bmL_2(\si,s)}{\delta},\bl_j(\rho)\right)\right.\\
&&\times\left.
F_{j,\bt}\left(\rho,\frac{\bmL_1(\si,\rho)}{\delta},
\frac{\bmL_2(\si,\rho)}{\delta},\bl_j(\si)\right)\zeta\right]
ds\,d\rho\nonumber\\
&&+
\frac{1}{\delta^2}\sum\limits_{i,\ga}\sum\limits_{j,\al,\bt}
\int\limits_t^u\int\limits_\si^s\int\limits_0^1 \bbE\left[
D_{\bbl_j,\al}G(\bl_1(\rho),\bl_2(\rho))
D_{\bby_i,\ga}D_{\bbl_j,\bt}F_{j,\al}\left(s,\frac{\bmR_1(v,\si,s)}{\delta},
\frac{\bmR_2(v,\si,s)}{\delta},\bl_j(\si)\right)\right.\nonumber\\
&&\times\left.
F_{j,\bt}\left(\rho,\frac{\bmL_1(\si,\rho)}{\delta},
\frac{\bmL_2(\si,\rho)}{\delta},\bl_j(\si)\right)
(y_{i,\ga}(s)-L_{i,\ga}(\si,s))\zeta\right] ds\,d\rho\,dv
\nonumber\\
&&
+
\frac{1}{\delta^2}\sum\limits_{i,\ga}\sum\limits_{j,\al,\bt}
\int\limits_t^u\int\limits_\si^s\int\limits_0^1 \bbE\left[
D_{\bbl_j,\al}G(\bl_1(\rho),\bl_2(\rho))
D_{\bby_i,\ga}D_{\bbl_j,\bt}F_{j,\al}\left(s,\frac{\by_1(s)}{\delta},
\frac{\by_2(s)}{\delta},\bl_j(\si)\right)\right.\nonumber\\
&&\times\left.
F_{j,\bt}\left(\rho,\frac{\bmR_1(v,\si,\rho)}{\delta},
\frac{\bmR_2(v,\si,\rho)}{\delta},\bl_j(\si)\right)
(y_{i,\ga}(s)-L_{i,\ga}(\si,\rho))\zeta\right]
ds\,d\rho\,dv.\nonumber
\end{eqnarray}
A simple argument using Lemma \ref{lm1}, (\ref{80703}) and
(\ref{80704}) shows that the second and third terms of
(\ref{60104}) are both of order of magnitude
$O(\delta^{1/2-3\ga_1})(u-t)\|\nabla
G\|_{L^\infty((\R^d)^2)}\bbE[\zeta]$.

The first term, on the other hand, can be handled with the help of
Lemma \ref{mix2} in the same fashion as we have dealt with the
term $J^{(2)}_{1,2,1}$, given by (\ref{70612}) of Section
\ref{seca41}, and
we obtain that
\begin{equation}
\label{80306}
|\bbE[I^{(2)}\zeta]|\leq C_{22}(\delta)(u-t)\|\nabla
G\|_{L^\infty((\R^d)^2)}\bbE[\zeta],
\end{equation}
where $\lim\limits_{\delta\rightarrow0}C_{22}(\delta)=0$.

Finally, concerning the limit of $\bbE[I^{(3)}\zeta]$ we note
that by Lemma \ref{lm1} we have
\begin{equation}
\label{60201} \bbE[I^{(3)}\zeta]\leq C_{23}(\delta)(u-t)\|\nabla
G\|_{L^\infty((\R^d)^2)}\bbE[\zeta]
 +\sum\limits_{i,j} {\cal I}_{i,j}
 \end{equation}
where $\lim\limits_{\delta\rightarrow0}C_{23}(\delta)=0$ and
\begin{eqnarray*}
&& {\cal I}_{i,j}:=\frac{1}{\delta}\sum\limits_{\al,\bt}
\int\limits_t^u\int\limits_\si^s\bbE\left\{
D_{\bbl_i,\bt}D_{\bbl_j,\al}G(\bl_1(\si),\bl_2(\si))\vphantom{\int}\right.\\
&&~~~~~\times\left. F_{j,\al}\left(s,\frac{\bmL_1(\si,s)}{\delta},
\frac{\bmL_2(\si,s)}{\delta},\bl_j(\si)\right)
F_{i,\bt}\left(\rho,\frac{\bmL_1(\si,\rho)}{\delta},
\frac{\bmL_2(\si,\rho)}{\delta},\bl_i(\si)\right)\zeta\right\}
ds\,d\rho,
\end{eqnarray*}
First, let $i\not=j$ and $2\delta^{1-\ga_1}M\leq (2q)^{-1}$.
Suppose also that $s\geq t^{(p)}_1$. We have then
\[
|\bmL_i(\si,s)-\bmL_j(\si,\rho)|\geq \frac{1}{q}-2M(s-\si)\geq
\frac{1}{2q}
\]
on the event (with fixed $\al,\bt$)
\[
\Theta_j\left(s,\frac{\bmL_1(\si,s)}{\delta},
\frac{\bmL_2(\si,s)}{\delta},\bl_j(\si)\right)
\Theta_i\left(s,\frac{\bmL_1(\si,\rho)}{\delta},
\frac{\bmL_2(\si,\rho)}{\delta},\bl_i(\si)\right)
\]\[
\times
D_{\bbl_i,\bt}D_{\bbl_j,\al}G(\bl_1(\si),\bl_2(\si))
|\bl_j(\si)||\bl_i(\si)|\not=0.
\]
When, on the other hand, $s,\rho\in[0,t^{(p)}_1]$, then we
conclude from (\ref{81101}) that
\[
|\bmL_i(\si,s)-\bmL_j(\si,\rho)|\geq \ga_N's\geq \ga_N'(s-\si).
\]
Therefore $|{\cal I}_{i,j}|$ can be then
estimated via Lemma \ref{mix2} and Lemma \ref{lm3} by 
\begin{equation}\label{Iij}
C_{23} D_1^2M^2\|D_{\bbl_1}D_{\bbl_2}G\|_{L^\infty}
\left[\delta^{-\ga_1}
\phi^{1/2}\left(\frac{\ga_N'}{\delta^{\ga_1}}\right)+\delta^{1-2\ga_1}\right]
\E[\zeta].
\end{equation}
It obviously vanishes, as $\delta\rightarrow0$. The second term in (\ref{Iij})
arises from the contribution of $s<t+{\delta}^{1-\gamma_1}$.

When $i=j$  we can use Lemma \ref{mix2} in order to obtain
\[
|{\cal I}_{i,i}|\leq C_{24}(u-t)\|\nabla^2
G\|_{L^\infty((\R^d)^2)}\bbE[\zeta].
\]
Summarizing, we conclude that
\[
|\bbE[I^{(3)}\zeta]|\leq C_{25}(u-t)\|\nabla^2
G\|_{L^\infty((\R^d)^2)}\bbE[\zeta],
\]
where $C_{25}$ can be chosen independently of $\delta$ and $G$.
Hence we conclude (\ref{73101}) and tightness
follows.

Suppose now that $Q$  is any limiting measure of $Q^{(\delta_n)}$
for a certain sequence $\delta_n\rightarrow0$, as $n\rightarrow
+\infty$. Coming back to (\ref{53101}) we conclude, using
calculation (\ref{70606})--(\ref{70607}), that the limit, as
$\delta\rightarrow0$, of the expression on the left hand side of
(\ref{53101}) equals to
\begin{equation}
\label{60110} \sum\limits_{i,\al}\int\left[\int\limits_t^u
a_{\al}^{(i)}(s) D_{\bbl_i,\al}G(K_1(s),K_2(s))|K_i(s)|
\overline{\Lambda}_{i}(s)\,\zeta(K)\,ds\,\right]Q(dK),
\end{equation}
where
\begin{eqnarray*}
&&\overline{\Lambda}_{i}(s):=\Lambda_i\left(s,Y_1(s),Y_2(s)
,K_1(s),K_2(s);K(\cdot),\hat{K}(\cdot)\right),\\
&& Y_i(s):=\bbx_i+\int\limits_0^s\hat{K_i}(\rho)\,d\rho,\quad
i=1,2,\\
&& a_{\al}^{(i)}(s):=-\sum\limits_{\ga}\int\limits_0^{+\infty}
 \rho_1 \partial_{\al,\ga,\ga}^3
 R\left(\rho_1\,\hat{K_i}(s)\right)\,d\rho_1.
\end{eqnarray*}

Similarly, we calculate the limit, as $\delta\rightarrow0$, of
$\bbE[I^{(3)}\zeta]$. We know that only the limits of the terms
${\cal I}_{i,i}$ contribute. A straightforward computation shows
that
\[
\lim\limits_{\delta\rightarrow0}\sum\limits_{i}{\cal I}_{i,i}=
\sum\limits_{i,\al,\bt}\int\left[\int\limits_t^u c_{\al,\bt}^{(i)}(s)
D_{\bbl_i,\al}D_{\bbl_i,\bt}G(K_1(s),K_2(s))
\overline{H}^{(i)}(s)\,\zeta(K)\,ds\,\right]Q(dK),
\]
where 
\[
c^{(i)}_{\al,\bt}(s):=-|K_{i}(s)|^2\int\limits_{0}^{+\infty}
\partial_{\al,\bt}^2R(\rho\hat{K_i}(s))\,d\rho,
\]
\[ \overline{H}^{(i)}(s):= \Theta_i^2\left(s,
Y_1(s),Y_2(s),K_i(s)\right).
\]
Summarizing, we have shown that any limiting measure $Q$ satisfies
(\ref{51401}).



\subsection{The removal of cut-offs and the proof of weak
 convergence of $(\bmk^{(\delta)}_1(\cdot),\bmk^{(\delta)}_2(\cdot))$}
Let $\mathfrak Q_{\bbk_1,\bbk_2}:=\mathfrak Q_{\bbk_1}\otimes
\mathfrak Q_{\bbk_2}$ be the law of two independent copies of the
diffusion given by (\ref{61102}) over ${\cal C}_2(k_1,k_2)$
starting respectively at $\bbk_1$ and $\bbk_2$. For a fixed $M$
let $\mathfrak Q_{\bbk_1}^{(M)}$ be the law over ${\cal C}_1$ of
any diffusion starting at a given $\bbk_1\in\R^d$ with the
generator ${\cal L}^{(M)}$ given by
\[
{\cal
L}^{(M)}F(\bbk)=\sum\limits_{p,q}a^{(M)}_{p,q}(\bbk)\partial_{k_p,k_q}^2F(\bbk)
+\sum\limits_{p}b^{(M)}_p(\bbk)\partial_{k_p}F(\bbk),\quad F\in
C^\infty_0(\R^d).
\]
Here $a^{(M)}_{p,q}(\cdot)$, $b^{(M)}_p(\cdot)$ are bounded and
twice continuously differentiable,
$a^{(M)}_{p,q}(\bbk)=|\bbk|^2D_{p,q}(\hat{\bbk})$,
$b^{(M)}_p(\bbk)=|\bbk|E_{p}(\hat{\bbk})$ for $M^{-1}\leq
|\bbk|\leq M$.  By virtue of Theorems 5.2.3 and 5.3.2 of
\cite{stvar} we conclude that $\mathfrak Q_{\bbk_1}^{(M)}$ is the
unique probability measure such that
\[
F(K(t))-F(\bbk_1)-\int\limits_0^t{\cal L}^{(M)}F(K(s))ds,\quad
t\geq0
\]
is an $\left({\cal M}_1^{0,t}\right)_{t\geq0}$-martingale for any
$F\in C_b^2(\R^d)$. We define $\mathfrak
Q_{\bbk_1,\bbk_2}^{(M)}:=\mathfrak Q_{\bbk_1}^{(M)}\otimes
\mathfrak Q_{\bbk_2}^{(M)}$.

Let us briefly describe the strategy of the proof of weak convergence
of $(\bmk^{(\delta)}_1(\cdot),\bmk^{(\delta)}_2(\cdot))$. First, for
any $K\in{\cal C}_2$ we define a certain stopping time $W(K;M,N,p,q)$,
see (\ref{xxx}). The crucial property of that time is that the
dynamics given by (\ref{eq1b}) agrees with the dynamics of the
truncated system (\ref{eq2}) up to $W(\cdot;M,N,p,q)$. We also show
that any limiting measure $Q(\cdot;M,N,p,q)$ satisfies, up to the
stopping time, the martingale problem associated with the diffusion
given by $\mathfrak Q_{\bbk_1,\bbk_2}$. This property allows to
identify $Q(\cdot;M,N,p,q)$ with $\mathfrak Q_{\bbk_1,\bbk_2}$ on the
$\si$--algebra ${\cal M}^{0,W}_2$ corresponding to the stopping time.
The final step is to show that for sufficiently large $N$, so that
(\ref{81101}) is satisfied, and sufficiently large $M$, as in
(\ref{80401}), the stopping time $W(\cdot;M,N,p,q)$ converges to
infinity in $\mathfrak Q_{\bbk_1,\bbk_2}$ as $q\rightarrow+\infty$ and
$p\rightarrow+\infty$ (in that order), see (\ref{80805}). The weak
convergence statement is a consequence of this property of the
stopping time and it is shown in the calculation following
(\ref{71601}).

\label{cutoff}

We introduce the following $({\cal M}^{0,t}_2)_{t\geq0}$--stopping
times. As before, for any $K=(K_1,K_2)$ such that $K(t)\not=\bze$
for all $t\geq0$ we define
\begin{equation}\label{extra}
 Y_j(t):=\int\limits_0^t\hat{K}_j(s)ds.
\end{equation}
 For such a $K$ we let
$S(N,p):=\lim\limits_{n\uparrow+\infty}S_n(N,p)$, where
\begin{eqnarray*}
&&S_n(N,p):=\inf\left\{t\geq0:\mbox{ for some }k\geq 0\mbox{ we
have }t\in\left[t_k^{(p)},t_{k+1}^{(p)}\right)\right. \\ &&\left.
~~~~~~~~~\mbox{ and } \hat{K}_i(t_j^{(p)})\cdot \hat
K_i(t)<1-\frac{2}{N+1}+\frac{1}{n},\mbox{ for some
}\,i\in\{1,2\}\mbox{ or }\,j\in\{k-1,k\}\right\}.
\end{eqnarray*}
If $K$ is such that it becomes $\bze$ for some $t$ we adopt of the
convention that $S(N,p)=+\infty$. We let further
$T(M):=\lim\limits_{n\uparrow+\infty}T_n(M)$, where
\[
T_n(M):=\inf\left\{t\geq0:
|K_i(t)|<\frac{1}{M}+\frac{1}{n},\mbox{ for some
}\,i\in\{1,2\},\right.\]
\[
\left.\mbox{ or }|K_i(t)|>M-\frac{1}{n},\mbox{ for some
}\,i\in\{1,2\}\right\}.
\]
Finally, for any $R_1,R_2>0$ and $K=(K_1,K_2)\in {\cal C}_2(R_1,R_2)$ we
let $U(p,q;K):=\lim\limits_{n\uparrow+\infty}U_n(p,q;K)$,
$V(p,q;K):=\lim\limits_{n\uparrow+\infty}V_n(p,q;K)$,
 where
\begin{eqnarray*}
&&U_n(p,q;K):=\inf\left\{t\geq0:\mbox{ for some }k\geq
1,\,i\in\{1,2\}\,\mbox{ we have
}u\in[0,t_{k-1}^{(p)}],\,t\in\left[t_k^{(p)},t_{k+1}^{(p)}\right)\right.\\
&&~~~~~~~~~~~~~~~~~ \left.\mbox{ such that }
\left|Y_i(t)-Y_i(u)\right|<\frac{1}{q}+\frac{1}{n}\,\right\},
\end{eqnarray*}
\[
V_n(p,q;K):=\inf\left\{t\geq\frac{1}{p}:
\inf\limits_{0\leq u\leq
t}\left|Y_1(t)-Y_2(u)\right|<\frac{1}{q}+\frac{1}{n},\,
\mbox{ or }\,\inf\limits_{0\leq u\leq
t}\left|Y_2(t)-Y_1(u)\right|<\frac{1}{q}+\frac{1}{n}\right\}.
\]
We adopt the convention that any of the above defined stopping
times is infinite if the respective set of times over which it is
determined is empty.

Suppose that $T_0>0$ is an arbitrary deterministic time. Let
\begin{equation}\label{xxx}
 W(M,N,p,q):= S(N,p)\wedge T(M)\wedge U(p,q)\wedge V(p,q)\mbox{ and
}
\end{equation}
\[
B(M,N,p,q):=\{ S(N,p)\wedge U(p,q)\wedge V(p,q)\leq T(M)\wedge
T_0\}.\] We have $B\in{\cal M}_2^{0,W}$. According to Theorem
6.1.2 of \cite{stvar} the measures $\mathfrak Q_{\bbk_1,\bbk_2}$,
$\mathfrak Q_{\bbk_1,\bbk_2}^{(M)}$, $Q(\cdot;M,N,p,q)$ agree,
when restricted to ${\cal M}_2^{0,W}$. In what follows we show
that
\begin{equation}
\label{80805}
 \lim\limits_{p\rightarrow+\infty}
 \lim\limits_{q\rightarrow+\infty}\mathfrak Q_{\bbk_1,\bbk_2}[W(M,N,p,q)<+\infty]=0.
\end{equation}
The condition
\begin{equation}
\label{80806}
T_0<W(\bmk^{(\delta)}_1(\cdot),\bmk^{(\delta)}_2(\cdot);M,N,p,q)=
W(\bl^{(\delta)}_1(\cdot),\bl^{(\delta)}_2(\cdot);M,N,p,q)
\end{equation}
implies
$(\bmk^{(\delta)}_1(s),\bmk^{(\delta)}_2(s))=(\bl^{(\delta)}_1(s),\bl^{(\delta)}_2(s))$
for $s\in[0,T_0]$. We will use both (\ref{80805}) and
(\ref{80806}) to establish weak convergence of the laws of
$(\bmk^{(\delta)}_1(\cdot),\bmk^{(\delta)}_2(\cdot))$ over
$C([0,T_0];(\R^d)^2)$.

We start with the following simple observation.
\begin{lemma}
With the choice of $M$ as in $(\ref{80401})$ we have
\label{lm6}
\[
\mathfrak Q_{\bbk_1,\bbk_2}[T(M)=+\infty]=1.
\]
\end{lemma}
\proof A simple calculation using It\^{o} formula and Remark
\ref{rm10} shows that $d|\bbk_j(t)|^2=0$, $j=1,2$ which proves the
lemma. \endproof

\begin{lemma}
\label{lm5}
Under the assumptions of Theorem $\ref{thm2-main}$ we have
\begin{equation}
\label{71901}
\lim\limits_{q\rightarrow+\infty}U(p,q)=+\infty,\forall\,p,\quad
\mathfrak Q_{\bbk_1,\bbk_2}-a.s.
\end{equation}
\end{lemma}
\proof The proof is essentially the repetition of the argument from
\cite{kp} pp. 60-61 so we only highlight its main points. It suffices
only to show that
\begin{equation}\label{72301}
 \lim\limits_{q\rightarrow+\infty}U^{(i)}(p,q)=+\infty,\forall\,p,\quad
\mathfrak Q_{\bbk_i}-a.s.,
\end{equation}
where
 $U^{(i)}(p,q):=\lim\limits_{n\uparrow+\infty}U_n^{(i)}(p,q)$,
\[
U_n^{(i)}(p,q):=\inf\left\{t\geq0:\mbox{ for some }k\geq 1,\,\mbox{ we have
}u\in[0,t_{k-1}^{(p)}],\,t\in\left[t_k^{(p)},t_{k+1}^{(p)}\right)\right.
\]
\[
\left.\mbox{ such that }
\left|\int_u^t\hat{K}_i(s)ds\right|<\frac{1}{q}+\frac{1}{n}\,\right\}.
\]
However, (\ref{72301}) can be proved with the help of the argument
contained in  pp. 60-61 \cite{kp} so we omit the details here. We
obtain from (\ref{72301})
\[
\lim\limits_{q\rightarrow+\infty}U(p,q)=+\infty,\forall\,p,\quad
\mathfrak Q_{\bbk_1,\bbk_2}-a.s.
\]
However $U(p,q)=U^{(1)}(p,q)\wedge U^{(2)}(p,q)$ and (\ref{71901})
follows.  \endproof

Let us denote by
\begin{equation}\label{extra1}
  \mathfrak Y_t^{(j)}:=\bigcup\limits_{0\leq s\leq
t}Y_j(s)
\end{equation}
 and by $B_{r}(\mathfrak Y_t^{(j)}):=[\bbx:\mbox{
dist}(\bbx,\mathfrak Y_t^{(j)})\leq r]$ the sausage, up to time
$t$, of diameter $r>0$ around trajectory $Y_j(\cdot)$.

The next lemma shows that $S(N,p)$ becomes infinite as
$p\to\infty$ for each $N$.
\begin{lemma}
\label{lmS}
We have
\begin{equation}
\label{eq-lmS}
\lim\limits_{p\rightarrow+\infty}S(N,p)=+\infty,\forall~ N,\quad
\mathfrak Q_{\bbk_1,\bbk_2}-a.s.
\end{equation}
\end{lemma}
\proof The conclusion of the lemma is a consequence of the uniform
continuity of paths of the diffusion on any finite time interval
$[0,T]$, which implies that
\[
\lim\limits_{p\rightarrow+\infty}\min\limits_{t_k^{(p)}\in[0,T]}
\min\limits_{t\in[t_k^{(p)},t_{k+1}^{(p)}]}\hat{K}_j(t)\cdot
\hat{K}_j(t_k^{(p)}) =1,\quad j=1,2.
\]
\endproof

Our next lemma shows that $V(p,q)$ becomes infinite with
$p,q\rightarrow+\infty$.
\begin{lemma}
\label{lm12}
Suppose that $N$ is as in $(\ref{81101})$ and $T_1,\eta>0$ are arbitrary.
Then,
one can find $p_0,q_0$ such that
\begin{equation}
\label{71902} \mathfrak Q_{\bbk_1,\bbk_2}[S(N,p)\wedge V(p,q)\leq
T_1\,]<\eta,
\quad \forall\,p\geq p_0,\,q\geq q_0.
\end{equation}
\end{lemma}
In order to prove this lemma we will need an auxiliary property of
$(K_1(\cdot),Y_1(\cdot))$.  Let $k_1=|\bbk_1|$. Note that the
process $(K_1(\cdot),Y_1(\cdot))$ is a diffusion on
$\R^d\times\R^d$, actually supported on
$S^{d-1}_{k_1}(\bze)\times\R^d$, over $({\cal T}_1,\mathfrak
Q_{\bbk_1})$.
 Its
generator  is given by
\[
{\cal N}(\bbk,\bbx):={\cal L}F(\bbk,\bbx)+\hat{\bbk}\cdot\nabla_\bbx
F(\bbk,\bbx).
\]
We denote by $P(t,\bbk,\bbx;\cdot)$ its  transition probability.
It satisfies the Fokker-Planck equation
\begin{equation}
\label{80901} \int\limits_0^{+\infty}\int\limits_{\R^d}
\int\limits_{\R^d}(\partial_t-{\cal
N})\vphi(t,\bbk',\bby)P(t,\bbk,\bbx;dt,d\bbk',d\bby)=0,\quad\forall\vphi\in
C_0^\infty((0,+\infty)\times \R^d\times \R^d).
\end{equation}

\begin{lemma}
\label{lm11}
Let $t>0$, $(\bbk,\bbx)\in S^{d-1}_{k}\times \R^d$ ($k=|\bbk|$).
Then, $P(t,\bbk,\bbx;\cdot)$ is absolutely continuous with respect to
the Lebesgue measure on $S^{d-1}_{k}\times \R^d$, with the transition
probability density $p(t,\bbk,\bbx,\cdot,\cdot)$ that is a
$C^\infty$-function. In particular, for any $T,K,\eta>0$ there exists
a constant $C>0$ such that
\begin{equation}\label{72601}
  \max\limits_{t\in[\eta,T]}\max\limits_{(\bbk,\bbx)\in
  S^{d-1}_k\times \overline{B}_K(\bze)
  }P(t,\bbk,\bbx;S^{d-1}\times A)\leq C|A|
\end{equation}
for any $A\subset \overline{B}_K(\bze)$ and $A\in{\cal B}(\R^d)$.
\end{lemma}
\proof
Let $k:=|\bbk|$ and
 $S_i^{(\pm)}:B_k^{d-1}(\bze)\rightarrow S^{d-1}$ be
given by
\[
S_i^{(\pm)}(\bbl):=\mathop{
\underbrace{(l_1,\cdots,\pm\sqrt{k^2-l^2},\cdots,l_{d-1})}}\limits_{i\mbox{-th
component}},\quad \bbl=(l_1,\cdots,l_{d-1})\in
B_k^{d-1}(\bze),\,l:=|\bbl|.\]
Define the measure $P_i^{(\pm)}(t,B\times
A):=P(t,\bbk,\bbx;S_i^{(\pm)}(B)\times A)$, $A\in{\cal B}(\R^d)$,
$B\in {\cal B}(B_k^{d-1}(\bze))$.  The conclusion of the lemma holds
if we can show that each measure $P_i$ possesses a $C^\infty$ smooth
density.  In what follows we consider only the case $i=d$ and denote
$S:=S^{(+)}_d$, $P_S:=P^{(+)}_{d}$.  Note that the respective measure
satisfies the equation \[ (\partial_t-\tilde{\cal N}^*)P_S(t,\cdot)=0
\]
in the distribution sense, with
\[
\tilde{\cal
N}F(\bbl,\bbx):=k^2\sum\limits_{p,q=1}^{d-1}\tilde{D}_{p,q}(\bbl)\partial_{l_p,l_q}^2F(\bbl,\bbx)+
k\sum\limits_{p=1}^{d-1}\tilde{E}_p(\bbl)\partial_{l_p}F(\bbl,\bbx)
\]
\[+\frac{1}{k}\sum\limits_{p=1}^{d-1}l_p\partial_{x_p}
F(\bbk,\bbx)+\sqrt{1-\left(\frac{l}{k}\right)^2}\partial_{x_d}F(\bbk,\bbx),
\] 
where
$\tilde{D}_{p,q}(\bbl)=D_{p,q}(k^{-1}\bbl,k^{-1}\sqrt{k^2-l^2})$,
$\tilde{E}_p(\bbl)=E_p(k^{-1}\bbl,k^{-1}\sqrt{k^2-l^2})$. It
suffices therefore to prove that $\partial_t-\tilde{\cal N}^*$ is
hypoelliptic in order to prove the Lemma. Note that
\[
(\partial_t-\tilde{\cal
N}^*)F=\sum\limits_{p=1}^dX_p^2F+X_0F+a(\bbl)F,\quad\forall\, F\in
C^\infty_0(B_k^{d-1})
\]
where
\begin{eqnarray*}
&&X_p(\bbl):=k\sum\limits_{q=1}^{d-1}\hat
D_{p,q}^{1/2}(\bbl)\partial_{l_q},\quad p=1,\cdots,d-1,\\
&&X_0(\bbl):=\partial_t-\frac{1}{k}\sum\limits_{q=1}^{d-1}l_q\partial_{x_q}-
\sqrt{1-\left(\frac{l}{k}\right)^2}\partial_{x_d}+
\sum\limits_{q=1}^{d-1}a_q(\bbl)\partial_{l_q}
\end{eqnarray*}
and $a(\cdot)$, $a_p(\cdot)$, $p=1,\cdots,d-1$ are
$C^\infty$-smooth functions. It suffices therefore to prove that
for any $(t,\bbl,\bbx)\in \R\times S^{d-1}\times \R^d$ the linear
subspace $\mathfrak L_{t,\bbl,\bbx}$ of the tangent space to
$\R\times S^{d-1}\times \R^d$, spanned by the vector fields
belonging to the Lie algebra $\mathfrak L$ generated by
$X_0,X_1,\cdots,X_d$, is of dimension $2d$. The $(d-1)\times
(d-1)$ matrix $\tilde{\bD}(\cdot):=[\tilde{D}_{p,q}(\bbl)]$, as
well as $\tilde{\bD}^{1/2}(\cdot)$, is non-degenerate in
$B^{d-1}_k(\bze)$ due to Proposition \ref{lm72601} (actually it
degenerates in the limit when $\bl$ approaches $\partial
B^{d-1}_k(\bze)$).
 Hence the vectors $\partial_{l_p}\in \mathfrak
L_{t,\bbl,\bbx}$, $p=1,\cdots,d-1$.

Note also that
\[
[X_0,X_p]=\sum\limits_{q=1}^{d-1}\hat
D_{p,q}^{1/2}(\bbl)\left(\partial_{x_q}+
\frac{l_q}{\sqrt{k^2-l^2}}\partial_{x_d}\right)+
\sum\limits_{q=1}^{d-1}b_q(\bbl)\partial_{l_q},
\]
where $b_p(\cdot)$, $p=1,\cdots,d-1$ are $C^\infty$-smooth
functions. Hence,
$\partial_{x_q}+l_q(k^2-l^2)^{-1/2}\partial_{x_d}\in \mathfrak
L_{t,\bbl,\bbx}$, $q=1,\cdots,d-1$. Furthermore,
\begin{eqnarray*}
&&\sum\limits_{p=1}^{d-1}[[X_0,X_p],X_p]=
-k\left[\mbox{tr}\tilde{\bD}(\bbl)+(\tilde{\bD}(\bbl)\bbl,\bbl)_{\R^d}
(k^2-l^2)^{-1}\right] (k^2-l^2)^{-1/2}\partial_{x_d}\\
&&~~~~~~~~~~~~~~~~~~~~~~~+\sum\limits_{q=1}^{d-1}
d_q(\bbl)\left(\partial_{x_q}+\frac{l_q}{\sqrt{k^2-l^2}}\partial_{x_d}\right)+
\sum\limits_{q=1}^{d-1}c_q(\bbl)\partial_{l_q},
\end{eqnarray*}
where $c_p(\cdot),d_p(\cdot)$, $p=1,\cdots,d-1$ are
$C^\infty$-smooth functions. We can conclude therefore that
$\partial_{x_d}\in \mathfrak L_{t,\bbl,\bbx}$, hence also
$\partial_{x_p}\in \mathfrak L_{t,\bbl,\bbx}$, $p=1,\cdots,d-1$
and finally we also get $\partial_t\in \mathfrak L_{t,\bbl,\bbx}$,
so that the proof of Lemma \ref{lm11} is complete. \endproof

{\bf{Proof of Lemma \ref{lm12}}.} Let
\[
A(N,p):=[S(N,p)\geq T_1+1].\] Choose $p$ sufficiently large so
that
\[
 \mathfrak Q_{\bbk_1,\bbk_2}[A(N,p)]\geq 1-\eta/2.
\]
 This can be done thanks to the continuity property of diffusion
paths. For any $(K_1(\cdot),K_2(\cdot))\in A(N,p)$ we have
\begin{equation}\label{extra2}
\left|Y_1\left(\frac{1}{p}\right)-Y_2\left(\rho\right)\right|\geq
\la_N(p),\mbox{ and
}\left|Y_2\left(\frac{1}{p}\right)-Y_1\left(\rho\right)\right|\geq
\la_N(p)
\end{equation}for all $\rho\in[0,1/p]$,
according to (\ref{81101b}) (see (\ref{extra}) for the definition
of $Y_i(\cdot)$, $i=1,2$ ). Recall also that $\mathfrak
Y_t^{(i)}(K_i)$, $i=1,2$ are defined by (\ref{extra1}).

Let
$V^{(1)}(p,q;K):=\lim\limits_{n\rightarrow+\infty}V_n^{(1)}(p,q;K)$,
where
\[
V_n^{(1)}(p,q;K):=\inf\left\{t\geq\frac{1}{p}:
\mbox{ dist}(Y_1(t),\mathfrak Y_t^{(2)})<\frac{1}{q}+\frac{1}{n}
\right\}\] and likewise we introduce
$V^{(2)}(p,q;K):=\lim\limits_{n\rightarrow+\infty}V_n^{(2)}(q;K)$,
with
\[
V_n^{(2)}(p,q;K):=\inf\left\{t\geq\frac{1}{p}:
\mbox{ dist}(Y_2(t),\mathfrak Y_t^{(1)})<\frac{1}{q}+\frac{1}{n}
\right\}\] Note that $V(p,q;K):=V^{(1)}(p,q;K)\wedge
V^{(2)}(p,q;K)$.

The conclusion of  Lemma \ref{lm12} is then a consequence of the
following.
\begin{lemma}
\label{lm12b} For any $N$ sufficiently large so that
$(\ref{81101})$ holds and $p\geq1$ we have
\begin{equation}
\label{71902b}
\lim\limits_{q\rightarrow+\infty}V^{(i)}(p,q;K)=+\infty\quad\mathfrak
Q_{\bbk_1,\bbk_2}-a.s.\mbox{ on }A(N,p),
\quad\forall\,\bbk_1,\bbk_2\not=\bze,\,\,i=1,2.
\end{equation}
\end{lemma}
\proof With no loss of generality we assume that $i=1$. 
Note  that obviously
\[
V^{(1)}(p,q';K)\geq V^{(1)}(p,q;K)\mbox{ for
}q'\geq q\mbox{ and all }K\in{\cal C}_2(k_1,k_2).
\] For any $K_2$ we
denote by
\[
A(N,p;K_2):=[K_1:\,(K_1,K_2)\in A(N,p)].
\] It suffices
 to show that for $\mathfrak Q_{\bbk_2}$--a.s. $K_2$ we have
\[
\lim\limits_{q\rightarrow+\infty}
V^{(1)}(p,q;K_1,K_2)=+\infty,\quad\mathfrak Q_{\bbk_1}-a.s.\mbox{
on }A(N,p;K_2).
\]
Let us denote
\[
B(t,\bbx;K_2):=\left[K_1:\,|Y_1(t;\bbx)-Y_2(\rho)|\geq
\la_N(p),\,\rho\in[0,1/p]\,\right].
\]
Note that $A(N,p;K_2)\subseteq B\left(\frac{1}{p},\bze;K_2\right)$,
according to (\ref{extra2}).

Let $T_1>0$ be arbitrary.  We show that
\begin{equation}\label{72302}
 \lim\limits_{q\rightarrow+\infty} \mathfrak Q_{\bbk_1}\left[\,V^{(1)}(p,q;\cdot,K_2)\leq T_1,
 \,B\left(\frac{1}{p},\bze;K_2\right)\,\right]=0,\quad \,
 \mathfrak Q_{\bbk_2}-\mbox{a.s. in }
 K_2.
\end{equation}

The expression under the limit in (\ref{72302}) can be estimated
by
\[
\mathfrak Q_{\bbk_1}\left[\,\inf\limits_{u\in[0,T_1]}\mbox{dist }(Y_1(u),
\mathfrak Y_{T_1}^{(2)})\leq
\frac{1}{q},\,B\left(\frac{1}{p},\bze;K_2\right)\,\right]
\]
\[=\mathop{\int\int}\limits_{S^{d-1}_{k_1}\times
[1/p\geq |\bbx|\geq \la_n(p)]}P\left(\frac{1}{p},\bbk_1,\bze,d\bbk,d\bbx\right)\mathfrak Q_{\bbk}
\left[\,\inf\limits_{u\in[0,T_1-1/p]}\mbox{dist }(Y_1(u;\bbx),
\mathfrak Y_{T_1}^{(2)})\leq
\frac{1}{q},\,B(0,\bbx;K_2)\,\right].
\]
Here  we used the Markov property of the process
$(K_1,Y_1)$.
(\ref{72302}) follows if we can show that
\begin{equation}
\label{91701}
 \lim\limits_{q\rightarrow+\infty} \mathfrak Q_{\bbk}
\left[\,\inf\limits_{u\in[0,T_1-1/p]}\mbox{dist }(Y_1(u;\bbx),
\mathfrak Y_{T_1}^{(2)})\leq
\frac{1}{q}\,\right]=0\,
\end{equation}
for every $\bbk\in S^{d-1}_{k_1}$ and $\bbx$ satisfying $1/p\geq
|\bbx|\geq\la_N(p)$, $ \mathfrak Q_{\bbk_2}-\mbox{a.s. in } K_2$.

Suppose first that   $\eta_1:=\frac{1}{2}\mbox{dist}(\bbx,\mathfrak
Y_{T_1}^{(2)})>0$. Then,
\begin{equation}\label{72303}
 \mathfrak Q_{\bbk}\left[\inf\limits_{0\leq u\leq
 \eta_1}\,\mbox{dist}\left(Y_1(u;\bbx),\mathfrak Y_{T_1}^{(2)}\right)\geq
 \frac{1}{q}\right]=1,\quad\forall\,q\geq 4\eta^{-1}.
\end{equation}

Note that the expression under the limit on the left hand side of
(\ref{91701}) can be estimated by
\begin{equation}
\label{72310} \mathfrak Q_{\bbk}\left[\inf\limits_{\eta_1\leq
j/q\leq T_1}\,\mbox{dist}\left(Y_1(j/q;\bx),\mathfrak
Y_{T_1}^{(2)}\right)\leq
 \frac{2}{q}\right]\end{equation}
 \[\leq (T_1+1)q\,\max\limits_{\eta_1\leq j/q\leq
 T_1}\mathfrak Q_{\bbk}\left[Y_1(j/q;\bbx)\in B_{2/q}(\mathfrak
 Y_{T_1}^{(2)})\right].
\]
The right hand side of (\ref{72310}) can be
 estimated, with the help of Lemma \ref{lm11}, by
 \[
 C(\eta_1)(T_1+1)q^{2-d},\quad\forall\,q\geq 4\eta^{-1}_1
 \]
 (recall that $Y_2(\cdot)$ is of $C^1$-class, with $|Y_2'(\cdot)|\leq 1$)
and (\ref{72302}) follows, provided we can prove that
\begin{equation}\label{72311}
  \mathfrak Q_{\bbk_2}\left[\mbox{dist}\left(\bbx,\mathfrak
  Y_{T_1}^{(2)}\right)=0\right]=0
\end{equation}
for $1/p\geq|\bbx|\geq \la_N(p)$
Recall that $|\bbx-Y_2(\rho)|\geq\la_N(p)$, $\rho\in[0,1/p]$.
For any $\rho>0$ we can estimate therefore the
left hand side of (\ref{72311}) by
\[
\frac{T_1+1}{\rho}\max_{1/p\leq j\rho\leq T_1} \mathfrak
Q_{\bbk_2}\left[\,|Y_2(j\rho)-\bbx|\leq 2\rho\,\right]\leq
C(p)(T_1+1)\rho^{d-1}
\]
for some constant $C(p)>0$ depending only on $p$.
Since the last inequality holds for all $\rho>0$ we conclude
(\ref{72311}).
\endproof

An immediate consequence of Lemmas \ref{lm6}, \ref{lm5} and
\ref{lm12} is the following.
\begin{corollary}
\label{cor10}
For any $M,\ep>0$ there exist find sufficiently large $p,q$ and $N$ so
that \[\mathfrak Q_{\bbk_1,\bbk_2}[B(M,N,p,q)]<\ep.\]
\end{corollary}

Choose any $T_0>0$ and $F$ a bounded and continuous functional over
${\cal C}_2$ that is ${\cal M}_2^{0,T_0}$-measurable.  We show that
\begin{equation}
\label{71601}
\limsup\limits_{\delta\rightarrow0}\bbE\left[F(\bmk^{(\delta)}_1(\cdot),\bmk^{(\delta)}_2(\cdot))\right]\leq
\int F(K(\cdot))\mathfrak Q_{\bbk_1,\bbk_2}(dK).
\end{equation}
This, in fact, implies weak convergence of the laws of
$(\bmk^{(\delta)}_1(\cdot),\bmk^{(\delta)}_2(\cdot))$
over ${\cal C}_2$, as $\delta\rightarrow0$.

Fix $\eta>0$ and choose
$M>0$ such that $M-1$ satisfies (\ref{80401}). Then, by virtue of
Lemma \ref{lm6}
\begin{equation}\label{71602}
\mathfrak Q_{\bbk_1,\bbk_2}[T(M-1)\leq T_0]=0.
\end{equation}
Let
 $p,q$ be such that
\begin{equation}
\label{71602b}
\mathfrak Q_{\bbk_1,\bbk_2}[B(M,N,p,q)]\leq \eta.
\end{equation}
Note that $\overline{B}(M-1,N-1,p,q-1)\subseteq B(M,N,p,q)$.

Let $\delta_n\rightarrow0$, then we can choose a subsequence, that
we still denote as $(\delta_n)$, such that the laws of
$(\bl^{(\delta_n)}_1(\cdot),\bl^{(\delta_n)}_2(\cdot))$ converge
over ${\cal C}_2$, as $n\rightarrow+\infty$, to a certain
$Q(\cdot;M,N,p,q)$. We have
\begin{eqnarray}
\label{71603}
&&\limsup\limits_{n\rightarrow+\infty}\bbE\left[F(\bmk^{(\delta_n)}_1(\cdot),\bmk^{(\delta_n)}_2(\cdot))\right]\\
&&\leq
\limsup\limits_{n\rightarrow+\infty}\bbE\left[F(\bl^{(\delta_n)}_1(\cdot),\bl^{(\delta_n)}_2(\cdot));
W(\bl^{(\delta_n)}_1(\cdot),\bl^{(\delta_n)}_2(\cdot);M-1,N-1,p,q-1)>
T_0\right]\nonumber\\
&&+\limsup\limits_{n\rightarrow+\infty}\left|\bbE\left[F(\bmk^{(\delta_n)}_1(\cdot),\bmk^{(\delta_n)}_2(\cdot));
W(\bl^{(\delta_n)}_1(\cdot),\bl^{(\delta_n)}_2(\cdot);M-1,N-1,p,q-1)
\leq T_0\right]\right|.\nonumber
\end{eqnarray}
The second term on the right hand side of (\ref{71603}) can be
estimated by
\begin{equation}
\label{71606}
 \|F\|_{L^\infty}\left(\limsup\limits_{n\rightarrow+\infty}\bbP\left[
T(\bl^{(\delta_n)}_1(\cdot),\bl^{(\delta_n)}_2(\cdot);M-1)\leq
T_0\right]
+\mathfrak Q_{\bbk_1,\bbk_2}\left[
\overline{B}(M-1,N-1,p,q-1)\,\right]\right)
\end{equation}
Note also that
\begin{eqnarray}
&&\!\!\!\!\!\limsup\limits_{n\rightarrow+\infty} \left|\bbP\left[
T(\bl^{(\delta_n)}_1(\cdot),\bl^{(\delta_n)}_2(\cdot);M-1)>T_0
\right]-\bbP\left[
W(\bl^{(\delta_n)}_1(\cdot),\bl^{(\delta_n)}_2(\cdot);M-1,N-1,p,q-1)>
T_0\right]\right|\nonumber
\\
&&\leq \limsup\limits_{n\rightarrow+\infty}
 \bbP\left[(\bl^{(\delta_n)}_1(\cdot),\bl^{(\delta_n)}_2(\cdot))\in
 \overline{B}(M-1,N-1,p,q-1)\right]\label{71607}\\
 &&
 \leq \mathfrak Q_{\bbk_1,\bbk_2}\left[
 \overline{B}(M-1,N-1,p,q-1)\right]\leq \eta\nonumber
\end{eqnarray}
and hence
\begin{eqnarray}
\label{71608} &&\limsup\limits_{n\rightarrow+\infty} \bbP\left[
T(\bl^{(\delta_n)}_1(\cdot),\bl^{(\delta_n)}_2(\cdot);M-1)\leq T_0
\right]\\
&& \!\!\!\!\!\!\!=1-\liminf\limits_{n\rightarrow+\infty}
\bbP\left[
T(\bl^{(\delta_n)}_1(\cdot),\bl^{(\delta_n)}_2(\cdot);M-1)> T_0
\right] \leq\mathfrak Q_{\bbk_1,\bbk_2}\left[
W(K;M-1,N-1,p,q-1)\leq T_0\right]+\eta.\nonumber
\end{eqnarray}
The first expression on the utmost right hand side of
(\ref{71608}) is less than or equal to
\begin{equation}
\label{71610}
\mathfrak Q_{\bbk_1,\bbk_2}\left[
T(K;M-1)\leq
T_0\right]+\mathfrak Q_{\bbk_1,\bbk_2}\left[
 \overline{B}(M-1,N-1,p,q-1)\right]\leq\eta
\end{equation}
according to (\ref{71602}) and (\ref{71602b}). Summarizing, the
expression in (\ref{71606}) can be estimated by $2\eta\|F\|_{L^\infty}$

The first term on the right hand side of (\ref{71603}) can be
estimated by
\[
\int F(K(\cdot))\bone_{[W(K;M,N,p,q)>T_0]} \mathfrak
Q_{\bbk_1,\bbk_2}(dK)\leq \int F(K(\cdot)) \mathfrak
Q_{\bbk_1,\bbk_2}(dK)
\]
\[
+\| F\|_{L^\infty}\mathfrak
Q_{\bbk_1,\bbk_2}[W(K;M,N,p,q)\leq T_0]\leq\int F(K(\cdot)) \mathfrak
Q_{\bbk_1,\bbk_2}(dK)+2\eta\| F\|_{L^\infty}.
\]
The last estimate follows from an analogous estimate to
(\ref{71610}). Summarizing, since $\eta>0$ is arbitrary   we
 conclude (\ref{71601}).

\begin{appendix}
\section{Proof of Lemma \ref{lem-drop-eps}}\label{app:lem-drop-eps}

We define
\begin{equation}
d_\eps^\delta(\vxi,\vx_0)=\bv_\eps^{\delta,B}-\tilde\bv_\eps^{\delta,B}=
\int
e^{i\bk\cdot(\vxi-\vy)}\left[W_\eps^\delta(\vx_0+\frac{\eps(\vxi+\vy)}{2},\bk)-
W_\eps^\delta(\vx_0,\bk)\right]\Gamma_0\bS_0(\vy)\frac{d\bk
d\vy}{(2\pi)^d}
\end{equation}
and split $d_\eps^\beta=\sum_{j=1}^3 d_\eps^{\delta,j}$ according
to the decomposition
\begin{eqnarray*}
&&W_\eps^\delta(\vx_0+\frac{\eps(\vxi+\vy)}{2},\bk)-
W_\eps^\delta(\vx_0,\bk)=
\big(W_\eps^\delta(\vx_0+\frac{\eps(\vxi+\vy)}{2},\bk)-
U^\delta(\vx_0+\frac{\eps(\vxi+\vy)}{2},\bk)\big)\\
&&\qquad\qquad +\big(U^\delta(\vx_0+\frac{\eps(\vxi+\vy)}{2},\bk)
-U^\delta(\vx_0,\bk)\big)
+\big(U^\delta(\vx_0,\bk)-W_\eps^\delta(\vx_0,\bk)\big).
\end{eqnarray*}
Here $U^\delta=\sum u_q^\delta\Pi_q$ is the semi-classical
approximation of $W_\eps^\delta$, with $u_q^\delta$ the solutions
of the Liouville equations. The last term may be estimated as
\begin{eqnarray*}
\int\|d_\eps^{\delta,3}(\vxi,\vx_0)\|^2 d\vx_0&\le&\int\left\|\int
e^{i\bk\cdot\vxi}[U^\delta(\vx_0,\bk)-W_\eps^\delta(\vx_0,\bk)]
\Gamma_0\hat\bS_0(\bk)\frac{d\bk}{(2\pi)^d}\right\|^2d\vx_0\\
&\le&C\|\bS_0\|_{L^2}^2\int\|U^\delta(\vx_0,\bk)-W_\eps^\delta(\vx_0,\bk)\|^2
{d\vx_0d\bk}\to 0
\end{eqnarray*}
as ${\cal K}_\mu\ni(\eps,\delta)\to 0$ with $C$ independent of $\xi$.
The Fourier transform of the first term in $\vx_0$ is
\begin{eqnarray*}
&&\hat d_\eps^{\delta,1}(\vxi;\bp)=\int
e^{-i\bp\cdot\vx_0+i\bk\cdot(\vxi-\vy)}
f_\eps^\delta(\vx_0+\frac{\eps(\vxi+\vy)}{2},\bk)\Gamma_0\bS_0(\vy)\frac{d\vy
d\bk d\vx_0}{(2\pi)^d}\\
&&= \int e^{i\bk\cdot(\vxi-\vy)+i\eps\bp\cdot(\vxi+\vy)/2}\hat
f_\eps^\delta(\bp,\bk)\Gamma_0\bS_0(\vy)\frac{d\vy
d\bk}{(2\pi)^d}
=\int e^{i(\bk+\eps\bp/2)\cdot\vxi}\hat
f_\eps^\delta(\bp,\bk)\Gamma_0\hat\bS_0(\bk-\frac{\eps\bp}{2})\frac{d\bk}{(2\pi)^d},
\end{eqnarray*}
where
$f_\eps^\delta(\vx,\bk)=W_\eps^\delta(\vx,\bk)-U^\delta(\vx,\bk)$.
Therefore we have using the Cauchy-Schwartz inequality
\begin{eqnarray*}
\int\|\hat d_\eps^{\delta,1}(\vxi;\bp)\|^2d\bp\le
C\|\bS_0\|_{L^2}\|f_\eps^\delta\|_{L^2}\to 0
\end{eqnarray*}
as ${\cal K}_\mu\ni(\eps,\delta)\to 0$ with $C$ independent of $\xi$.
Finally, the Fourier transform of $d_\eps^{\delta,2}$ is
\begin{eqnarray*}
\hat d_\eps^{\delta,2}(\vxi;\bp)&=&\int
e^{-i\bp\cdot\vx_0+i\bk\cdot(\vxi-\vy)}
[U^\delta(\vx_0+\frac{\eps(\vxi+\vy)}{2},\bk)-U^\delta(\vx_0,\bk)]\Gamma_0
\bS_0(\vy)\frac{d\vy
d\bk d\vx_0}{(2\pi)^d}\\
&=&\int
e^{i\bk\cdot(\vxi-\vy)}\left[e^{i\eps\bp\cdot(\vxi+\vy)/2}-1\right]\hat
U^\delta(\bp,\bk)\Gamma_0\bS_0(\vy)\frac{d\vy d\bk}{(2\pi)^d}\\
&=&\int e^{i\bk\cdot\vxi}\hat
U^\delta(\bp,\bk)\Gamma_0\left[ e^{i\eps\bp\cdot\vxi/2}
 \hat\bS_0(\bk-\frac{\eps\bp}{2})-\hat\bS_0(\bk)\right]\frac{d\bk}{(2\pi)^d}.
\end{eqnarray*}
We write $e^{i\eps\bp\cdot\vxi/2}=(e^{i\eps\bp\cdot\vxi/2}-1)+1$ and
decompose $\hat d_\eps^{\delta,2}(\vxi;\bp)$ as
$I_1(\vxi;\bp)+I_2(\vxi;\bp)$ accordingly.  We have for the second
term
\begin{eqnarray*}
&&\int\|I_2(\vxi;\bp)\|^2d\bp\le
C\int\left(\int \|\hat
U^\delta(\bp,\bk)\|\|\hat\bS_0(\bk-\frac{\eps\bp}{2})-
\hat\bS_0(\bk)\|\frac{d\bk}{(2\pi)^d}\right)^2d\bp\\
&&\le C\int\left(\int\|\hat
U^\delta(\bp,\bk)\|^2d\bk\right)\left(\int\|\hat\bS_0(\vl-\frac{\eps\bp}{2})-
\hat\bS_0(\vl)\|^2{d\vl}\right)d\bp.
\end{eqnarray*}
Note that
\[
\int\|\hat\bS_0(\vl-\frac{\eps\bp}{2})-\hat\bS_0(\vl)\|^2{d\vl}\le
\eps^2\|\bp\|^2\int\left(\int_0^1\|\nabla\bS_0(\vl-\eps\bp
s)\|^2ds\right)d\vl\le\eps^2\|\bp\|^2\|\nabla\bS_0\|^2_{L^2}
\]
and hence
\begin{eqnarray*}
&&\int\|I_2(\vxi;\bp)\|^2d\bp\le\eps^2\|\nabla\bS_0\|^2_{L^2}
\|U^\delta\|^2_{H^1}\to 0
\end{eqnarray*}
as ${\cal K}_\mu\ni(\eps,\delta)\to 0$ according to Lemma
\ref{lemma-liov-regular}.  It remains to bound the $L^2$ norm of
$I_1(\bp;\vxi)$. We derive two estimates according as $\xi$ is small
or large. The first estimate is
\begin{displaymath}
  \begin{array}{rcl}
|I_1(\bxi;\bp)| &\leq & C \dint e^{i\bk\cdot\bxi}
|\hat U^\delta(\bp,\bk)| \eps |\bp\cdot\bxi| |\hat \bS_0(\bk-\frac{\eps\bp}{2})|
   \frac{d\bk}{(2\pi)^d} \\
 &\leq & C \eps |\xi| \|\hat U^\delta(\bp,\bk)\|_{L^2_{\bk}(\bp)|\bp|}
  \|\bS_0\|_2,
  \end{array}
\end{displaymath}
so that
\begin{displaymath}
  \dint |I_1(\bxi;\bp)|^2d\bp \leq \eps^2|\bxi|^2 
    \|U^\delta(\bp,\bk)\|^2_{H^1(\Rm^d_{\bx}; L^2(\Rm^d_{\bk}))}
    \|\bS_0\|^2_2.
\end{displaymath}
At the same time using integrations by parts we get 
\begin{displaymath}
  I_1(\bxi;\bp) = \dint \dfrac{i}{|\bxi|}e^{i\bk\cdot\bxi}
    \hat\bxi\cdot\nabla_{\bk} \big(\hat U^\delta(\bp,\bk)
   \Gamma_0 \hat \bS_0(\bk-\frac{\eps\bp}{2}) \big)
    (e^{i\eps\bp\cdot\vxi/2}-1)\frac{d\bk}{(2\pi)^d}.
\end{displaymath}
This shows that 
\begin{displaymath}
  \dint |I_1(\bxi;\bp)|^2d\bp \leq \dfrac{C}{|\bxi|^2}
  \|U\|_{L^2(\Rm^d_{\bx}; H^1(\Rm^d_{\bk}))}^2 \| (1+|\bx|^2)^{1/2}\bS_0 \|^2_2.
\end{displaymath}
With these estimates, we obtain that
\begin{displaymath}
  \dint |I_1(\bxi;\bp)|^2d\bp \leq C \min (h_\eps^\delta|\bxi|^2,|\bxi|^{-2})
\end{displaymath}
with $h^\delta_\eps\to0$ as ${\cal K}_\mu\ni(\eps,\delta)\to0$.  This
implies that $\int |I_1(\bxi;\bp)|^2d\bp\to0$, hence $\int |\hat
d^{\delta,2}_\eps(\bxi;\bp)|^2d\bp\to0$ as ${\cal
  K}_\mu\ni(\eps,\delta)\to0$ uniformly with respect to $\bxi\in\Rm^d$
and concludes the proof of Lemma \ref{lem-drop-eps}.

\section{Proof of Lemma \ref{evol-weps}}\label{app:lem-evol-weps}

We may recast (\ref{symm-version-intro}) as
\begin{equation}\label{c-0}
\pdr{\bv_\eps^\delta}{t}+
c_\delta(\vx)D^j\pdr{\bv}{x^j}+\pdr{c_\delta}{x^j}\left[
\ve_{j}\otimes\ve_{d+1}\right]\bv_\eps^\delta=0.
\end{equation}
Thanks to calculations of the form (\ref{eq:diffadelta}), this is
equivalent to the equation
\begin{eqnarray}\label{eq:waves}
\eps\pdr{\bv_\eps^\delta}{t} + P_\eps^{\delta,W}(\bx,\eps D_\bx)
\bv_\eps^\delta=0,
\end{eqnarray}
where the symbol $P_\eps^{\delta}$ is given by (\ref{P-eps-delta}).
We recall that the pseudo-differential Weyl operator $P^W(\bx,\eps D)$
associated to a symbol $P(\vx,\bk)$ is defined by Weyl's quantization
rule
\begin{equation}\label{eq:PepsW}
P^W(\bx,\eps D_\bx) u =\dint_{\Rm^{2d}}
e^{i(\bx-\vy)\cdot\bk} P(\dfrac{\bx+\vy}{2},\eps\bk) u(\by)
\dfrac{d\vy d\bk}{(2\pi)^d}.
\end{equation}
The fact that (\ref{eq:waves}) is equivalent to (\ref{c-0})
is verified by a straightforward calculation:
\begin{eqnarray*}
&&P_0^{\delta,W}(\vx,\eps D)\bv_\eps^\delta(\vx)=\int e^{i(\vx-\vy)\cdot\bxi}
P_0^\delta\left(\frac{\vx+\vy}{2},\eps\bxi\right)\bv_\eps^\delta(\vy)
\frac{d\bxi d\vy}{(2\pi)^d}\\
&&=i\eps\int e^{i(\vx-\vy)\cdot\bxi}c_\delta\left(\frac{\vx+\vy}{2}\right)
D^j\xi_j\bv_\eps^\delta(\vy)
\frac{d\bxi d\vy}{(2\pi)^d}=\eps\int c_\delta\left(\frac{\vx+\vy}{2}\right)
D^j\bv_\eps^\delta(\vy)
\left(-\pdr{}{y_j}\delta(\vx-\vy)\right)d\vy\\
&&=\eps\pdr{}{y_j}\left.\left[c_\delta\left(\frac{\vx+\vy}{2}\right)
D^j\bv_\eps^\delta(\vy)\right]\right|_{\vy=\vx}
=\eps c_\delta(\vx)D^j\pdr{\bv_\eps^\delta(\vx)}{x_j}
+\frac {\eps}2\pdr{c_\delta(\vx)}{x_j}D^j
\bv_\eps^\delta(\vx)\\
&&=-\eps \pdr{\bv_\eps^\delta}{t}
+\frac{\eps}{2}\pdr{c_\delta(\vx)}{x_j}[-\ve_j\otimes\ve_{d+1}+
\ve_{d+1}\otimes\ve_{j}]\bv_\eps^\delta=-\eps \pdr{\bv_\eps^\delta}{t}
-\eps P_1^\delta(\vx)\bv_\eps^\delta(\vx)
\end{eqnarray*}
and now (\ref{eq:waves}) follows because
$P_1^{\delta,W}(\vx)\bv_\eps^\delta(\vx)=
P_1^\delta(\vx)\bv_\eps^\delta(\vx)$ since $P_1^\delta(\vx)$ is
independent of $\bk$.

The associated Cauchy problem for the Wigner transform $\tilde W_\eps^\delta$
with a fixed $\zeta$ is given by
\begin{equation}
  \label{eq:wigner}
  \begin{array}{l}
\eps \pdr{\tilde W_\eps^\delta}{t} +
\tilde W[P_\eps^{\delta,W}(\bx,\eps D_\bx)\bv_\eps^\delta,\bv_\eps^\delta]
+\tilde W[\bv_\eps^\delta,P_\eps^{\delta,W}(\bx,\eps D_\bx)\bv_\eps^\delta]=0\\
  \tilde W_\eps^\delta(0,\bx,\bk))=\tilde W_\eps^\delta(0,\bx,\bk;\zeta),
  \end{array}
\end{equation}
where the Wigner transform of two different fields is defined by
\begin{equation}\label{eq:defwignerbis}
\tilde W[\phi_\eps,\psi_\eps](\bx,\bk)= \dint_{\Rm^d} e^{i\bk\cdot\vy}
\phi_\eps(\bx-\dfrac{\eps\vy}{2})\psi_\eps^*(\bx+\dfrac{\eps\vy}{2})
\frac{d\vy}{(2\pi)^d}.
\end{equation}
We deduce from the definitions of $\tilde W_\eps$ and $P^W_\eps$ that
\begin{eqnarray}
&&\quad\tilde W[P_\eps^{\delta,W}(\bx,\eps D_\bx)\bv_\eps^\delta,
\bv_\eps^\delta](\bx,\bk)
\nonumber\\
&&= \dint e^{i\bk\cdot\vy}
   (P^W_\eps(\bx,\eps D_\bx)\bv_\eps^\delta)(\bx-\dfrac{\eps\vy}{2})
    \bv_\eps^{\delta*}(\bx+\dfrac{\eps\vy}{2})\frac{d\vy}{(2\pi)^d}\nonumber\\
&&= \dint e^{i\bk\cdot\vy}  e^{i(\bx-\frac{\eps\vy}{2}-\bz)\cdot\bq}
  P_\eps^\delta(\dfrac{\bx-\frac{\eps\vy}{2}+\bz}{2},\eps\bq)
\bv_\eps^\delta(\bz) \bv_\eps^{\delta*}(\bx+\dfrac{\eps\vy}{2})
\frac{d\vy d\bz d\bq}{(2\pi)^{2d}}\nonumber\\
&&=
  \dint e^{i\bk\cdot\vy}  e^{i(\bx-\frac{\eps\vy}{2}-\bz)\cdot\bq}
   P_\eps^\delta(\dfrac{\bx-\frac{\eps\vy}{2}+\bz}{2},\eps\bq)
 e^{-i\frac{\bp}{\eps}\cdot(\bx+\frac{\eps\vy}{2}-\bz)}
   W_\eps^\delta(\dfrac{\bx+\frac{\eps\vy}{2}+\bz}{2},\bp)
\frac{d\bp d\vy d\bz d\bq}{(2\pi)^{2d}} \label{wpw}\\
&&=\dint e^{i\vy\cdot(\bk-\bp)} e^{i 2(\bx-\bz)\cdot(\bq-\bp/\eps)}
P_\eps^\delta(\bz-\dfrac{\eps\vy}{2},\eps\bq) W_\eps^\delta(\bz,\bp)
\frac{d\bz d\bp d\vy d\bq}{(\pi\sqrt{2})^{2d}} \nonumber\\
&&=\!\dint\! P_\eps^\delta(\vy,\bq) W_\eps^\delta(\bz,\bp)
e^{\frac{2i}{\eps}((\bp-\bk)\cdot\vy+
(\bq-\bp)\cdot\bx+(\bk-\bq)\cdot\vz}
\frac{d\bz d\bp d\vy d\bq}{(\pi{\eps})^{2d}}=
\!\dint\! P_\eps^\delta(\vy,\bq) W_\eps^\delta(\bz,\bp)e^{i\phi}
\frac{d\bz d\bp d\vy d\bq}{(\pi{\eps})^{2d}}.\nonumber
\end{eqnarray}
Moreover, the matrix $\tilde W_\eps^\delta$ is self-adjoint, while
$W[f_\eps,g_\eps]= W_\eps^*[g,f]$ for any pair of functions $f$ and
$g$, and the symbol $P_\eps$ is skew-symmetric. Thus (\ref{wpw}) and
(\ref{eq:wigner}) imply that the pure Wigner transform $\tilde W_\eps$
satisfies (\ref{wigeps-evol}), and hence so does $W_\eps$.  Moreover,
the function $\phi$ satisfies an anti-symmetry relation
\begin{displaymath}
  \phi(\bx,\bz,\bk,\bp;\vy,\bq)=-\phi(\bz,\bx,\bp,\bk;\vy,\bq).
\end{displaymath}
Then, using the fact that $W_\eps$ is self-adjoint we obtain
\begin{eqnarray*}
&&\dint_{\Rm^{2d}} \tr{({\cal L}_\eps^\delta W_\eps^\delta)
W_\eps^{\delta*}}d\bx d\bk\\
&&=
\dint_{\Rm^{6d}} \tr{P_\eps^\delta(\vy,\bq)
W_\eps^\delta(\bz,\bp) W_\eps^\delta(\bx,\bk)
e^{i\phi}-W_\eps(\bz,\bp)P_\eps^\delta(\vy,\bq)
W_\eps^\delta(\bx,\bk)e^{-i\phi}}
\frac{d\vx d\bk d\vy d\vz d\bp d\bq}{(\pi\eps)^{2d}}\\
&&= \dint_{\Rm^{6d}}\tr{P_\eps^\delta(\vy,\bq) W_\eps^\delta(\bz,\bp)
W_\eps^\delta(\bx,\bk)e^{i\phi} - P_\eps^\delta(\vy,\bq) W_\eps^\delta(\bz,\bp)
W_\eps^\delta(\bx,\bk) e^{i\phi}}=0,
\end{eqnarray*}
where we interchanged $\bx\!\leftrightarrow\!\bz$ and
$\bk\!\leftrightarrow\!\bp$ in the second term on the last line, and
used the anti-symmetry of $\phi$. This implies conservation of the
$L^2$-norm (\ref{L2-preserve}). Note that (\ref{trace-preserve})
follows immediately from (\ref{wigeps-evol}) and the proof of Lemma
\ref{evol-weps} is complete.

\section{Regularity of the Liouville equations}
\label{sec:reg-liouv}

We prove Lemma \ref{lemma-liov-regular} in this Appendix.  We recall
that the functions $u_q^\delta$ satisfy the evolution equations
\begin{eqnarray}
  \label{eq:liouv}
 &&   \pdr{u_{q}^\delta}{t} + \{\lambda_q^\delta,u_{q}^\delta\} = 0\\
 && u_{q}^0=u_q(t=0)=\hbox{Tr}[\Pi_q W_0\Pi_q].\nonumber
\end{eqnarray}
These equations can be solved by following the Hamiltonian flow
generated by $\lambda_q^\delta$. More precisely, let us define for $T$,
$\bx$, $\bk$ given, the trajectories
\begin{eqnarray}
      \dr{\vX(t)}{t} &=& -\pdr{\lambda_q^\delta}{\bk}(\bX(t),\bK(t)) ,\qquad
    \vX(0)=\bx \nonumber\\
    \dr{\bK(t)}{t} &=& \pdr{\lambda_q^\delta}{\bx}(\bX(t),\bK(t)),\qquad
    \bK(0)=\bk.  \label{eq:traject}
\end{eqnarray}
Then solution of (\ref{eq:liouv}) is given by
\begin{equation}
  \label{eq:W0char}
  u_{q}^\delta(T,\bx,\bk)= u_q^0(\vX(T,\vx,\bk),\bK(T,\vx,\bk)).
\end{equation}
The flow (\ref{eq:odes}) preserves the Hamiltonian
$\lambda_q^\delta(\vx,\bk)$ and the initial data $u_q^0$ is supported on
a compact set $S$. Therefore the set
\[
{\cal S}=\bigcup_{t\ge 0,\delta\in(0,1]}\hbox{supp}~u_q^\delta(t,\vx,\bk)
\]
is compact because the speed $c_\delta(\vx)$ is uniformly bounded from
above and below. Furthermore
\[
\nabla u_q^\delta=
D^{\delta*}\nabla u_q^0,~~\|u_q(t)\|_{\dot H^1}\le \|u_{q}^0\|_{\dot H^1}
\|D^\delta(t)\|_\infty
\]
where $D^\delta(t,\vx,\bk)$ is the Jacobian matrix,
$D^{\delta,i}_{j}=\partial Z_i^\delta/\partial z_j$, with $\det
D^\delta(t)\equiv 1$, $\bZ=(\bX,\bK)$, and $\bz=(\bx,\bk)$.  To
simplify notation, we do not write explicitly the dependence of
$D^\delta$ and its derivatives with respect to the eigenvalue label
$q$ in the sequel.  Here we define
\[
\|D^\delta\|_\infty=
\sup_{(\vx,\bk)\in{\cal S}}
\left(\hbox{Tr}[D^\delta(\vx,\bk)D^{\delta*}(\vx,\bk)]\right)^{1/2}.
\]
More generally, given a tensor $T_{j_1j_2\dots j_m}$ we denote
\[
\|T\|_\infty=\sup_{(\vx,\bk)\in{\cal S}}
\left(\sum_{j_1,\dots,j_m}|T_{j_1\dots j_m}|^2\right)^{1/2}.
\]
We will also use the matrix norm $|A|$ that is dual to the Euclidean
norm on ${\mathbb R}^d$ and is equal to the square root of the largest
eigenvalue of the matrix $AA^*$, and denote
\[
|A|_\infty=\sup_{(\vx,\bk)\in{\cal S}}|A(\vx,\bk)|.
\]
Furthermore, we have
\[
\frac{\partial^2u_q^\delta}{\partial z_j\partial z_p}=
\frac{\partial^2 Z_m^\delta}{\partial z_j\partial z_p}\pdr{u_q^0}{z_m}+
\pdr{Z_m^\delta}{z_j}\pdr{Z_r^\delta}{z_p}
\frac{\partial^2u_q^0}{\partial z_m\partial z_r}
\]
so that
\begin{eqnarray*}
&&\sum_{j,p}\left|\frac{\partial^2u_q^\delta}
{\partial z_j\partial z_p}\right|^2\le
2\frac{\partial^2 Z_m^\delta}{\partial z_j\partial z_p}\pdr{u_q^0}{z_m}
\frac{\partial^2 Z_l^\delta}{\partial z_j\partial z_p}\pdr{u_q^0}{z_l}+
2\pdr{Z_m^\delta}{z_j}\pdr{Z_r^\delta}{z_p}
\frac{\partial^2u_q^0}{\partial z_m\partial z_r}
\pdr{Z_l^\delta}{z_j}\pdr{Z_s^\delta}{z_p}
\frac{\partial^2u_q^0}{\partial z_l\partial z_s}\\
&&\le
2\sum_{m,j,p}\left|\frac{\partial^2 Z_m^\delta}
{\partial z_j\partial z_p}\right|^2
\|\nabla u_q^0\|^2+
2\sum_{l,s}\left|\frac{\partial^2u_q^0}{\partial z_l\partial z_s}\right|^2
\sum_{m,j}|D_{mj}^\delta|^2
\end{eqnarray*}
and hence
\[
\|u_q^\delta(t)\|_{\dot H^2}\le 2\|u_{q}^0\|_{\dot H^1}
\|D_2^\delta(t)\|_{\infty}+
2\|u_{q}^0\|_{\dot H^2}\|D^\delta(t)\|_{\infty}^2,
\]
with $D_{2,jl}^{\delta,m}=\partial^2Z_m^\delta/\partial z_j\partial z_l$.
We observe that
\begin{eqnarray*}
&&\frac{\partial^3 u_q^\delta}{\partial z_j\partial z_p\partial z_s}=
\frac{\partial^3 Z_m^\delta}{\partial z_j\partial z_p\partial z_s}
\pdr{u_q^0}{z_m}+
\frac{\partial^2 Z_m^\delta}{\partial z_j\partial z_p}
\pdr{Z_l^\delta}{z_s}\frac{\partial^2u_q^0}{\partial z_m\partial z_l}+
\frac{\partial^2 Z_m^\delta}{\partial z_j\partial z_s}
\pdr{Z_r^\delta}{z_p}\frac{\partial^2u_q^0}{\partial z_m\partial z_r}\\
&&~~~~~~~~~~~~~~~+
\pdr{Z_m^\delta}{z_j}\frac{\partial^2 Z_r^\delta}{\partial z_p\partial z_s}
\frac{\partial^2u_q^0}{\partial z_m\partial z_r}
+\pdr{Z_m^\delta}{z_j}\pdr{Z_r^\delta}{z_p}\pdr{Z_l^\delta}{z_s}
\frac{\partial^3 u_q^0}{\partial z_m\partial z_r\partial z_l}.
\end{eqnarray*}
Therefore we have
\begin{eqnarray*}
&&\sum_{j,p,s}\left|
\frac{\partial^3 u_q^\delta}{\partial z_j\partial z_p\partial z_s}\right|^2\le
5\sum_{m,j,p,s}\left|
\frac{\partial^3 Z_m}{\partial z_j\partial z_p\partial z_s}\right|^2
\|\nabla u_q^0 \|^2+15\|D(t)\|_\infty^2\|D_2(t)\|_\infty^2\sum_{m,n}
\left|\frac{\partial^2u_q^0}{\partial z_m\partial z_n}\right|^2\\
&&~~~~~~~~~~~~~~~~~~~~~~~+5
\|D(t)\|_\infty^6\sum_{m,r,l}
\left|\frac{\partial^3 u_q^0}{\partial z_m\partial z_r\partial z_l}\right|^2
\end{eqnarray*}
so that
\[
\|u_q^\delta(t)\|_{\dot H^3}\le
3\left[\|u_{q}^0\|_{\dot H^1}\|D_3^\delta(t)\|_{\infty}+
3\|u_{q}^0\|_{\dot H^2}\|D_2^\delta(t)\|_{L^\infty}\|D^\delta(t)\|_{\infty}+
\|u_{q}^0\|_{\dot H^3}\|D^\delta(t)\|_{\infty}^3\right]
\]
with $D_{3,jlp}^{\delta,m}=
\partial^3Z_m^\delta/\partial z_j\partial z_l\partial z_p$.

It thus remains to estimate the matrices $D^\delta$, $D^\delta_2$, and
$D^\delta_3$.  The matrix $D^\delta$ satisfies the differential
equation
\[
\frac{dD^\delta}{dt}=F^\delta D^\delta,~~~F^\delta=\left(\begin{matrix}
-\dfrac{\partial^2\lambda_q^\delta}{\partial k_i\partial x_j}&
-\dfrac{\partial^2\lambda_q^\delta}{\partial k_i\partial k_j} \cr
\dfrac{\partial^2\lambda_q^\delta}{\partial x_i\partial x_j}&
\dfrac{\partial^2\lambda_q^\delta}{\partial x_i\partial k_j}\cr
                            \end{matrix}\right),~~D^\delta(0)=I.
\]
Therefore we have
\begin{eqnarray*}
&&\frac{d}{dt}\hbox{Tr}[D^\delta D^{\delta*}]=
2\hbox{Tr}[F^\delta D^\delta D^{\delta*}]=
2\sum F_{il}^\delta D^{\delta,l}_{m}D^{\delta,m}_{i}\\
&&\le 2\sum F_{il}^\delta\left(\sum_{k}|D^{\delta,l}_{k}|^2\right)^{1/2}
\left(\sum_{p}|D^{\delta,p}_{i}|^2\right)^{1/2}
\le 2|F|\hbox{Tr}[D^\delta D^{\delta*}]
\end{eqnarray*}
so that
$\|D^\delta(t)\|_{\infty}\le \exp\left(|F^\delta|_{\infty}t\right)$ and hence
\[
\|u_q^\delta(t)\|_{\dot H^1}\le \|u_{q}^0\|_{\dot H^1}
\exp\left(|F^\delta|_{\infty}t\right).
\]
Differentiating (\ref{eq:odes}) once again we obtain
\[
\frac{d D_{2,jk}^{\delta,i}}{dt}=
\pdr{F_{il}^\delta}{z_m}D_{k}^{\delta,m}D_{j}^{\delta,l}+
F_{il}D_{2,jk}^{\delta,l},
\]
so that along each characteristic
\[
\frac{1}{2}\frac{d}{dt}\|D_{2}^\delta\|^2=
\pdr{F_{il}^\delta}{z_m}D_{k}^{\delta,m}
D_{j}^{\delta,l}D_{2,jk}^\delta+
F_{il}^\delta D_{2,jk}^{\delta,l}D_{2,jk}^{\delta,i}
\le\|F_2^\delta\|_\infty\|D^\delta\|^2\|D_2^\delta\|+
|F^\delta|_\infty\|D_2^\delta\|^2,
\]
where $F_{2,ijk}^\delta=\partial F_{ij}^\delta/\partial
z_k$. Furthermore, initially at $t=0$ we have $D_2^\delta(0)=0$. Therefore we
obtain
\[
\|D_2^\delta(t)\|_{L^\infty}\le\frac{\|F_2^\delta\|_\infty}{|F^\delta|_\infty}
\exp(2|F^\delta|_\infty t).
\]
and thus
\[
\|u_q^\delta(t)\|_{\dot H^2}\le
2\left(\frac{\|F_2^\delta\|_{\infty}}{|F^\delta|_{\infty}}
\|u_q^0\|_{\dot H^1}+\|u_q^0\|_{\dot H^2}\right)\exp(2|F^\delta|_{\infty}t)\le
2\left(\frac{\|F_2^\delta\|_{\infty}}{|F^\delta|_{\infty}}+1\right)
\|u_q^0\|_{H^2}\exp(2|F^\delta|_{\infty}t).
\]
Similarly, the tensor $D_3^\delta$ satisfies the ordinary differential
equation
\[
\frac{d D_{3,jkm}^{\delta,i}}{dt}=F_{3,ilnp}^\delta D_{m}^{\delta,p}
D_{k}^{\delta,n}D_{j}^{\delta,l}+
F_{2,iln}^\delta D_{2,km}^{\delta,n}D_{j}^{\delta,l}
+F_{2,iln}^\delta D_{k}^{\delta,n}D_{2,jm}^{\delta,l}+
F_{2,iln}^\delta D_{m}^{\delta,n}D_{2,jk}^{\delta,l}+
F_{il}^\delta D_{3,jkm}^{\delta,l}
\]
so that along each characteristic
\begin{eqnarray*}
&&\frac{1}{2}\frac{d}{dt}\|D_{3}^\delta\|^2
=F_{3,ilnp}^\delta D_{m}^{\delta,p}D_{k}^{\delta,n}D_{j}^{\delta,l}
D_{3,jkm}^{\delta,i}+
F_{2,iln}^\delta D_{2,km}^{\delta,n}D_{j}^{\delta,l}D_{3,jkm}^{\delta,i}\\
&&+F_{2,iln}^\delta D_{k}^{\delta,n}D_{2,jm}^{\delta,l}D_{3,jkm}^{\delta,i}+
F_{2,iln}^\delta D_{m}^{\delta,n}D_{2,jk}^{\delta,l}D_{3,jkm}^{\delta,i}
+F_{il}^\delta D_{3,jkm}^{\delta,l}D_{3,jkm}^{\delta,i}\\
&&\le
\|F_3^\delta\|\|D^\delta\|^3\|D_3^\delta\|+
3\|F_2^\delta\|\|D^\delta\|\|D_2^\delta\|\|D_3^\delta\|+
|F^\delta|\|D_3^\delta\|^2,
\end{eqnarray*}
where $F_{3,ijkn}^\delta=\partial F_{2,ijk}^\delta/\partial z_n$, and
at $t=0$ we have $D_3(0)=0$. Therefore we obtain
\[
\|D_{3}^\delta(t)\|_\infty\le
\left(\frac{\|F_3^\delta\|_\infty}{|F^\delta|_\infty}+
\frac{3\|F_2^\delta\|_\infty^2}{|F^\delta|_\infty^2}\right)
\exp(3|F^\delta|_\infty t)
\]
and thus
\begin{eqnarray*}
&&\|u_q^\delta(t)\|_{\dot H^3}\le
\left[\left(\frac{\|F_3^\delta\|_\infty}{|F^\delta|_\infty}+
\frac{3\|F_2^\delta\|_\infty^2}{|F^\delta|_\infty^2}\right)\|u_q^0\|_{\dot H^1}
+3\frac{\|F_2^\delta\|_\infty}{|F^\delta|_\infty}\|u_q^0\|_{\dot H^2}+
\|u_q^0\|_{\dot H^3}\right]\exp(3|F^\delta|_\infty t)\\
&&~~~~~~~~~~~~~\le 6\left(\frac{\|F_3^\delta\|_\infty}{|F^\delta|_\infty}+
\frac{\|F_2^\delta\|_\infty^2}{|F^\delta|_\infty^2}+1\right)
\|u_q^0\|_{H^3}\exp(3|F^\delta|_\infty t).
\end{eqnarray*}
This completes the proof of Lemma \ref{lemma-liov-regular} because
$\bar\gamma_\delta=|F^\delta|_\infty$.

\end{appendix}

\end{document}